\documentclass[english,master]{diploma}
\usepackage[autostyle=true]{csquotes} 
\usepackage[backend=biber, style=iso-numeric, alldates=iso]{biblatex} 
\usepackage{dcolumn} 
\usepackage[cpp]{diplomalst}
\usepackage{amsmath}
\usepackage{tikz}
\usepackage{braket}
\usepackage{amssymb}

\usepackage{acronym}

\usepackage{qcircuit }
\usepackage{booktabs}
\usepackage{todonotes}

\usepackage{graphicx}
\usepackage{subcaption}
\usepackage{float}
\usepackage{svg}

\usepackage{hyperref}
\usepackage{cleveref}
\usetikzlibrary{positioning}
\crefname{figure}{Fig.}{Figs.}
\Crefname{figure}{Fig.}{Figs.}

\crefname{table}{Tab.}{Tabs.}
\Crefname{table}{Tab.}{Tabs.}

\crefname{section}{Sec.}{Secs.}
\Crefname{section}{Sec.}{Secs.}

\crefname{equation}{Eq.}{Eqs.}
\Crefname{equation}{Eq.}{Eqs.}

\usetikzlibrary{arrows.meta}
\tikzset{
    block/.style={rectangle, draw, fill=blue!10, text width=25em, text centered, rounded corners, minimum height=3em},
    arrow/.style={thick,->,>=Stealth}
}

\ThesisAuthor{Bc. Silvie Illésová}

\ThesisSupervisor{prof. RNDr. Marek Lampart, Ph.D.}

\CzechThesisTitle{Využití kvantových vrstev v klasických neuronových sítích}

\EnglishThesisTitle{Leveraging Quantum Layers in Classical Neural Networks}

\SubmissionYear{2025}

\ThesisAssignmentFileName{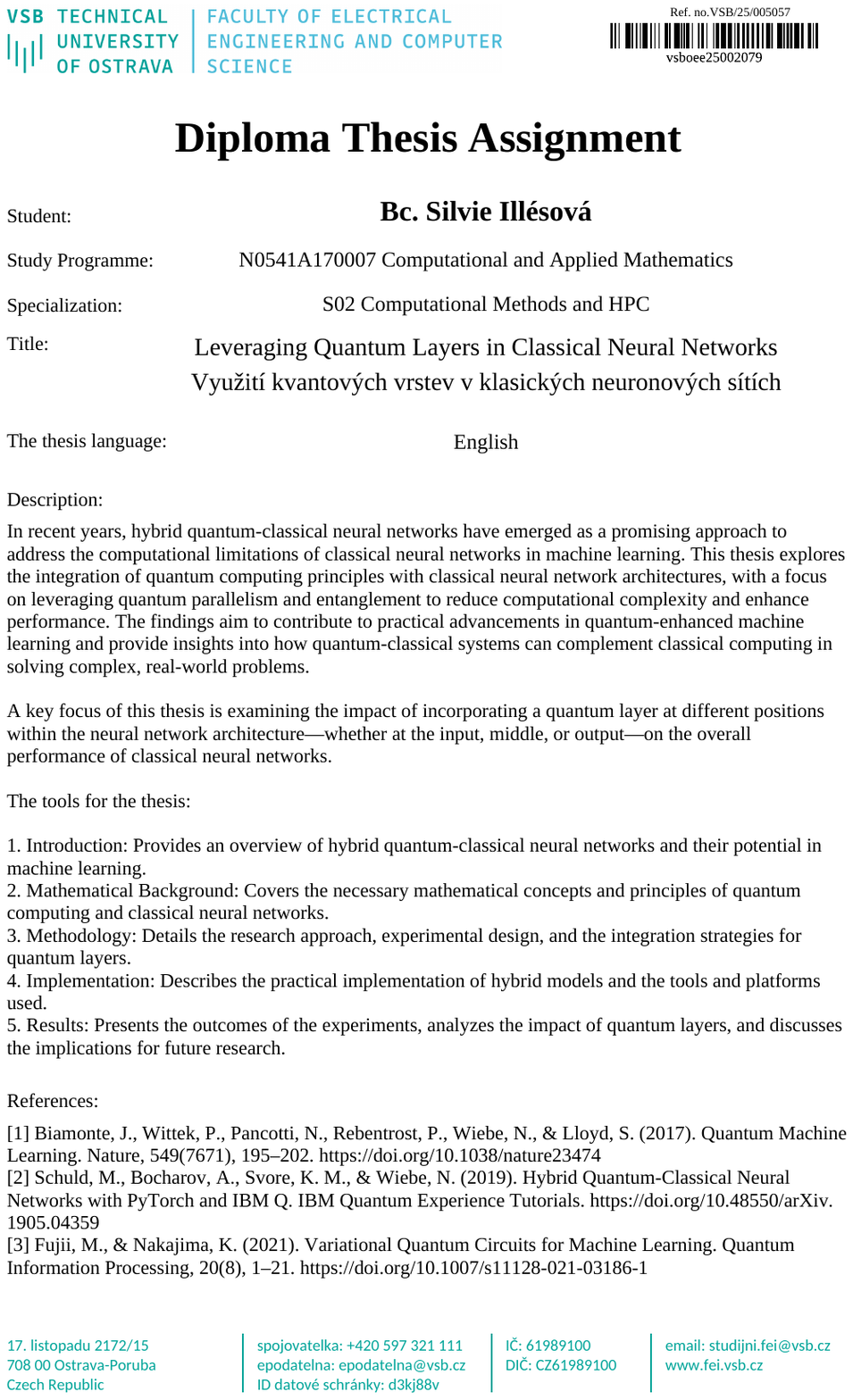}

\Acknowledgement{I would like to thank my family for their unwavering support.  Their constant presence and care made this accomplishment possible. I am particularly grateful to Ing. Martin Beseda, Ph.D., for his exceptional guidance, encouragement, and generous support throughout this project.}

\CzechAbstract{Hybridní kvantově-klasické neuronové sítě představují slibnou oblast ve snaze o vylepšení modelů strojového učení. Tato diplomová práce se zabývá integrací kvantových vrstev do klasických architektur konvolučních neuronových sítí s cílem využít kvantové provázání a mapování příznaků pro zvýšení schopností učení. Je zde představena podrobná metodologie konstrukce a trénování těchto hybridních modelů s využitím frameworků PyTorch a Qiskit Machine Learning. Experimenty zkoumají vliv zařazení kvantových vrstev na různá místa neuronové sítě. Výsledky naznačují, že kvantové komponenty mohou i s omezeným počtem qubitů přinášet významné transformace, což motivuje k dalšímu výzkumu škálovatelného kvantového strojového učení. Kompletní implementace je veřejně dostupná a budoucí práce se zaměří na rozšíření experimentálních vyhodnocení a publikaci dalších výsledků.}

\CzechKeywords{hybridní neuronové sítě, kvantové výpočty, variační kvantové obvody, kvantové mapování příznaků}

\EnglishAbstract{Hybrid quantum-classical neural networks represent a promising frontier in the search for improved machine learning models. This thesis explores the integration of quantum layers within classical convolutional neural network architectures, aiming to leverage quantum entanglement and feature mapping to enhance learning capabilities. A detailed methodology for constructing and training such hybrid models is presented, using PyTorch and Qiskit Machine Learning frameworks. Experiments investigate the performance impact of inserting quantum layers at different stages of the neural network pipeline. The results suggest that quantum components can introduce meaningful transformations even with a limited number of qubits, motivating further research into scalable quantum machine learning. The full implementation is made publicly available, and future work will focus on expanding experimental evaluations and publishing additional findings.}
\EnglishKeywords{hybrid neural networks, quantum computing, variational quantum circuits, quantum feature mapping}

\newcolumntype{d}[1]{D{.}{.}{#1}}

\begin{document}

\MakeTitlePages

\listoffigures
\clearpage

\listoftables
\clearpage

\chapter*{List of Symbols and Abbreviations}
\addcontentsline{toc}{chapter}{List of symbols and abbreviations}
\begin{acronym}
    \acro{cnot}[CNOT]{Controlled-NOT}
    \acro{pca}[PCA]{Principal Component Analysis}
    \acro{qnn}[QNN]{Quantum Neural Network}
    \acro{qml}[QML]{Quantum Machine Learning}
    \acro{nisq}[NISQ]{Noisy Intermediate-Scale Quantum}
    \acro{vqcs}[VQCs]{Variational Quantum Circuits}
    \acro{ml}[ML]{Machine Learning}
    \acro{cnns}[CNNs]{Convolutional Neural Networks}
    \acro{vqe}[VQE]{Variational Quantum Eigensolver}
    \acro{vha}[VHA]{Variational Hamiltonian Ansatz}
    \acro{saoovqe}[SA-OO-VQE]{State-Averaged Orbital-Optimized VQE}
    \acro{hqcnn}[HQCNN]{Hybrid Quantum-Convolutional Neural Network}
\end{acronym}
\newpage

\chapter*{Introduction}\addcontentsline{toc}{chapter}{\protect\numberline{}Introduction}
The rapid advancement of quantum computing in recent years has opened up new possibilities for enhancing classical machine learning algorithms. Quantum phenomena such as superposition, entanglement offer fundamentally new ways to represent and process information, which can potentially address some of the limitations inherently present in classical neural networks. Among the emerging approaches, hybrid quantum-classical neural networks have attracted significant attention, combining the strengths of both computational paradigms to build more powerful models.

This thesis explores the integration of quantum layers into classical neural network architectures, specifically convolutional neural networks. The goal is to investigate whether leveraging quantum operations within the structure of established machine learning models can enhance their representational capabilities, reduce computational complexity, or improve generalization and flexibility, especially for challenging learning tasks.

A particular focus of this work is to examine the impact of incorporating a quantum layer at different points within the neural network—at the input, intermediate, or output stage—and to evaluate how these placements affect the model's performance. To this end, hybrid architectures are designed where the classical feature extraction stages are followed by quantum variational circuits, constructed using feature maps and parameterized ansatzes tailored for machine learning tasks.

The research is conducted using Python as the primary programming language, employing the PyTorch framework for the classical neural network components and Qiskit's Machine Learning modules for the quantum elements. The complete implementation is made available as an open-source release on Zenodo and Gitlab, promoting reproducibility and enabling future research in hybrid quantum-classical learning.

This thesis is structured as follows: \cref{chap:theory} presents the theoretical background necessary for understanding quantum computing, classical machine learning, and hybrid models. \cref{chap:method} describes the methodology and the specific hybrid architectures designed and implemented. \cref{chap:res} presents the experimental results and evaluates the performance of the hybrid models compared to classical baselines. Finally, conclusions and perspectives for future research are discussed in \cref{chap:conc}.
\chapter{Theory}\label{chap:theory}
This chapter provides the theoretical background necessary for understanding the principles underpinning this work. It covers fundamental concepts in quantum computing, including qubits, quantum circuits, and measurement, as well as the basics of classical machine learning. Special emphasis is placed on hybrid quantum-classical models, quantum feature maps, and variational quantum circuits, which form the core components of the hybrid architectures investigated in this thesis.
\section{Quantum Computing Paradigm}

Quantum computing is a novel technology that stems from the principles of quantum mechanics. Whereas in classical computing, we rely on binary bits as the atomic unit, which means, that at the lowest level, we have either $0$ or $1$, as our computational base, in quantum computing we utilize quantum bits (qubits). These can exist in a superposition of the two edge cases simultaneously. This characteristic gives us several advantages, which are discussed in the following sections. But first, we shall focus on the several computational paradigms that have emerged. That said, the most notable are gate-based quantum computing, quantum annealing, and boson sampling. And each of these paradigms is better suited for different classes of problems, and also, uses distinct technological solutions.

\subsection{Gate-Based Quantum Computing}
The most widely considered quantum computing paradigm is that of gate-based quantum computing, which is also sometimes referred to as the quantum circuit model. To make a parallel to classical computing, in logical circuits, we use logical gates to act on the input, and after the series of them, we obtain results. In gate-based quantum computing, we employ quantum gates to manipulate our qubits and obtain the desired result. Quantum gates are unitary operations, and the space of qubits is a Hilbert space. The gates allow us to construct complex quantum circuits, that we use to perform arbitrary quantum computations.

For the gate-based paradigm, we can see the power of quantum computing in examples of quantum algorithms that outperform their classical counterpart. If we have a look at Shor's algorithm \cite{monz2016realization, yimsiriwattana2004distributed}, it offers os an exponential speedup for the case of integer factorization, and Grover's algorithm \cite{long2001grover,jozsa1999searching,mandviwalla2018implementing} on the other hand provides us with a quadratic speedup for search problems. We also know, that gate-based computers are universal, from which we can summarize, that in fact, we can simulate any model given a sufficient number of qubits and also with a great enough number of quantum gates.

In the last decade, a significant progess in the developement of gate-based hardware has been made. Several companies, including IBM \cite{ibm_quantum}, Google \cite{google_quantum_ai} and Rigetti \cite{rigetti_computing}, have demonstrated quantum computers with tens, hundreds, even more than a thousand qubits.  Nevertheless, practical large-scale implementations face considerable challenges, including qubit decoherence, gate infidelity, and the overhead of quantum error correction \cite{corcoles2019challenges}.

\subsection{Quantum Annealing}

The next quantum paradigm that we introduce is the quantum annealing paradigm \cite{hauke2020perspectives, yulianti2022implementation}. This one is distinct in a way that it focuses primarily on solving optimization problems. At the core of it are the principles of adiabatic quantum computation, where we start with a quantum system that is prepared in the ground state of the Hamiltonian \cite{kato1951fundamental}, and then we slowly evolve the system into the ground state of a more complex Hamiltonian in which the problem solution is encoded \cite{albash2018adiabatic}. And, according to the adiabatic theorem, if said evolution is sufficiently slow and the gap between the ground state and the first excited state remains large enough during the whole evolution, then the system will remain in the ground state for the whole process, thus allowing us to obtain the optimal solution.

The approach of this paradigm is, in particular, effective for combinatorial optimization problems, such as the Ising model minimization \cite{kadowaki1998quantum} or the problem of traveling salesman \cite{martovnak2004quantum}. Because of this, we have several applications in the fields like machine learning \cite{nath2021review}, material science \cite{camino2023quantum} or operations research \cite{yarkoni2022quantum}. For the quantum annealers themselves, D-Wave \cite{dwave} has developed many commercially available with thousands of qubits, offering a platform for real-world applications.

The drawback of this approach is that unlike the gate-based paradigm it is not universal, and remains tailored to the specific problem sets. In addition, the question of the extent of the computational advantages that annealers are capable of giving us over the classical algorithms still remains open.

\subsection{Boson Sampling}

The paradigm of Boson sampling \cite{gard2015introduction} is yet another non-universal quantum computing model. Here we operate with the principles of linear optics, which involves putting indistinguishable bosons, most typically single photons, through a linear interferometer that is composed of beam splitters and phase shifters. There the bosons undergo quantum interference and we measure the distribution of the detections at the interferometer's output.

The core of this idea is the fact that calculating the output distribution of such a system is computationally challenging for classical computers because they need to calculate the permanent of large matrices. This problem is known to be $\#P-hard$ \cite{papadimitriou2003computational}, where $\#P-hard$ refers to problems that are at minimum as hard as the most difficult in $\#P$. The $\#P$ consists of counting the number of solutions to NP class problems. The  $\#P-hard$  problems are usually more difficult to solve than Np-complete ones and it is thought that they are not solvable in polynomial time. So even though boson sampling does not allow us to make general-purpose calculations, it is theoretically capable of quantum supremacy, meaning that on specific tasks, it is capable of outperforming classical machines.

The biggest appeal of boson sampling thus lies in the relative experimental simplicity and also, in its ability to benchmark quantum advantage. As was demonstrated in 2020 \cite{zhong2020quantum}, using boson sampling we can solve tasks that are believed to be infeasible in a realistic amount of time for classical supercomputers.

\section{Qubit}
A quantum bit, or qubit in short, is the fundamental unit of quantum information, having the same role as a classical bit in classical computing. The main difference is that qubits can exist in a linear combination of two discrete states. This phenomenon is known as quantum superposition. This is the property that allows us to encode information and work with it in a way, that is impossible on classical systems. This means that the superposition is the basis for quantum computing speedups \cite{renner2022computational}.

\subsection{Dirac Notation}
When describing the qubit mathematically, we are oftentimes working with the Dirac notation \cite{borrelli2010dirac}, which is also known as the bra-ket notation. This formalism allows us to efficiently write down quantum states. We call a quantum state a ket, which is a complex vector from Hilbert space. For one single qubit, we can define the computational basis states as $\ket{0}$ and $\ket{1}$. With the basis states defined, we can write any pure state of a single qubit as 
\begin{equation}\label{eq:qubitlinearcomb}
    \ket{\psi}=\alpha\ket{0}+\beta\ket{1},
\end{equation}
where $\alpha, \beta \in \mathbb{C}$. The \cref{eq:qubitlinearcomb} shows us that a pure quantum state is a linear combination of basis states. We also know, that
\begin{equation}
    |\alpha|^2 + |\beta|^2  = 1,
\end{equation}

Which is the normalization condition for those two complex numbers.

Here $|\alpha|^2$ and $|\beta|^2$ are the probabilities of measuring our qubit in either one of the basis states. The Dirac notation in the end simplifies the representation of quantum states and operations.

\subsection{Superposition}
One of the key features that distinguishes qubits from their classical counterparts is the superposition \cite{grover1998advantages}. As we said before, a qubit can exist in a state, that is a combination of the basis states. This allows us to process different possibilities in parallel until we perform a measurement and the quantum state collapses in either $\ket{0}$ or $\ket{1}$. 

The most often-used example of one qubit superposition is a quantum state of
\begin{equation}
    \ket{\psi} = \frac{1}{\sqrt{2}}\left(\ket{0} + \ket{1}\right)
\end{equation}
which is an equal superposition of both basis states, because the probability of measuring the quantum state in $\ket{0}$ is equal to the probability of the quantum state collapsing to $\ket{1}$ as shown here

\begin{equation}
    p_{\ket{0}} = \left|\frac{1}{\sqrt{2}}\right|^2 = \frac{1}{2} = p_{\ket{1}}.
\end{equation}

However, to fully exploit the power of quantum computing, entanglement \cite{horodecki2009quantum} is essential. Entanglement is a uniquely quantum phenomenon where the state of one qubit cannot be described independently of the state of another, no matter the distance between them. 

\subsection{Bloch Sphere}

\begin{figure}
    \centering
    \includegraphics[width=0.5\linewidth]{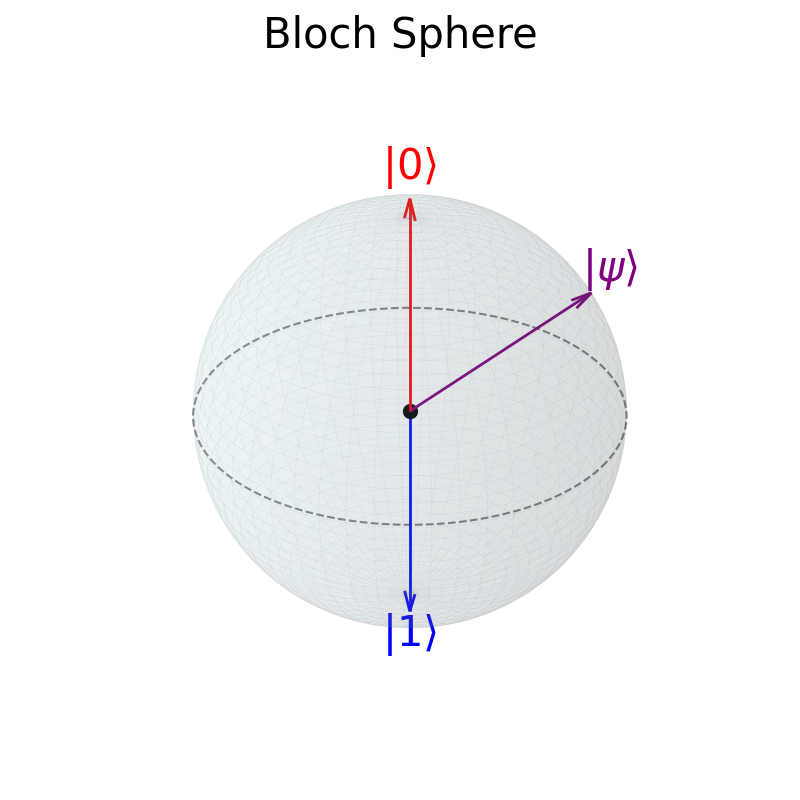}
    \caption{Bloch Sphere}
    \label{fig:bloch_sphere}
\end{figure}

One of the ways that we can represent geometrically our single qubit's quantum state is to use the Bloch sphere \cite{glendinning2005bloch}. The Bloch sphere maps the pure quantum state to the point on the surface of a unit sphere in three dimensions. With this representation, $\ket{0}$ is mapped to the north pole of the sphere, and in turn, $\ket{1}$ is assigned to the sphere's southern pole. All other points on the sphere correspond to different superpositions. The Bloch sphere along with the state in superposition is shown in \cref{fig:bloch_sphere}.

Mathematically we can write any pure single qubit state as

\begin{equation}
    \ket{\psi} = \cos\left(\frac{\theta}{2}\right) \ket{0} + e^{i\phi} \sin\left(\frac{\theta}{2}\right) \ket{1},
\end{equation}

where $\theta \in [0, \pi]$ and $ \phi \in [0, 2\pi)$.

This particular representation is useful, when we need to visualize how single qubit gates rotate our quantum state. It is also a good tool for demonstrating quantum decoherence and the effects of noise in our quantum system.

\section{Quantum Operators}
When working with qubits, we use quantum operators as a mathematical framework for different computations. The quantum operators are represented as matrices, and they act on quantum states within the aforementioned Hilbert Space. These operators enable us to transform the quantum states. In practice, each operator is associated with a quantum gate, which is a component that realizes the particular operation in the quantum circuit.

\subsection{Single-Qubit Operators}
For transformations that affect individual qubits, we say that we are deploying single-qubit operators. These are operating on a two-dimensional complex vector space. The most basic single-qubit operators are the so-called Pauli operators \cite{pauli2012general}. In matrix form, we call the Pauli matrices, $X$, $Y$, and $Z$ respectively. These matrices correspond to bit-flip, bit-and-phase-flip, and phase-flip operations. The Pauli-X gate can find its classical counterpart in the logical NOT gate because it swaps from one basis state to the other.

The Pauli matrices are defined as
\begin{align}\label{eq:paulis}
X &= \begin{pmatrix}
0 & 1 \\
1 & 0
\end{pmatrix},\\
Y &= \begin{pmatrix}
0 & -i \\
i & 0
\end{pmatrix},\\
Z &= \begin{pmatrix}
1 & 0 \\
0 & -1
\end{pmatrix}.
\end{align}

With the Pauli matrices \cref{eq:paulis} we can manipulate both the amplitude and the phase of a single qubit state.

Another important single-qubit operator is the Hadamard operator defined as

\begin{equation}
    H = \frac{1}{\sqrt{2}} \begin{pmatrix}
        1 & 1\\
        1 & -1
    \end{pmatrix}.
\end{equation}

The Hadamard operator is necessary for quantum parallelism as it places our single qubit into a superposition of the basis states $\ket{0}$ and $\ket{1}$. T
As for other single-qubit gates, we will introduce two other gates that introduce global and relative phase shifts respectively, their matrix representations are

\begin{align}
    S &= \begin{pmatrix}
        1 & 0\\
        0 & i
    \end{pmatrix},\\
    T  &= \begin{pmatrix}
        1 & 0\\
        0 & e^{i\frac{\pi}{4}}
    \end{pmatrix}.
\end{align}

\subsection{Multi-qubit operators}
As we already know, single-qubit gates allow us to manipulate the quantum state of the single-qubit, but having a quantum machine only capable of execution of single-qubit gates would not be enough for the computer to be universal. For that, we also need multi-qubit operators. These can act on two or more qubits simultaneously and without them, we would not be able to implement any execute any complex quantum algorithm.

The one multi-qubit operator that is necessary for gate-based computing is the \ac{cnot} operator. When using this operator, we talk about applying it to two qubits, the control, and the target qubit, and the operator acts on the target qubit and flips the state if and only if the control qubit is currently in state $\ket{1}$ otherwise, it leaves the target qubit unchanged.
Now we can have a look at the matrix form of \ac{cnot} operator
\begin{equation}
    CNOT = \begin{pmatrix}
        1 & 0 & 0 & 0 \\
        0 & 1 & 0 & 0 \\
        0 & 0 & 0 & 1 \\
        0 & 0 & 1 & 0 
    \end{pmatrix}.
\end{equation}

The important piece of information is that this is the operator that entangles the quantum states of two qubits.

\subsection{Unitary and Hermitian Matrices}

Both the single-qubit and multi-qubit quantum operators can be represented as unitary matrices. Matrix $U$ is unitary if the following unitary condition holds:
\begin{equation}\label{eq:unitarycond}
    U^\dagger U = U U^\dagger = I, 
\end{equation}
in \cref{eq:unitarycond} $U^\dagger$ is the Hermitian conjugate of matrix $U$ and $I$  is the identity matrix.

We work with unitary operators, because they ensure the preservation of the total probability, meaning that the norm of the state vector remains the same through the entire calculation. It further guarantees that the process of quantum evolution is reversible.

Now, many of the observables in quantum mechanics are represented by Hermitian operators, which have real eigenvalues. Hermitian matrix $H$ satisfies the following condition.

\begin{equation}
    H = H^\dagger.
\end{equation}

And while quantum gates must be unitary, we can create them from our Hermitian observables by exponentiation of them \cite{schwinger1960unitary}. An example of the process is the phase rotation around the Z-axis of the Bloch sphere. This can be achieved by
\begin{equation}
    R_z(\theta) = e^{-i\theta Z/2} = \cos\left(\frac{\theta}{2}\right) I - i \sin\left(\frac{\theta}{2}\right) Z.
\end{equation}

Thus, we see that Hermitian matrices form generators of unitary evolution. And by this process, we can ink quantum gates straightforwardly with the physical observables from quantum mechanics.

\subsection{Universal Gate Sets}
In reality, quantum computers are technologically limited and are able to realize aa finite set of quantum gates. And as we need for the quantum computer to be universal, we need for this basis gate set to be also universal.
So we define the universal gate set \cite{brylinski2002universal, sawicki2022universality} as a finite collection of quantum gates from which we can approximate any unitary operation and any number of qubits to any precision that we desire.

The universality of our gate set is crucial because when designing complex quantum algorithms, we need to have the capability to create complex unitary transformations using our simpler and well-defined components. To be considered a universal gate set, a gate set must satisfy the following two conditions: it must allow for all arbitrary single-qubit operations and also it must include at least one gate that is capable of creating entanglement between two qubits. These two conditions allow us to both manipulate singular qubits, as well as to create entanglement between any of our qubits.

\section{Quantum Circuits}
Quantum circuits \cite{chiribella2008quantum} allow us to visualize quantum algorithms in a model, that shadows how classical logical circuits are designed. It consists of qubits, and quantum gates, that are applied to the qubits in a given sequence. This model is central to the theory of quantum computing and also provides a practical way to implement quantum algorithms on real quantum hardware.

\subsection{Depth, Width, and Complexity}

Now that we know that quantum circuits consist of qubits and quantum gates, let's have a look at three fundamental attributes that characterize them. These attributes are the width of quantum circuits, their depth, and complexity.

The width of the quantum circuit refers to the number of qubits used during the calculation and circuits with different widths are shown in \cref{fig:width-circuits}. 

\begin{figure}
\centering

\begin{minipage}[b]{0.3\textwidth}
\centering
\[
\Qcircuit @C=1em @R=1em {
\lstick{\ket{q_0}} & \gate{H} & \qw 
}
\]
\caption*{(a) Width-1: Hadamard gate}
\end{minipage}
\hfill
\begin{minipage}[b]{0.3\textwidth}
\centering
\[
\Qcircuit @C=1em @R=1em {
\lstick{\ket{q_0}} & \gate{H} & \ctrl{1}  & \qw \\
\lstick{\ket{q_1}} & \qw      & \targ      & \qw 
}
\]
\caption*{(b) Width-2: Bell state}
\end{minipage}
\hfill
\begin{minipage}[b]{0.35\textwidth}
\centering
\[
\Qcircuit @C=0.8em @R=1em {
\lstick{\ket{q_0}} & \gate{H} & \ctrl{1} & \ctrl{2} & \qw      & \qw      & \qw \\
\lstick{\ket{q_1}} & \qw      & \gate{R_2} & \qw    & \gate{H} & \ctrl{1} & \qw \\
\lstick{\ket{q_2}} & \qw      & \qw      & \gate{R_3} & \qw      & \gate{R_2} & \gate{H} \\
}
\]
\caption*{(c) Width-3: 3-Qubit QFT}
\end{minipage}

\caption{Examples of quantum circuits with increasing width.}
\label{fig:width-circuits}
\end{figure}

When using real quantum machines, the width of the circuit is limited by the total number of qubits of said machine.

The depth of the quantum circuit refers to the number of layers of quantum gates that are in the circuit. Gate that operate on individual qubits or different subsets of qubits, can be executed in parallel and thus they belong to the same layer. The example of three circuits with three qubits and different depths is depicted in \cref{fig:depth-comparison}.

\begin{figure}
\centering
\begin{minipage}[b]{0.3\textwidth}
\centering
\[
\Qcircuit @C=1em @R=1em {
\lstick{\ket{q_0}} & \gate{H} & \qw \\
\lstick{\ket{q_1}} & \gate{H} & \qw \\
\lstick{\ket{q_2}} & \gate{H} & \qw \\
}
\]
\caption*{Depth-1: Parallel Hadamard gates}
\label{fig:depth1}
\end{minipage}
\hfill
\begin{minipage}[b]{0.3\textwidth}
\centering
\[
\Qcircuit @C=1em @R=1em {
\lstick{\ket{q_0}} & \gate{H} & \ctrl{1} & \ctrl{2} & \qw \\
\lstick{\ket{q_1}} & \qw      & \targ    & \qw      & \qw \\
\lstick{\ket{q_2}} & \qw      & \qw      & \targ    & \qw \\
}
\]
\caption*{Depth-2: GHZ state preparation}
\label{fig:depth2}
\end{minipage}
\hfill
\begin{minipage}[b]{0.3\textwidth}
\centering
\[
\Qcircuit @C=0.8em @R=1em {
\lstick{\ket{q_0}} & \gate{H} & \ctrl{1} & \ctrl{2} & \qw      & \qw      & \qw \\
\lstick{\ket{q_1}} & \qw      & \gate{R_2} & \qw    & \gate{H} & \ctrl{1} & \qw \\
\lstick{\ket{q_2}} & \qw      & \qw      & \gate{R_3} & \qw      & \gate{R_2} & \gate{H} \\
}
\]
\caption*{Depth-4: 3-qubit QFT}
\label{fig:depth4}
\end{minipage}

\caption{Three different 3-qubit quantum circuits of increasing depth.}
\label{fig:depth-comparison}
\end{figure}

The depth of the circuit comes into play when we begin to consider decoherence and sampling noise, as circuits with larger depths are more prone to errors. Also, if we can do so, reducing the depth of the circuit helps to minimize the errors that accumulate in it over time \cite{temme2017error}.
 
When talking about overall complexity of the circuit \cref{fig:depth3-complexity}, we refer to the gate count, which is the number of all the gates that are applied to the qubits in the circuit.

\begin{figure}
\centering

\begin{minipage}[b]{0.3\textwidth}
\centering
\[
\Qcircuit @C=1em @R=1em {
\lstick{\ket{q_0}} & \gate{H} & \ctrl{1} & \qw \\
\lstick{\ket{q_1}} & \qw      & \targ    & \gate{Z} \\
}
\]
\caption*{Low complexity: 3 gates }
\end{minipage}
\hfill
\begin{minipage}[b]{0.3\textwidth}
\centering
\[
\Qcircuit @C=1em @R=1em {
\lstick{\ket{q_0}} & \gate{H} & \ctrl{1} & \gate{T} \\
\lstick{\ket{q_1}} & \gate{S} & \targ    & \gate{Z} \\
}
\]
\caption*{Medium complexity: 5 gates}
\end{minipage}
\hfill
\begin{minipage}[b]{0.3\textwidth}
\centering
\[
\Qcircuit @C=1em @R=1em {
\lstick{\ket{q_0}} & \gate{H} & \gate{H} & \gate{T} \\
\lstick{\ket{q_1}} & \gate{S} & \gate{H}    & \gate{Z}\\   
}
\]
\caption*{Higher complexity: 6 gates}
\end{minipage}

\caption{Three circuits with the same depth and width, but increasing complexity.}
\label{fig:depth3-complexity}
\end{figure}

 These metrics—depth, width, and size—are all important in determining whether a quantum algorithm is efficient and implementable.

\subsection{Compilation}
When we talk about compilation \cite{maronese2022quantum, botea2018complexity} in the context of quantum computing, we refer to the process of the translation of high-level quantum algorithm into the sequence of low-level operation that our chosen quantum hardware platform is capable of executing. It is important to note that quantum programs are often written using abstract gate sets, that may not directly correspond to the basis set of our specific hardware, or we want to run the same quantum program on multiple machines with different basis sets, this specific part is called the transpilation \cite{younis2022quantum} of quantum circuit.

Therefore, the compilation process must take into account, the constraint of the selected hardware, which includes the native gate set, but also limited connectivity between qubits due to the specific quantum chip architecture and also different gate fidelities. The primary goal we have during compilation is to optimize the circuit for running on the selected quantum hardware, and during this optimization, we consider several different aspects, such as total gate count, circuit depth, and also the expected accumulated errors.

During the process of compilation, several different and sophisticated techniques are employed. Into these we count, circuit rewriting \cite{maslov2005quantum}, qubit routing \cite{wagner2023improving}, but also noise-aware optimization \cite{alam2020noise, murali2019noise}.  All of these techniques are typically employed together to achieve an executable quantum circuit, that takes into account the specifics of the physical quantum hardware.

\subsection{Gate Decomposition}

Another process that we use with circuit compilation is gate decomposition \cite{mottonen12006decompositions}. This involves the expression of complex quantum operations as a sequence of simple, elementary gates. As quantum hardware has a limited number of basis gates, it is necessary to decompose arbitrary unitary operations into some combination of the available primitives. 

While effective gate decomposition is crucial for making quantum algorithms executable on real devices, it is not always a straightforward process. But still, several standard methods have been developed for this task. Usually, they consist of the usage of Solovay-Kitaev theorem \cite{ozols2009solovay} for approximation of general unitaries with a small set of basis gates, and application of specific decompositions like the Quantum Shannon Decomposition \cite{krol2022efficient} or Cartan Decomposition \cite{mansky2023near} for structured circuits.

What we must realize, is that the efficiency with which the gate decomposition is done has a direct impact on the performance and feasibility of quantum algorithms, especially in the current era of noisy intermediate-scale quantum devices. Thus, different ways to decompose gates quickly are still being developed \cite{rosa2025optimizing,liu2023qcontext,vale2023circuit}.

\section{Mapping Operators to Qubits}

Current quantum computers operate on qubits, which are two-level quantum systems, but in many use cases we want to simulate fermionic operators \cite{szalay2021fermionic} that obey different anti-commutation relations. This is particularly necessary if we want to simulate physical systems. So if we want to simulate a fermionic system on a quantum computer, we first need to map the fermionic operator describing our system, onto qubit operators composed of Pauli matrices. Fermionic operators consist of creation $a_i^\dagger$ and annihilation $a_i$ operators and during their transformation we must ensure, that the anti-commutation relations of fermions,
\begin{equation}
    \{ a_i, a_j^\dagger \} = \delta_{ij}, \quad \{ a_i, a_j \} = \{ a_i^\dagger, a_j^\dagger \} = 0,
\end{equation}
are preserved. 

Several efficient mapping strategies exist, as this process is essential as it determines the overall circuit complexity and the scalability of quantum simulations.

\subsection{Parity Mapping}

The first mapping that we will talk about is called Parity mapping \cite{ender2023parity}. It encodes the occupation of fermionic mode \cite{vidal2021quantum} into the evenness of oddness of occupation numbers across a subset of modes. So instead of storing occupation numbers directly, the qubits store just the parity information. So when the occupation of mode $i$ is inferred from parity modes $0, 1, \dots, i$, we gain a typical parity-mapped operator for the fermionic creation operator in the form of
\begin{equation}
    a_i^\dagger \sim \left( \prod_{j=0}^{i-1} Z_j \right) X_i,
\end{equation}
where $X_i$ and $Z_i$ are Pauli matrices that act on qubits $i$ and $j$ respectively. 
One of the advantages of parity mapping is that it reduces operator locality,

\subsection{Jordan-Wigner Mapping}

One of the simplest transformations available is the Jordan-Wigner mapping \cite{veyrac2024geometric}. This is the most intuitive mapping as we represent the fermionic creation and annihilation operators as
\begin{align}
a_i^\dagger &= \left( \prod_{j=0}^{i-1} Z_j \right) \frac{X_i - iY_i}{2}, \\
a_i &= \left( \prod_{j=0}^{i-1} Z_j \right) \frac{X_i + iY_i}{2},    
\end{align}
where $X_i$, $Y_i$ and $Z_j$ are Pauli matrices. While the constructed operators clearly preserve fermionic anti-commutation relations, the circuit depth scales linearly with system size, which is not optimal for many applications.

\subsection{Bravyi-Kitaev Mapping}
The last mapping to be introduced is the Bravyi-Kitaev transformation \cite{seeley2012bravyi}, which optimizes between occupation number and parity encoding. This approach uses a binary tree structure to encode occupation and parity information in a way that is balanced. The fermionic operators are mapped to qubit operators which require a smaller number of qubits, than when doing the mapping via Jordan-Wigner transformation and typically the number of qubits scales logarithmically with the number of modes. While the explicit formula for Bravyi-Kitaev mapping is quite complex, in the end, it leads to the following form of Pauli strings
\begin{equation}
   a_i^\dagger \sim \text{(sum of products of Pauli-} X, Y, Z \text{ operators)}, 
\end{equation}
where the exact structure depends on the binary representation of $i$. The main advantage of this approach is that it results in a smaller number of quantum gates, which helps greatly when we want to minimize quantum errors, which happen during circuit evaluation on quantum computers.

\section{Hardware Architectures}

Just for the quantum machines following the gate-based paradigm, several various physical platforms have been developed. The individual architectures vary deeply and in this section, we will introduce a few of the most common ones. Each platform comes with a distinct set of advantages but also its own challenges and the type of platform we choose to utilize affects not only the scalability of our calculations but also error rates, gate fidelities, and coherence times. All of this is crucial for practically realizing our quantum calculation.

\subsection{Superconducting Qubit Systems}
The most widely used quantum computing platform utilized superconducting qubits \cite{kjaergaard2020superconducting}, which operate at millikelvin temperatures so that they can create anharmonic energy levels from which we define the quantum states $\ket{0}$ and $\ket{1}$. The most common designs are based on Josephson junction \cite{makhlin2001quantum}. 
Superconducting qubits give us the ability to perform gate operations quickly, in the order of tens of nanoseconds, at the cost of relatively short coherence times, typically tens to hundreds of microseconds. Another disadvantage is that they require complex cryogenic setup to maintain their superconductivity.

\subsection{Ion Trap Systems}

The second type of common architecture utilizes ion traps \cite{kielpinski2002architecture,stick2006ion}. These computers encode qubits in the internal electronic states of ions. For practical usage, we mostly tend to use $\text{Ca}^+$, $\text{Yb}^+$ or $\text{Ba}^+$. We confine these ions using electromagnetic fields, which are generated by radio-frequency and static electric fields.

In this architecture, the quantum gates are performed by laser beams that induce coherent transitions between the different qubit states. This approach gives us long coherence times, up to minutes or hours, and high fidelity of individual gate operations. However, the main problem is the fact that the scaling of the quantum computer to the larger number of qubits is problematic, mainly due to the issue of cross-talk between individual ions, when we perform multi-qubit operations.

\subsection{Photonic Quantum Computing}
Another particle we can use is the photon, and when we use it in quantum computers we call it photonic quantum computing \cite{takeda2019toward}. In this hardware, we use photons in the optical or near-infrared spectrum to carry the quantum information. Qubit states are often encoded in degrees of freedom, such as polarization, time-bin mode, or path of the photon.

The main benefit is that the decoherence is implicitly low as photons interact weakly with their environment. The problem arises, when we want to implemet two-photon gate, which often requires the utilization of probabilistic methods or ancillary resources and thus making any larger-scale calculation technologically demanding.

\subsection{Neutral Atom  Qubits}
The last type of architecture that we will introduce is the neural atom quantum computers \cite{schrader2004neutral}. Here we trap individual atoms, typically of rubidium or cesium in an optical lattice, or in tweezers formed by precisely focused laser beams. Qubits are then encoded in the hyperfine states of the atom, and interactions are mediated by Rydberg excitations, which enable us to perform high-fidelity two-qubit gates. This platform offers great scalability and gives us the ability to reconfigure qubit arrays dynamically.

\section{Measurement}
The term of measurement \cite{jauch1964problem} is the most fundamental aspect of quantum computing as it allows us to cross from the real quantum to the classical world by extracting classical information from the quantum state. In quantum algorithms, we usually perform measurement at the end of the circuit, but when the architecture allows for it we can also perform mid-circuit measurements, which we use mainly for error correction or feedback control. To truthfully interpret the results of quantum measurement, we first need to understand its principles and limitations.

\subsection{Quantum Measurement Postulate}

Firstly let us state the quantum measurement postulate \cite{segal1947postulates}, which tells us that the act of measurement causes the collapse of a quantum system from superposition to one of the eigenstates of the observable we measure. So if a quantum system is in the following quantum state
\begin{equation}
   \ket{\psi} = \sum_i c_i \ket{i}, 
\end{equation}
where $\ket{i}$ are the eigenstates of the measured observable and $c_i$ are complex amplitudes, then measurement gives us as the result the $i-th$ eigenstate with probability
\begin{equation}
    P(i) = |c_i|^2.
\end{equation}

Due to the collapse, immediately after the measurement, we know that the quantum system is found in state $\ket{i}$. From this postulate, we see that measurement outcomes are inherently probabilistic, and their probabilities are determined by the quantum state's amplitudes prior to the measurement itself.

It is also important to mention that in most quantum calculations, the measurement basis is the computational basis \cite{vedral1998basics}, where the measured eigenstates correspond to $\ket{0}$ and $\ket{1}$. 

\subsection{Error Sources in Quantum Measurement}
The measurement of quantum state is susceptible to various types of errors, that affect the reliability of the measured results. The first type of error is the so-called readout error, which occurs when the classical computer is interpreting the quantum signal and during this process, it makes an error and incorrectly assigns the measured outcome to a different eigenstate. For example, $\ket{0}$ can be mistaken as $\ket{1}$ due to imperfect interpretation.

The second type of error is cause by cross-talk. When working with multi-qubit systems, the measurement done on single qubit can influence the measurement output of qubits in its neighborhood, especially when the qubits are physically close together, or there are imperfection in the execution of measurement pulse.

The last type of error that we need to take into the account is the environmental noise. From cosmic ray and thermal fluctuations to vibrations and external electromagnetic fields, all of these effects can perturb our quantum system during the measurement and distort our result.

If we want to mitigate those errors \cite{cai2023quantum}, we usually employ calibration routines, such as measurement error mitigation \cite{funcke2022measurement}, or we make repeated measurements to improve the statistical confidence of the obtained results.

\section{Quantum Advantage and Reversible Computing}
Quantum computing has the potential to solve specific computational problems exponentially faster than classical computers, and this phenomenon is called quantum advantage. And when we talk about that, we also need to mention the principles of reversible computations, that will help us to understand why is quantum computing oftentimes energetically cheaper.

\subsection{Quantum Advantage}

Quantum advantage \cite{huang2022quantum,daley2022practical}, or as it was originally known quantum supremacy, is a name for a point at which a quantum computer performs a task that is not feasible for classical supercomputer, when we of course consider that the task needs to be completed in reasonable time. The fact that such an advantage can exist stems from key features of quantum mechanics, such as superposition, entanglement, and interference, which all allow quantum computers to explore and process larger spaces faster than their classical counterparts are able to do.

A quantum system, that consists of $n$ qubits, can represent $2^n$ complex amplitudes at once, which enables several quantum algorithms to achieve significant speedups compared to classical calculations. Each demonstration of quantum advantage is considered a great milestone in the world of quantum computing.

\subsection{Reversible Computing}

When we have a computational model, where every single step done during the calculation is logically reversible, which means that every input can be recovered from the output in a unique way, then we speak about reversible computing \cite{toffoli1980reversible}. In quantum computing, all operations are inherently reversible, because all of the gates correspond to unitary transformations.

The fact that a calculation is reversible has deep implications in terms of energy efficiency. When we consider classical computing, which is irreversible because if we know that the about of the summation is $a$ we are unable to identify which two numbers $b$ and $c$ were the inputs, we inevitably erase some information, which leads to thermodynamics. This fact is stated by Landauer's principle \cite{bennett2003notes}.

On the other hand, when we have reversible computation, we avoid this energy loss. Because of this, quantum computing is an attractive paradigm also for those who strive to achieve low-power calculations.

\subsection{Landauer’s Principle}

Landauer’s principle describes the link between thermodynamics and information theory. It states, that the erasure of one bit of information in a computational device leads to the necessary dissipation of the minimum amount of energy in the form of heat release to the environment. This is given by
\begin{equation}
   E_{\text{min}} = k_B T \ln 2, 
\end{equation}
where $k_B$ is Boltzmann's constant and $T$ is the temperature of the environment.

This ejection of energy into the environment happens because erasure is a logically irreversible operation, thus reducing the number of accessible microstates of the system and entropy. Landauer’s principle highlights that information processing is not just an abstract mathematical operation but also a physical process constrained by the laws of thermodynamics.

\section{Classical Machine Learning}
\ac{ml} is a subfield of the field of artificial intelligence, which is focused mainly on the development of algorithms that are capable of learning patterns and making decisions based on the data that is provided to them. While in traditional programming, we have a set of explicit instructions, that are subsequently executed, in \ac{ml} we expect the system to infer rules and relationships from a series of examples, most commonly called training data. \ac{ml} has lots of applications across diverse fields such as natural language processing \cite{olsson2009literature}, finance \cite{dixon2020machine}, medicine \cite{deo2015machine}, chemical calculations \cite{janet2020machine}, and many more.

\subsection{Supervised vs. Unsupervised Learning}

\ac{ml} tasks are commonly categorized based on the structure of the data and its availability. If the data that we are processing has corresponding labels and these labels are used during the training of the model, then we are speaking about supervised learning \cite{cunningham2008supervised}. Here each input $\mathbf{x}_i$ has corresponding output label $y_i$ and the goal is to learn mapping $f: \mathbf{x} \mapsto y$ that has the ability to generalize to unseen data. The most common tasks that are classified as supervised learning include classification and regression. The algorithms that are used to perform supervised learning are for example support vector machines \cite{hearst1998support}, decision trees \cite{charbuty2021classification} and neural networks \cite{gurney2018introduction}.

On the other hand, if the data is unlabeled, we are in the field of unsupervised learning \cite{barlow1989unsupervised}. Here the main goal is to discover hidden relationships and underlying structures in the data, by inferring from inputs $\{ \mathbf{x}_i \}$.  Common tasks of unsupervised learning include clustering \cite{rokach2005clustering} and dimensionality reduction \cite{van2009dimensionality}.

We will also mention that semi-supervised \cite{learning2006semi} and reinforcement learning \cite{kaelbling1996reinforcement} represent additional \ac{ml} paradigms, where the main difference is the level of supervision or the fact that the feedback mechanism is different. But for these cases, we will not go into further detail.

\subsection{Overfitting and Regularization}

When the \ac{ml} model is learning, it can learn not only the underlying patterns but also the noise and specifics of the training data, when this occurs we are talking about overfitting \cite{ying2019overview} of our model. This fact leads to a decrease in the model's ability to generalize, and mathematically we can talk about minimization of the training error at the expense of the increase of the generalization error. 

To combat overfitting, we can employ regularization techniques \cite{tian2022comprehensive}, that introduce additional information or specific constraints into the training process of our model. Some of the common regularization methods are L1 Regularization (Lasso) \cite{vidaurre2013survey}, where we add a penalty proportional to the absolute value of the model's coefficients 
\begin{equation}
     \mathcal{L}_{\text{L1}} = \mathcal{L}_{\text{original}} + \lambda \sum_i |\theta_i|.   
\end{equation}
The next one is for example L2 Regularization (Ridge) \cite{obi2023review}, where the penalty is proportional to the square of the same coefficients 
\begin{equation}
        \mathcal{L}_{\text{L2}} = \mathcal{L}_{\text{original}} + \lambda \sum_i \theta_i^2.
\end{equation}
The last example is the Dropout \cite{baldi2013understanding}, which randomly discards units and their connections during the learning process to prevent them from adapting too much to the training dataset, this process i usually defined by parameter $\lambda$.

\subsection{Convolutional Neural Networks}
\ac{cnns} \cite{li2021survey} are specialized classes of neural networks, which are particularly effective for data processing in which the input data is in grid-like structure, such as images, heatmaps, or matrices. \ac{cnns} work with three main ideas. The first one is that the neurons in a layer are connected just to a small subset of localized neurons in the previous layer, which allows the spatial hierarchy of features in the data. The next one is the concept of shared weights, or convolutions, which means, that the same set of weights is applied to different regions of the input data, acting like a kernel or filter. This drastically reduces the number of parameters and proves to improve the translation of invariance. The last idea is that there are pooling layers, which reduce the spatial dimensions while preserving all of the important features.

Mathematically, a convolution operation applied to an input $x$ with a kernel $w$ can be written as
\begin{equation}
    (x * w)(i, j) = \sum_{m} \sum_{n} x(i+m, j+n) \, w(m, n),
\end{equation}
where $(i,j)$ are indexes the position in the output.

\ac{cnns} are the state-of-the-art models for various tasks including object detection \cite{zhiqiang2017review}, semantic segmentation \cite{guo2018review} and image classification \cite{lu2007survey}.

\section{Linear Separability}

The concept of linear separability \cite{elizondo2012linear} is fundamental in the field of machine learning and classification theory. Particularly in the case of algorithms like support vector machines \cite{noble2006support}, logistic regression \cite{lavalley2008logistic} and perceptrons \cite{minsky1988perceptrons}, but not limited to those.

\subsection{Definition}

We say that a dataset is linearly separable if there exists at least a single hyperplane that can perfectly separate the data into distinct classes without any cases of misclassifications. Mathematically, we say that if we are given a dataset $\{(\mathbf{x}_i, y_i)\}_{i=1}^N$ with the inputs $\mathbf{x}_i \in \mathbb{R}^n$ and binary labels $y_i \in \{-1, +1\}$, then the data ate linearly separable if and only if there exists a weight vector $\mathbf{w} \in \mathbb{R}^n$ and bias $b \in \mathbb{R}$ such that the following condition holds
\begin{equation}
  y_i (\mathbf{w}^\top \mathbf{x}_i + b) > 0 \quad \forall i = 1, \dots, N.  
\end{equation}
The decision boundary is then the hyperplane defined by
\begin{equation}
    \mathbf{w}^\top \mathbf{x} + b = 0.
\end{equation}
Which tells us that the points for which the inequality is positive belong in first class and the ones where the inequality is negative belong into the other class.

\subsection{Examples of Separable and Non-Separable Data}
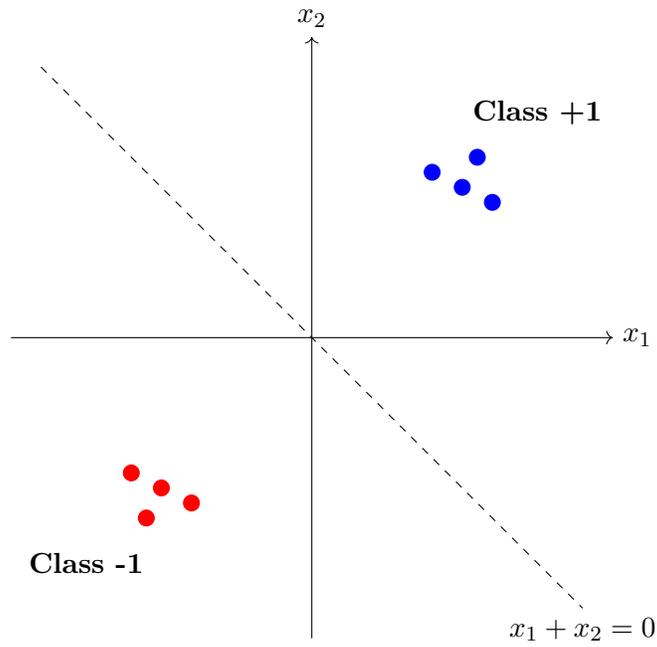
\begin{figure}
    \centering
    \begin{tikzpicture}[scale=2]
    
        \draw[->] (-2,0) -- (2,0) node[right] {\(x_1\)};
        \draw[->] (0,-2) -- (0,2) node[above] {\(x_2\)};
        
        \filldraw[blue] (1,1) circle (1.5pt);
        \filldraw[blue] (1.2,0.9) circle (1.5pt);
        \filldraw[blue] (0.8,1.1) circle (1.5pt);
        \filldraw[blue] (1.1,1.2) circle (1.5pt);
        
        \filldraw[red] (-1,-1) circle (1.5pt);
        \filldraw[red] (-1.2,-0.9) circle (1.5pt);
        \filldraw[red] (-0.8,-1.1) circle (1.5pt);
        \filldraw[red] (-1.1,-1.2) circle (1.5pt);
        
        \draw[dashed] (-1.8,1.8) -- (1.8,-1.8) node[below] {\(x_1 + x_2 = 0\)};
        
        \node at (1.5, 1.5) {\textbf{Class +1}};
        \node at (-1.5, -1.5) {\textbf{Class -1}};
        
    \end{tikzpicture}
    \caption{Linearly separable data: two clusters separated by the hyperplane \(x_1 + x_2 = 0\).}
    \label{fig:linear_separable}
\end{figure}

To better understand linear separability, let's have a look at some examples. For separable data, we can consider two-dimensional data that consists of two clusters, one around the point $(1,1)$, and the other cluster around $(-1, -1)$. From this, we can clearly deduce that a line described by 
\begin{equation}
    x_1 + x_2 = 0
\end{equation}
is able to separate the classes perfectly. This is shown in \cref{fig:linear_separable}.

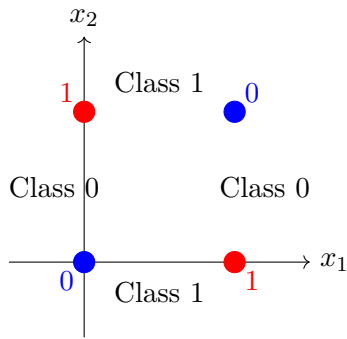
\begin{figure}
    \centering
    \begin{tikzpicture}[scale=2]
    
        \draw[->] (-0.5,0) -- (1.5,0) node[right] {\(x_1\)};
        \draw[->] (0,-0.5) -- (0,1.5) node[above] {\(x_2\)};
        
        \filldraw[blue] (0,0) circle (2pt) node[below left] {\(0\)};
        \filldraw[blue] (1,1) circle (2pt) node[above right] {\(0\)};
        
        \filldraw[red] (1,0) circle (2pt) node[below right] {\(1\)};
        \filldraw[red] (0,1) circle (2pt) node[above left] {\(1\)};
        
        \node at (0.5,1.2) {Class 1};
        \node at (0.5,-0.2) {Class 1};
        \node at (-0.2,0.5) {Class 0};
        \node at (1.2,0.5) {Class 0};

    \end{tikzpicture}
    \caption{XOR dataset: an example of non-linearly separable data. Points \((0,0)\) and \((1,1)\) belong to class 0, while \((1,0)\) and \((0,1)\) belong to class 1.}
    \label{fig:xor_non_separable}
\end{figure}

As for non-separable data, a classic example is the exclusive OR problem. The dataset consists of the following points
\begin{equation}
       (0,0) \to 0,\quad (1,1) \to 0,\quad (1,0) \to 1,\quad (0,1) \to 1, 
\end{equation}
and this dataset can't be separated by any straight line. The example is displayed in \cref{fig:xor_non_separable}.

For separable data, we can make do with Linear classifiers, which are able to perform optimally on such data. But for non-separable data, we have to use more complicated models, such as  multi-layer perceptrons or kernel-based SVMs. 
 
\section{Quantum Machine Learning}

\ac{qml} \cite{ciliberto2018quantum} is a multidisciplinary field in which the principles of quantum computing are combined with machine learning. One of the goals of \ac{qml} is to leverage quantum resources to achieve either speedup or different kinds of improvements during the training of a model. Several approaches have already been proposed, which all combine quantum resources and classical computing in different ways.

\subsection{Classical-Quantum Paradigms}

When we consider the context of \ac{qml}, four different types of paradigms arise, depending on if we are using classical or quantum data or model.

The first case is when we are using both classical data and classical model. This is the traditional machine learning approach, and thus is not interesting for us now.

The second case is when we are using quantum data, possibly obtained via quantum experiments, but this data is processed using a classical computer.  

The third way is the most interesting for us at the moment, and that is the combination of classical data and quantum model. In this case, the whole model is executed on quantum hardware, but the data remain classical and must be mapped to the circuits, or observables.

The last paradigm consists of the usage of quantum data that is directly fed to a quantum machine, this approach should in theory preserve the possibility of quantum advantage throughout the whole computation process.

The two hybrid models are the most interesting for today's science as they are available to be executed even when we are limited by \ac{nisq} devices \cite{de2022survey}.

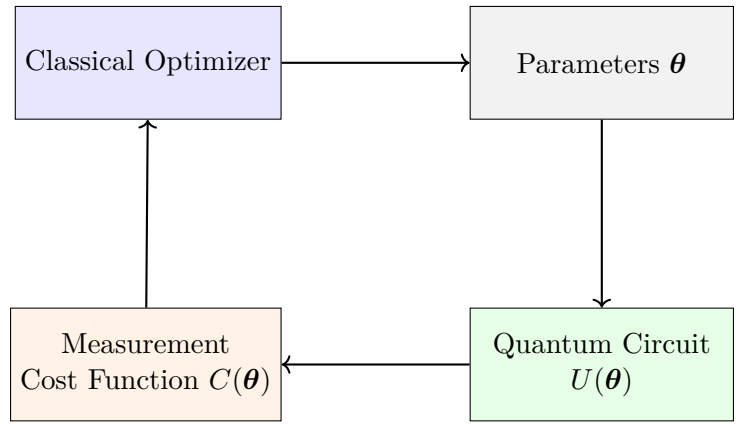
\begin{figure}
    \centering
    \begin{tikzpicture}[scale=1, every node/.style={align=center}]
    
        \node[draw, rectangle, minimum width=3.5cm, minimum height=1.5cm, fill=blue!10] (optimizer) {Classical Optimizer};
        \node[draw, rectangle, minimum width=3.5cm, minimum height=1.5cm, right=2.5cm of optimizer, fill=gray!10] (params) {Parameters \(\boldsymbol{\theta}\)};
        \node[draw, rectangle, minimum width=3.5cm, minimum height=1.5cm, below=2.5cm of params, fill=green!10] (circuit) {Quantum Circuit\\\( U(\boldsymbol{\theta}) \)};
        \node[draw, rectangle, minimum width=3.5cm, minimum height=1.5cm, left=2.5cm of circuit, fill=orange!10] (measure) {Measurement \\ Cost Function \( C(\boldsymbol{\theta}) \)};
        
        \draw[->, thick] (optimizer) -- (params);
        \draw[->, thick] (params) -- (circuit);
        \draw[->, thick] (circuit) -- (measure);
        \draw[->, thick] (measure) -- (optimizer);

    \end{tikzpicture}
    \caption{Square layout of a Variational Quantum Circuit training loop. Parameters are optimized based on measurement outcomes via a classical optimizer.}
    \label{fig:vqc_square_training}
\end{figure}

\subsection{Variational Quantum Circuits for Machine Learning}

\ac{vqcs} is a powerful yet simple class of models that are used in \ac{qml}. These combine classical optimization and quantum circuit, which is parameterized by a set of classical parameters $\boldsymbol{\theta}$, which control the rotation angle of quantum gates
\begin{equation}
    U(\boldsymbol{\theta}) = \prod_{i} U_i(\theta_i),
\end{equation}
where $U_i(\theta_i)$ represents a gate dependent on a variational parameter $\theta_i$.

When we want to train the following circuit we need to follow these subsequent steps. At first, we must initialize input parameters $\boldsymbol{\theta}$, and then we prepare the initial state which consists either of quantum data or classical data embedded in the quantum device. Then we apply our parametrized variational circuit $U(\boldsymbol{\theta})$. When this is done we perform the measurement of an observable and calculate the cost function. After that, we update the $\boldsymbol{\theta}$ parameters using data provided by the classical optimizer. This whole loop is shown in \cref{fig:vqc_square_training}.

\ac{vqcs} are really versatile and can be used for regression, classification, generative modeling, and even reinforcement learning tasks. Thus it is considered one of the most promising approaches with respect to practical quantum machine learning tasks in the \ac{nisq} era.

\subsection{Quantum Feature Maps and Data Encoding}

In the previous section we mentioned that we can encode classical data onto the quantum machine, and construct initial circuit out of them. To do so we use quantum feature maps, which are mechanisms, that tells us how to encode classical data into high-dimensional Hilbert spaces, that correspond to quantum calculations. This has a potential to give us more power with pattern recognition than we have in classical models.

If we are given classical data point $\mathbf{x}$, and feature map $\Phi(\mathbf{x})$, then we can use the feature map to embed the data point into quantum state $\ket{\Phi(\mathbf{x})}$, by doing parametrized unitary transformations
\begin{equation}
    \ket{\Phi(\mathbf{x})} = U(\mathbf{x})|0\rangle^{\otimes n},
\end{equation}
where $U(\mathbf{x})$ is and unitary operation that depends on the classical data point.

We can also talk about different types of data encoding, which include angle encoding \cite{ovalle2023quantum}, where the input data is mapped to rotation angles of single-qubit gates, or amplitude encoding \cite{weigold2020data}, where the data is directly encoded into the quantum state, but in this case, we have to remember that it is necessary to normalize the data. The last data encoding strategy that we will mention is so-called qubit encoding \cite{tudisco2022encoding}, where the inputs are mapped directly to computational basis states, this approach typically requires the largest number of qubits, so it is the least used one.

\section{Considerations for Quantum Layer Placement in Hybrid Neural Architectures}

In hybrid quantum-classical neural networks, the position of the quantum layer within the architecture plays a crucial role in shaping the behavior and effectiveness of the model \cite{cerezo2022challenges}. Depending on where the quantum component is inserted—whether at the input, within hidden layers, or near the output—it can influence the flow of information, the type of representations formed, and the overall computational dynamics of the network.

One approach is to place the quantum layer at the input stage, where it acts as an initial feature transformation block. In this configuration, the input data is encoded into quantum states and processed by a quantum circuit before entering the classical portion of the network. This may allow for more expressive representations to be constructed early on, potentially enhancing the learning capacity of subsequent layers.

Alternatively, the quantum layer can be positioned in the middle of the network, between classical hidden layers. In this arrangement, the quantum component operates on intermediate feature representations, introducing non-classical transformations at a stage where the data has already been partially abstracted. This can help the network capture complex patterns that might be less accessible to purely classical architectures.

Finally, the quantum layer can be placed near the output of the model, where it functions as a classifier or final decision-making mechanism. In this case, the classical layers are responsible for extracting features from the input data, and the quantum circuit processes the resulting high-level representation to produce a prediction. This setup may be advantageous when leveraging quantum measurement outcomes as part of the output logic.

Each of these placements carries distinct implications in terms of data encoding complexity \cite{de2022survey}, circuit depth, training stability, and interpretability. Selecting the appropriate location for the quantum layer is therefore a critical architectural decision, as it directly affects the hybrid model’s behavior, capabilities, and efficiency.

\section{Leveraging Quantum Properties in Hybrid Models}

The integration of quantum layers into classical neural network architectures is motivated by the potential to exploit uniquely quantum mechanical phenomena—most notably, quantum parallelism and entanglement. These properties offer a promising foundation for reducing computational complexity and enhancing model performance, particularly in tasks involving high-dimensional data and complex pattern recognition.

Quantum parallelism \cite{markidis2024quantum} refers to the ability of quantum systems to exist in superpositions of many states simultaneously. A quantum circuit can, in principle, process an exponential number of states in parallel, enabling certain computations to be performed more efficiently than their classical counterparts. When embedded into a neural architecture, a quantum layer can act as a powerful transformation mechanism, rapidly projecting classical data into a higher-dimensional feature space. This process may allow the network to capture subtle patterns or separations in the data that would require more depth or capacity in a purely classical network.

Entanglement \cite{horodecki2009quantum}, on the other hand, introduces non-classical correlations between qubits that can be harnessed to represent complex dependencies within the data. In the context of neural networks, these correlations can enrich the expressivity of the model by enabling joint representations that are not easily decomposed into independent components. By incorporating entangled states within the quantum layer, the model gains the ability to capture interactions and structures that might otherwise be inaccessible or costly to represent classically.

Together, these properties suggest that the correct placement and design of a quantum layer can offer both a computational and representational advantage. By leveraging quantum parallelism and entanglement, hybrid models may achieve more compact architectures, faster convergence, or improved generalization—especially in scenarios where classical approaches struggle with scalability or feature complexity.

\section{Variational Quantum Eigensolver}

\ac{vqe} \cite{tilly2022variational} is a hybrid quantum-classical algorithm, whose goal is to find the ground state of the energy of a system. This type of algorithm is well suited for the \ac{nisq} era, as it is quite well suited for noisy computers. As the main block consists of variational quantum circuits, we can consider it as a quantum machine learning algorithm. 

\subsection{Variational Principle}
The \ac{vqe} is built around one of the principles from quantum mechanics, which states that for any trial state $\ket{\psi(\boldsymbol{\theta})}$ the following condition holds
\begin{equation}
    E(\boldsymbol{\theta}) = \frac{\langle \psi(\boldsymbol{\theta}) | \hat{H} | \psi(\boldsymbol{\theta}) \rangle}{\langle \psi(\boldsymbol{\theta})|\psi(\boldsymbol{\theta})\rangle} \geq E_0,
\end{equation}
where $\hat{H}$ is the Hamiltonian of the system,  $E(\boldsymbol{\theta})$ is the expected energy of the trial state, and $E_0$ is the true energy of the ground state. By minimizing $E(\boldsymbol{\theta})$ over parameters $\boldsymbol{\theta} \in \mathbb{R}^n$, we are able to approximate the ground state and its corresponding energy. This is called the Variational principle \cite{ekeland1974variational}.

\subsection{Mathematical Description of \ac{vqe}}

\begin{figure}
    \centering
    \begin{tikzpicture}[node distance=2cm, every node/.style={align=center, font=\small}]
    
        \node[draw, rectangle, minimum width=2.8cm, minimum height=1.2cm, fill=blue!10] (ansatz) {Prepare Ansatz\\\( |\psi(\boldsymbol{\theta})\rangle \)};
        \node[draw, rectangle, minimum width=2.8cm, minimum height=1.2cm, below=of ansatz, fill=green!10] (measure) {Measure\\Energy \( E(\boldsymbol{\theta}) \)};
        \node[draw, rectangle, minimum width=2.8cm, minimum height=1.2cm, right=of measure, fill=orange!10] (optimize) {Classical Optimizer\\Update \( \boldsymbol{\theta} \)};
        
        \draw[->, thick] (ansatz) -- (measure);
        \draw[->, thick] (measure) -- (optimize);
        \draw[->, thick] (optimize) |- (ansatz);

    \end{tikzpicture}
    \caption{The Variational Quantum Eigensolver loop.}
    \label{fig:vqe_loop}
\end{figure}
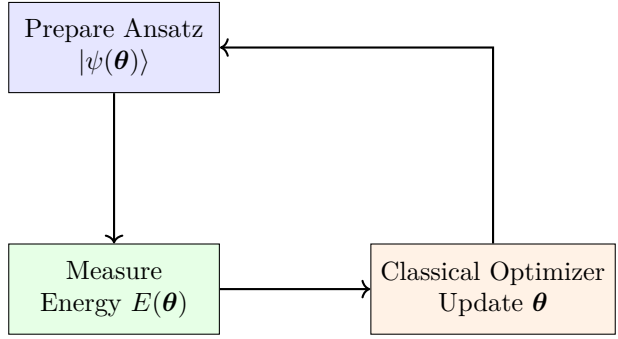

Now let's have a look at how we can execute \ac{vqe}. At first, we assume that we already have a well-defined Hamiltonian operator of our system as well as a set of initial parameters, then we have to choose the right parametrized quantum circuit, which is oftentimes called ansatz. This circuit is defined as 
\begin{equation}
    \psi(\boldsymbol{\theta})\rangle = U(\boldsymbol{\theta}) |0\rangle.
\end{equation}
After this, we evaluate the expectation value
\begin{equation}
        E(\boldsymbol{\theta}) = \langle 0 | U^\dagger(\boldsymbol{\theta})| \hat{H} |U(\boldsymbol{\theta}) | 0 \rangle
\end{equation}
on a quantum computer, and we follow with the usage of the classical optimizer, which will determine the ansatz parameters for the following iterations. This loop is shown in \cref{fig:vqe_loop}.

As physical Hamiltonians are decomposed into sums of Pauli operators,
\begin{equation}
    \hat{H} = \sum_i h_i P_i,
\end{equation}
where $P_i$ are tensor products of Pauli matrices and $h_i$ are real coefficients, the quantum computer can easily measure the expectation values of every $P_i$ independently and calculate their weighted sum classically. 

\subsection{Ansatz Selection}

Well-chosen ansatz is the crucial part of \ac{vqe}, as it affects both the efficiency and success of the algorithm. We can typically divide ansatzes into three main categories. The first one consists of hardware-efficient ansatzes \cite{leone2024practical}, which are parametrized circuits that are specifically designed to work well on quantum hardware and they usually consist of single-qubit rotations followed by entangling gates. These are easy to implement, also we can logically increase the number of points of freedom, but they may suffer quite a lot from barren plateaus.

The second ansatz category is the category of Unitary Coupled Cluster ansatzes \cite{romero2018strategies}, these are inspired by my quantum chemistry methods and are used in energy calculations of molecular systems. As they have real-world meaning, they are well suited for chemical problems. If we take for example ansatz with single and double excitation, it takes the following form
\begin{equation}
        |\psi(\boldsymbol{\theta})\rangle = e^{T(\boldsymbol{\theta}) - T^\dagger(\boldsymbol{\theta})} | \phi_0 \rangle,
\end{equation}
    where $T(\boldsymbol{\theta})$ contains excitation operators acting on a reference state $|\phi_0\rangle$.

The last category contains so-called Problem-Specific Ansatzes \cite{matsuo2023enhancing}, and those are tailored to the specific problem, exploiting its known structure, and symmetries, and having a lower number of parameters.

To select an ideal ansatz for our problem, we need to balance desired expressibility, and trainability with hardware compatibility.

\subsection{Applications}
\ac{vqe} has many different applications in a multitude of different fields. For example, we can see it being used in quantum chemistry, where it is used to estimate ground state energies of molecules \cite{fedorov2022vqe, goings2023molecular}.

Or in material science, where \ac{vqe} is used to study different properties of condensed matter system \cite{bauer2020quantum}, like the modeling of the Hubbard model \cite{alvertis2024classical}. 

We can also reformulate certain classical optimization problems, for example, the Max-Cut \cite{li2023accelerating} problem into Hamiltonian and the use \ac{vqe} to find its solutions. 

In fundamental physics we use \ac{vqe} to investigate small nuclear systems \cite{romero2022solving} and lattice gauge theories \cite{paulson2021simulating}.

VQE remains one of the most promising near-term quantum algorithms, offering a pathway to practical quantum advantage despite hardware limitations.

\section{Variational Hamiltonian Ansatz}

The \ac{vha} \cite{wiersema2020exploring} is a specifically structured ansatz for variational quantum algorithms. It is designed to leverage the form of the system's Hamiltonian to improve its efficiency. It is one type of problem-specific ansatzes. i

\subsection{Mathematical Description}

Given a Hamiltonian $\hat{H} = \sum_j \hat{H}_j$ decomposed into local terms, the \ac{vha} prepares a quantum state through sequential application of exponentiated Hamiltonian components:
\begin{equation}
    |\psi(\boldsymbol{\theta})\rangle = \prod_{k=1}^p \left( \prod_{j} e^{-i \theta_{k,j} H_j} \right) |0\rangle^{\otimes n},
\end{equation}
where $p$ is the number of layers and $\theta_{k,j}$ are variational parameters.

\subsection{Advantages over Hardware-Efficient Ansatz}

Compared to generic hardware-efficient ansatzes, \ac{vha} allows us to better align with the problem's Hamiltonian, which reduces the barren plateaus in the landscape. Also, we can have fewer parameters, which simplifies the optimization process, also the fact that the parameters have physical meaning tends to lead to more stable training.

\section{State-Averaged Orbital-Optimized VQE}

The \ac{saoovqe} \cite{yalouz2021state, beseda2024state} is an expandion of the \ac{vqe}. This version allows the calculation of multiple quantum states, while also optimizing the underlying molecular orbitals, leading to improved accuracy and convergence.

\subsection{State Averaging}

In \ac{saoovqe}, instead of optimizing a single eigenstate, the cost function is based on a weighted average of multiple energy eigenstates. If \(\{|\psi_k(\boldsymbol{\theta})\rangle\}\) represent different eigenstates, the objective function becomes:
\[
E_{\text{SA}}(\boldsymbol{\theta}) = \sum_k w_k \langle \psi_k(\boldsymbol{\theta}) | \hat{H} | \psi_k(\boldsymbol{\theta}) \rangle,
\]
Where \(w_k\) are positive weights summing to one. This approach ensures simultaneous optimization across multiple states. This is important when we want to model excited states but want to avoid bias toward one particular state.
\subsection{Orbital Optimization}

Orbital optimization refines the molecular orbitals of the Hamiltonian. In \ac{saoovqe}, variational parameters also control orbital rotations through unitary transformations of the molecular orbitals:
\[
C = e^{\kappa - \kappa^\dagger},
\]
where \(\kappa\) is an anti-Hermitian matrix of orbital rotation parameters. This optimization leads to more compact and accurate wavefunctions.

\subsection{Advantages over State-Specific VQE}

When we compare \ac{saoovqe} with State-Specific \ac{vqe}, we realize that there are several advantages, The first is that we obtain the energy of the excited stated and ground state simultaneously, without the need to repeat the optimization, There is no bias concerning the states, and also the numerical optimization is easier as this approach avoids difficult landscapes.

\ac{saoovqe} is especially valuable in quantum chemistry applications where excited state properties are crucial.

\section{Causality}

To understand what is causality \cite{pearl2009causality} is one of the fundamentals of scientific reasoning, as it describes the underlying mechanism that drives the observed phenomena. Especially in machine learning and data science, we must be able to distinguish between causal relationships and statistical associations, to be able to build robust predictive models and to inform the decision-making process correctly.

\subsection{Definition of Causality}
Causality describes a relationship between two events where one event, called the cause, directly influences another event, the effect. Formally we can say, that if an intervention on variable $X$ creates a change in another variable $Y$, then we say that $X$ causally affects $Y$

\subsection{Causal Inference vs. Correlation}

While correlation \cite{yule1897theory} is the measure of strength and direction of an association between two variables, it does not imply a causal relationship as the two variables may be correlated because of a direct causal link, or a common cause, that is influencing both of them, or merely by coincidence.

On the other hand, causal inference wants to determine whether a change in one variable results directly in the change in another variable independently on confounding factors.
The difference between these two is illustrated in \cref{fig:causality_vs_correlation}.
\begin{figure}
    \centering
    \begin{tikzpicture}[scale=1.5, every node/.style={align=center}]
    
        \node[draw, rectangle, minimum width=1.5cm, minimum height=1cm, fill=blue!10] (cause) at (0,0) {Variable \(X\)};
        \node[draw, rectangle, minimum width=1.5cm, minimum height=1cm, fill=blue!10] (effect) at (3,0) {Variable \(Y\)};
        
        \draw[->, thick] (cause) -- (effect) node[midway, above] {Causal Link};
        
        \node[draw, rectangle, minimum width=1.5cm, minimum height=1cm, fill=red!10] (corr1) at (0,-3) {Variable \(X\)};
        \node[draw, rectangle, minimum width=1.5cm, minimum height=1cm, fill=red!10] (corr2) at (3,-3) {Variable \(Y\)};
        \node[draw, rectangle, minimum width=1.5cm, minimum height=1cm, fill=yellow!10] (confounder) at (1.5,-1.5) {Confounder \(Z\)};
        
        \draw[->, thick] (confounder) -- (corr1);
        \draw[->, thick] (confounder) -- (corr2);
        
        \node at (1.5, 1) {\textbf{Causality}};
        \node at (1.5, -0.7) {};
        \node at (1.5, -4) {\textbf{Correlation due to Confounder}};
        
    \end{tikzpicture}
    \caption{Illustration of causality (direct link between \(X\) and \(Y\)) versus correlation (common cause \(Z\) influencing both \(X\) and \(Y\)).}
    \label{fig:causality_vs_correlation}
\end{figure}

\subsection{Detection of Causal Inference}

To detect causal relations, randomized controlled trials, instrumental variable methods, graphical models like directed acyclic graphs or we can try to train a model, that will have the ability to predict causality.

When we employ Randomized Controlled Trials \cite{stolberg2004randomized}, which is oftentimes considered the gold standard in the field of causal detection, we randomly divide subjects into treatment and control groups and systematically eliminate any confounding biases.

If we decide to use Instrumental Variables \cite{bowden1990instrumental}, we introduce a variable that influences just the treatment and not the outcome directly, which allows us to identify causal effects even when unobserved confounders are present.

Or we can try to use Structural Causal Models \cite{bongers2021foundations}, which encode causal assumptions by using a set of structural equations and directed graphs, which then enable both the estimation of the causal effect and also allow counterfactual reasoning.

If we want to use a different approach to causality detection, we can try some of the Causal Discovery Algorithms \cite{malinsky2018causal}, like PC (Peter-Clark) algorithm \cite{nur2025causal}, FCI (Fast Causal Inference) \cite{strobl2018fast}, and LiNGAM (Linear Non-Gaussian Acyclic Model) \cite{shimizu2006linear} and attempt to infer causal structures from the data.

When we correctly detect causal relationship, it allows us to make more reliable predictions, and prepare more robust policy-making system, as opposed to models, that is based just on correlations.

\chapter{Methodology and Implementation} \label{chap:method}
In this section, we focus on describing the main objectives of this thesis, the selected research methodology, and the practical aspects of the code implementation. The structure of the proposed hybrid neural network model is introduced, followed by an explanation of how classical and quantum components are integrated to achieve the goals of this work.
\section{Motivation}

Quantum computing has emerged as a promising field with significant potential to address computationally complex problems that we are currently unable to solve on classical computers. Among its vast applications, the integration of quantum computing with classical neural networks is a particularly interesting subfield, which shows promising advances in both the theory and also in the practical capabilities. 

This work builds upon developments in variational quantum algorithms, particularly within the context of the \ac{vqe}. On one hand, the \ac{saoovqe} software package is being developed, where quantum machine learning is used to determine the energies of molecules and new diabatization schemes are developed \cite{illesova2025transformation}, and on the other hand, numerical optimization analysis is being performed in \ac{vqe} with applied \ac{vha}. Where the effects of quantum noise are being studied. 
 These investigations focus on understanding the robustness, convergence, and reliability of different optimization methods in noisy environments, a necessary consideration for \ac{nisq} applications.

Given these foundations, the natural progression of research extends toward hybrid quantum-classical machine learning. By combining classical models with quantum circuits, hybrid systems aim to leverage the strengths of both paradigms. Such systems offer the potential to enhance learning capabilities, improve generalization, and explore quantum advantages within practical machine-learning tasks.

The primary motivation for this thesis is to investigate how different architectural and design choices within hybrid quantum-classical neural networks — specifically feature mapping strategies and quantum ansatz configurations — affect the learning process, model stability, generalization, and final performance.  For this, a classification problem, where we are determining the direction of a causal relationship in a series of images, was chosen and inspiration was taken from previous research done with just classical machine learning \cite{yuan2020causal}.

\section{Methodology}
The methodology chosen for the scope of this thesis was systematically designed to explore how the performance of hybrid quantum-classical machine learning models is affected by changes in the quantum layer. So the research mainly focuses on two things. The first one is the impact of the complexity of the chosen ansatz, where particular focus is put on ansatz depth and what multiple repetitions of ansatz do with respect to the model's training. The second aspect is to figure out what is the influence of feature mapping strategies on the learning dynamics of hybrid models.

The approach consists of several different stages.

\subsection{Data Preparation} 
The first step is to prepare the data. We have obtained the data from the Kaggle database available at \url{https://www. kaggle.com/c/cause-effect-pairs}. In our case, the input data consists of two-dimensional matrices, representing relationships between variable pairs. Each element has size $8\times8$ elements. The dataset is labeled into three classes, corresponding to positive causality $+1$, negative causality $-1$, and no causality $0$.
Prior to training, a normalization step is applied to each heatmap to scale its values between $0$ and $1$.

\subsection{Model Architecture} 
The \ac{hqcnn} was constructed consisting of a classical convolutional neural network (CNN) with a quantum neural network (QNN). The classical CNN extracts features from the heatmaps, reduces their dimensionality, and feeds them into a parameterized quantum circuit. The output of the quantum circuit is processed by a classical classification layer.

\subsection{Quantum Circuit Design}
The quantum component consists of a parameterized quantum circuit constructed with two key elements. The first one is the quantum feature encoding circuits, which are used to encode classical inputs onto quantum circuits. Several types of feature mapping will be tested in this work. The second part is the ansatz, and because the problem has no chemical interpretation, a hardware-efficient ansatz was chosen specifically the TwoLocal ansatz. This part of the quantum circuit tested, how the difference in depth affects the model.

The quantum circuit was interfaced with PyTorch through the Qiskit Machine Learning TorchConnector, enabling end-to-end gradient-based training.

\subsection{Training Procedure} 
The training was conducted using the following standardized procedure. Cross-entropy loss was employed for multi-class classification. As an optimization method Stochastic Gradient Descent with Nesterov momentum and a small weight decay was used to optimize model parameters. A batch size of 64 and an initial learning rate of 0.01 were utilized.

\subsection{Evaluation and Metrics} 
To thoroughly characterize the models, a comprehensive set of evaluation metrics was computed. Training and validation accuracies over epochs, Generalization gap (difference between training and validation accuracy),  Early learning slope (rate of accuracy improvement in the early stages),  Overfitting drop (difference between peak and final validation accuracy),  Fluctuation metrics (standard deviation and mean of local accuracy changes),  Stability ratio (relationship between fluctuations in training and validation),  Silhouette score (clustering quality after PCA dimensionality reduction),  Fisher Discriminant Ratio (class separability after model output). 

\ac{pca} was also applied at multiple stages — after the classical feature extractor, after feature mapping, and after quantum processing — to visualize and better understand the evolution of data separability across the network.

\subsection{Experimental Design} 
Each experimental configuration was independently trained and evaluated: Different ansatz depths were compared while keeping other parameters fixed. Multiple feature mappings were tested across otherwise identical model setups. Each model was trained for sufficient epochs to ensure convergence, with performance metrics and diagnostic plots saved for analysis. 

This systematic methodology enabled an in-depth exploration of how architectural choices in hybrid quantum-classical models affect learning dynamics, generalization, and overall model performance.

\section{Hybrid Neural Network}

The \ac{hqcnn} model is a hybrid neural network architecture that combines classical convolutional neural networks with a quantum neural network to perform classification tasks. The model is parameterized by the number of output classes and the number of qubits.

The architecture begins with a classical feature extractor consisting of three convolutional blocks. Each block includes a convolutional layer with a $3\times3$ kernel and padding, followed by a ReLU activation function, a max-pooling layer with a $2\times2$ window and stride of 2, and a dropout layer with a dropout rate of 0.5. The number of feature channels increases through the network from 16 to 32 and then to 64, allowing the network to capture more complex representations of the input data.

After feature extraction, the resulting feature maps are flattened and passed through a fully connected layer that reduces the dimensionality to match the number of qubits required by the quantum layer. This step ensures that the input to the quantum circuit is appropriately sized.

The quantum neural network is constructed using a parameterized quantum circuit composed of a \texttt{ZZFeatureMap} for encoding classical input features into quantum states, followed by a \texttt{TwoLocal} ansatz with linear entanglement and one repetition. The quantum circuit is integrated into the PyTorch framework using the \texttt{TorchConnector} and the \texttt{EstimatorQNN} module, allowing backpropagation through the quantum layer. The \texttt{EstimatorQNN} is configured with input gradients enabled to facilitate training.

The output from the quantum circuit is then passed through a classical linear layer that maps the quantum output to the final class scores. The final classifier is a single fully connected layer that outputs a vector with dimensionality equal to the number of target classes.

The model is designed to benefit from both classical convolutional feature extraction and quantum-enhanced representation learning, potentially offering improved performance in settings where quantum computation can capture complex correlations within the data.

The whole \ac{hqcnn} is described in\cref{fig:hybridqnn}.

\begin{figure}[h]
\centering
\begin{tikzpicture}[node distance=1.8cm and 2cm, auto]

\node[block] (input) {Input Image};
\node[block, below=of input] (cnn) {CNN Feature Extractor \\ (Conv + ReLU + MaxPool + Dropout)};
\node[block, below=of cnn] (flatten) {Flatten};
\node[block, below=of flatten] (reduce) {Dimension Reduction \\ (Linear Layer)};
\node[block, below=of reduce] (qnn) {Quantum Neural Network \\ (Feature Map + Ansatz)};
\node[block, below=of qnn] (classifier) {Classifier \\ (Linear Layer)};
\node[block, below=of classifier] (output) {Output Class Scores};

\draw[arrow] (input) -- (cnn);
\draw[arrow] (cnn) -- (flatten);
\draw[arrow] (flatten) -- (reduce);
\draw[arrow] (reduce) -- (qnn);
\draw[arrow] (qnn) -- (classifier);
\draw[arrow] (classifier) -- (output);

\end{tikzpicture}
\caption{Architecture of the Hybrid Quantum CNN Model}
\label{fig:hybridqnn}
\end{figure}

\section{Feature Mapping}
In hybrid quantum-classical machine learning models, the feature map is a critical component that defines how classical input data is encoded into quantum states. The design of the feature map directly affects the model’s ability to create useful representations within the quantum Hilbert space, influencing the expressive power of the quantum circuit and, ultimately, the model’s classification performance.

In this thesis, several types of feature maps were systematically investigated to understand their impact on the learning dynamics and generalization abilities of hybrid models. Each feature map encodes the input into a quantum circuit differently, varying in complexity, entanglement structure, and number of repetitions.

The feature maps explored can be categorized as follows. 

\subsection{ZZ Feature Maps}
The ZZ feature maps encode input features through two-qubit Pauli-Z interactions. These mappings introduce entanglement between qubits and are often used to capture higher-order correlations among input features.

\begin{figure}[H]
    \centering
    \includegraphics[width=0.8\textwidth]{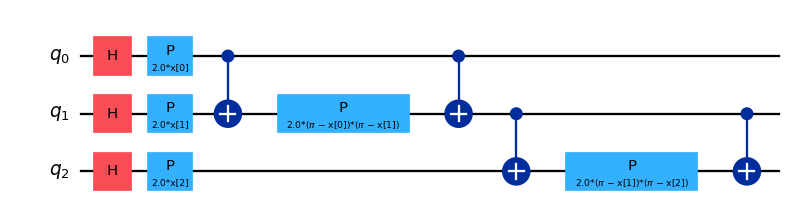}
    \caption{ZZ feature map with 1 repetition (linear entanglement).}
    \label{fig:zz_1rep_no_entanglement}
\end{figure}

\begin{figure}[H]
    \centering
    \includegraphics[width=0.8\textwidth]{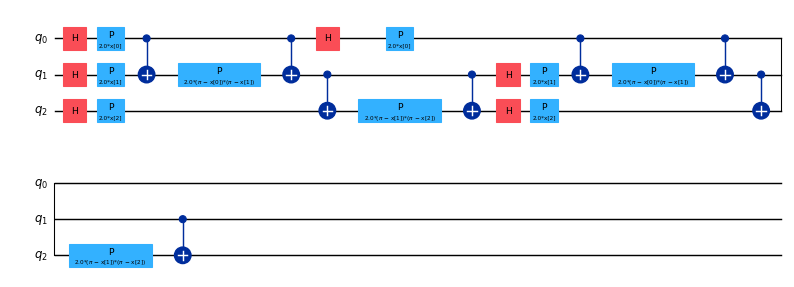}
    \caption{ZZ feature map with 2 repetitions (linear entanglement).}
    \label{fig:zz_2rep_linear_entanglement}
\end{figure}

\begin{figure}[H]
    \centering
    \includegraphics[width=0.8\textwidth]{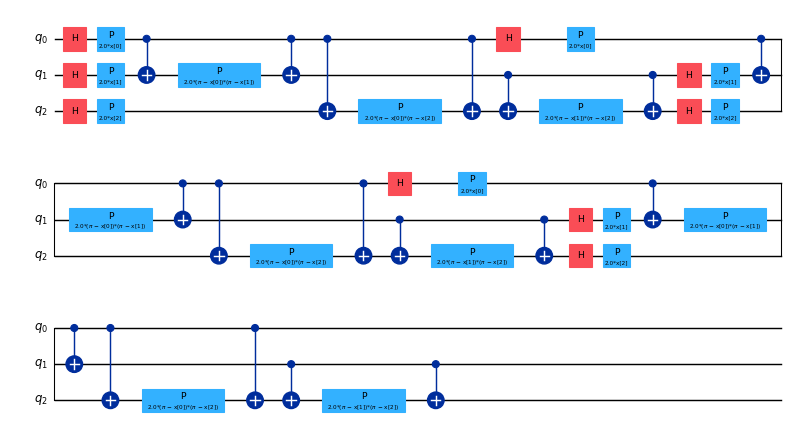}
    \caption{ZZ feature map with 3 repetitions (full entanglement).}
    \label{fig:zz_3rep_full_entanglement}
\end{figure}

Three different ZZ feature map configurations were tested. A ZZ feature map with one repetition and linear entanglement, visualized in \cref{fig:zz_1rep_no_entanglement}. A ZZ feature map with two repetitions and linear entanglement, shown in \cref{fig:zz_2rep_linear_entanglement}. A ZZ feature map with three repetitions and full entanglement, depicted in \cref{fig:zz_3rep_full_entanglement}.

These variations were intended to study how increasing the depth and entanglement complexity affects the model’s representational capacity and learning stability.

\subsection{Z Feature Maps} 
The Z feature maps use only single-qubit rotations around the Z-axis for encoding. These mappings are simpler and introduce no entanglement, relying solely on local transformations.

\begin{figure}[htbp]
    \centering
    \includegraphics[width=0.4\textwidth]{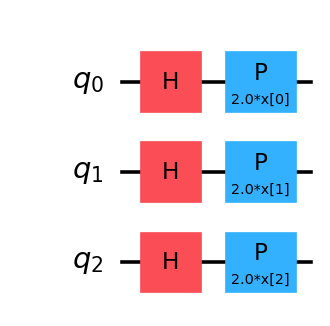}
    \caption{Z feature map with 1 repetition.}
    \label{fig:z_1rep}
\end{figure}

\begin{figure}[htbp]
    \centering
    \includegraphics[width=0.8\textwidth]{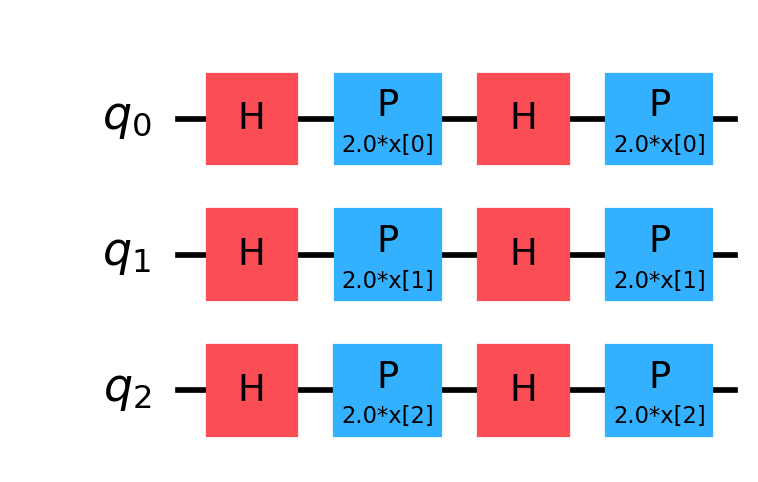}
    \caption{Z feature map with 2 repetitions.}
    \label{fig:z_2rep}
\end{figure}

\begin{figure}[htbp]
    \centering
    \includegraphics[width=0.8\textwidth]{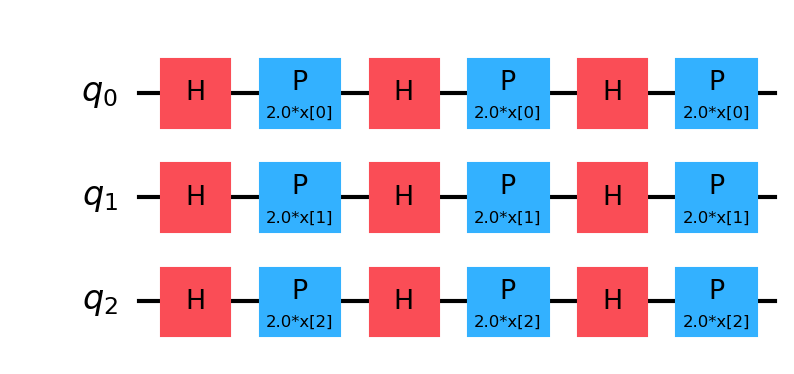}
    \caption{Z feature map with 3 repetitions.}
    \label{fig:z_3rep}
\end{figure}
Three versions of the Z feature maps were evaluated. A Z feature map with one repetition, illustrated in \cref{fig:z_1rep}. A Z feature map with two repetitions, shown in \cref{fig:z_2rep}. A Z feature map with three repetitions, displayed in \cref{fig:z_3rep}. 

Testing multiple repetitions allowed the investigation of whether deeper local transformations alone can sufficiently capture feature complexity without relying on entanglement.

\subsection{Pauli Rotations-Based Feature Maps}
More sophisticated feature maps were constructed using Pauli rotations around multiple axes (X, Y, Z) and different forms of entanglement. These mappings aim to introduce richer non-linear transformations into the quantum state.

\begin{figure}[htbp]
    \centering
    \includegraphics[width=0.8\textwidth]{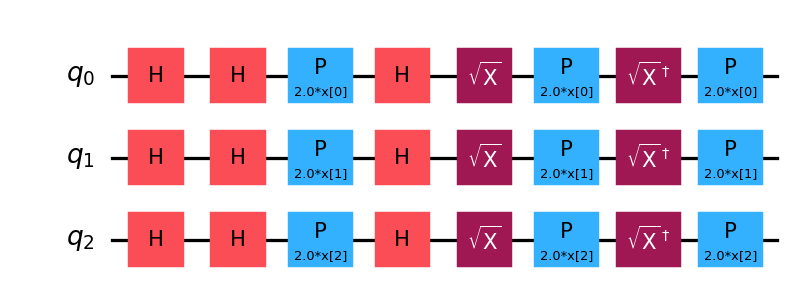}
    \caption{Pauli XYZ feature map with 1 repetition.}
    \label{fig:pauli_xyz_1rep}
\end{figure}

\begin{figure}[htbp]
    \centering
    \includegraphics[width=0.8\textwidth]{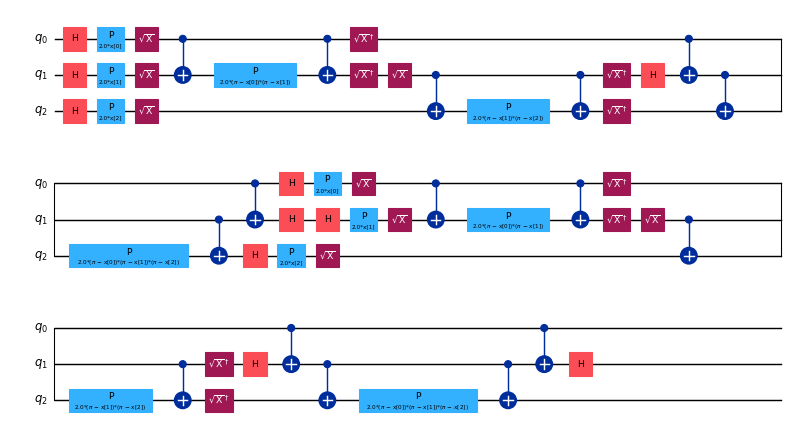}
    \caption{Pauli Z-YY-ZXZ feature map with linear entanglement.}
    \label{fig:pauli_z-yy-zxz_linear_ent}
\end{figure}

\begin{figure}[htbp]
    \centering
    \includegraphics[width=0.8\textwidth]{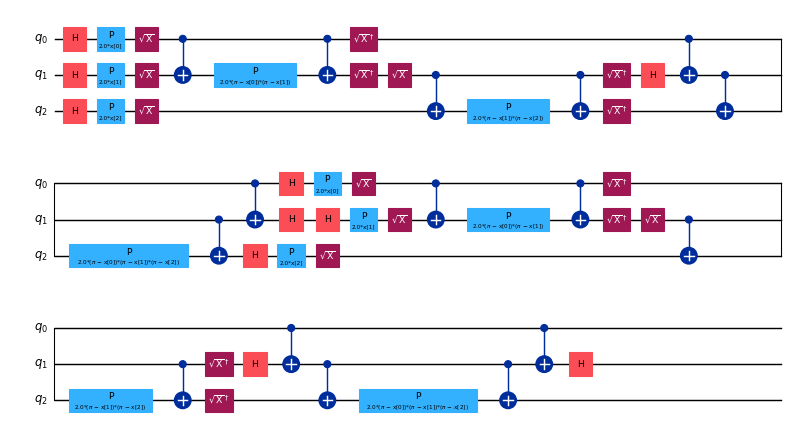}
    \caption{Pauli Z-YY-ZXZ feature map with 2 repetitions.}
    \label{fig:pauli_z-yy-zxz_2rep}
\end{figure}

Three Pauli rotations-based feature maps were tested. A Pauli XYZ feature map with one repetition, shown in \cref{fig:pauli_xyz_1rep}. A Pauli Z-YY-ZXZ feature map with linear entanglement, visualized in \cref{fig:pauli_z-yy-zxz_linear_ent}. A Pauli Z-YY-ZXZ feature map with two repetitions, depicted in \cref{fig:pauli_z-yy-zxz_2rep}. 

These mappings were selected to assess the benefit of using more expressive quantum feature maps that leverage multi-axis rotations and structured entanglement patterns.

\subsection{Summary} The diversity of feature maps evaluated — from simple Z-rotations to complex Pauli-based circuits — enabled a comprehensive study of how feature encoding choices influence hybrid model behavior. By systematically varying the depth, entanglement, and rotation complexity, it became possible to isolate and understand the role of the feature map in shaping the learning dynamics, stability, and generalization ability of the hybrid quantum-classical models.

\section{Implementation Details}

The ac{hqcnn}  model was implemented using the Python programming language, leveraging several specialized machine learning and quantum computing libraries. The classical neural network components, including the convolutional layers, dropout, and fully connected layers, were built using \texttt{PyTorch}, a popular deep learning framework known for its flexibility and dynamic computation graph.

For the quantum neural network components, the implementation utilized the \texttt{Qiskit Machine Learning} library, which provides high-level tools for constructing and training quantum neural networks. In particular, the \texttt{EstimatorQNN} module was employed, enabling integration between parameterized quantum circuits and PyTorch's automatic differentiation engine. The quantum circuits were composed using \texttt{Qiskit}'s quantum circuit construction utilities, featuring a \texttt{ZZFeatureMap} for data encoding and a \texttt{TwoLocal} ansatz for parameterized transformations.

The integration between the classical and quantum parts was managed by the \texttt{TorchConnector}, which allowed the quantum computations to be treated as trainable layers within the overall neural network, ensuring seamless backpropagation.

\section{Code Availability}

The complete source code supporting this work is publicly available on Zenodo and can be accessed via the following DOI:
\begin{center}
    \begin{quote}
\href{https://doi.org/10.5281/zenodo.15309749}{\fbox{\textsf{DOI: 10.5281/zenodo.15309749}}}
\end{quote}
\end{center}

and also on Gitlab:
\begin{center}
\begin{quote}
\href{https://gitlab.com/illesova.silvie.scholar/leveraging-quantum-layers-in-classical-neural-networks}{
  \begin{tabular}{c}
    \includegraphics[height=1cm]{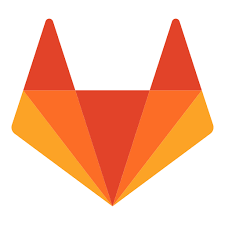} \\
    \fbox{\textsf{https://gitlab.com/illesova.silvie.scholar/leveraging-quantum-layers-in-classical-neural-networks}}
  \end{tabular}
}
\end{quote}
\end{center}

The repository includes the full implementation of the \ac{hqcnn} model, scripts for training and evaluation, and instructions for reproducing the experiments. By releasing the code, we aim to promote transparency, reproducibility, and further research in the integration of quantum computing with classical deep learning methods.

\chapter{Results} \label{chap:res}

In this chapter, we will have a look at the result of the analysis that was performed on the hybrid neural network described in the previous section. At first, the input data will be briefly introduced, then follows the analysis of how the size of the quantum layer affects the whole training of the model, and at last a large analysis of different feature mapping, which was also introduced in the previous section, is shown.

\section{Input Data}
The training data for this study comprises of two-dimensional heatmaps, describing the relationship between two variables. Each heatmap is $8 \times 8$ matrix, where each element describes the value of the relationship between the two variables.
These patterns are used in hybrid machine learning to train a model capable of distinguishing between different causality directions, or telling us that there is no causality. 

\begin{figure}[H]
    \centering
    \begin{subfigure}[b]{0.3\textwidth}
        \includegraphics[width=\textwidth]{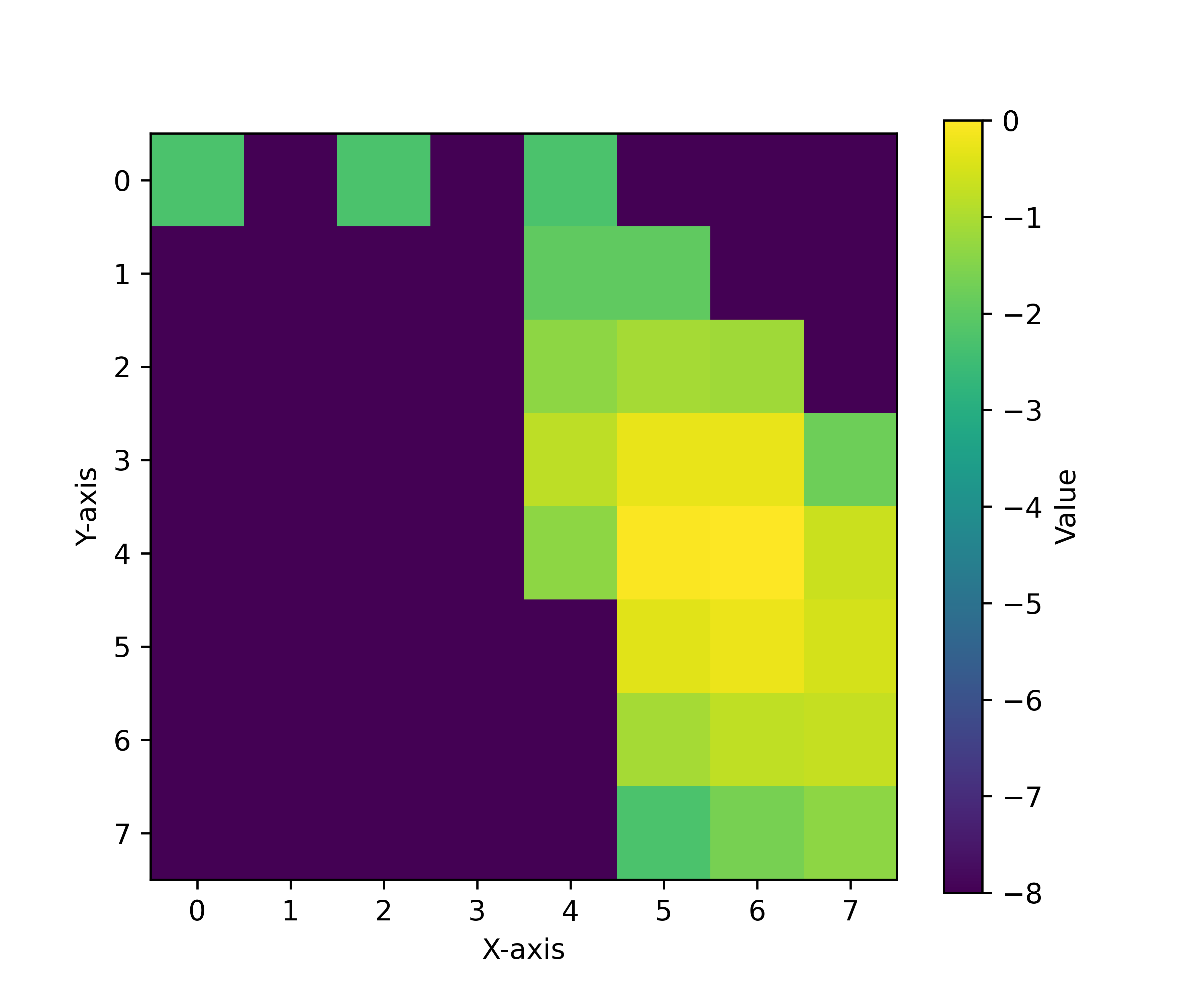}
        \caption{Causality Direction: 1}
    \end{subfigure}
    \hfill
    \begin{subfigure}[b]{0.3\textwidth}
        \includegraphics[width=\textwidth]{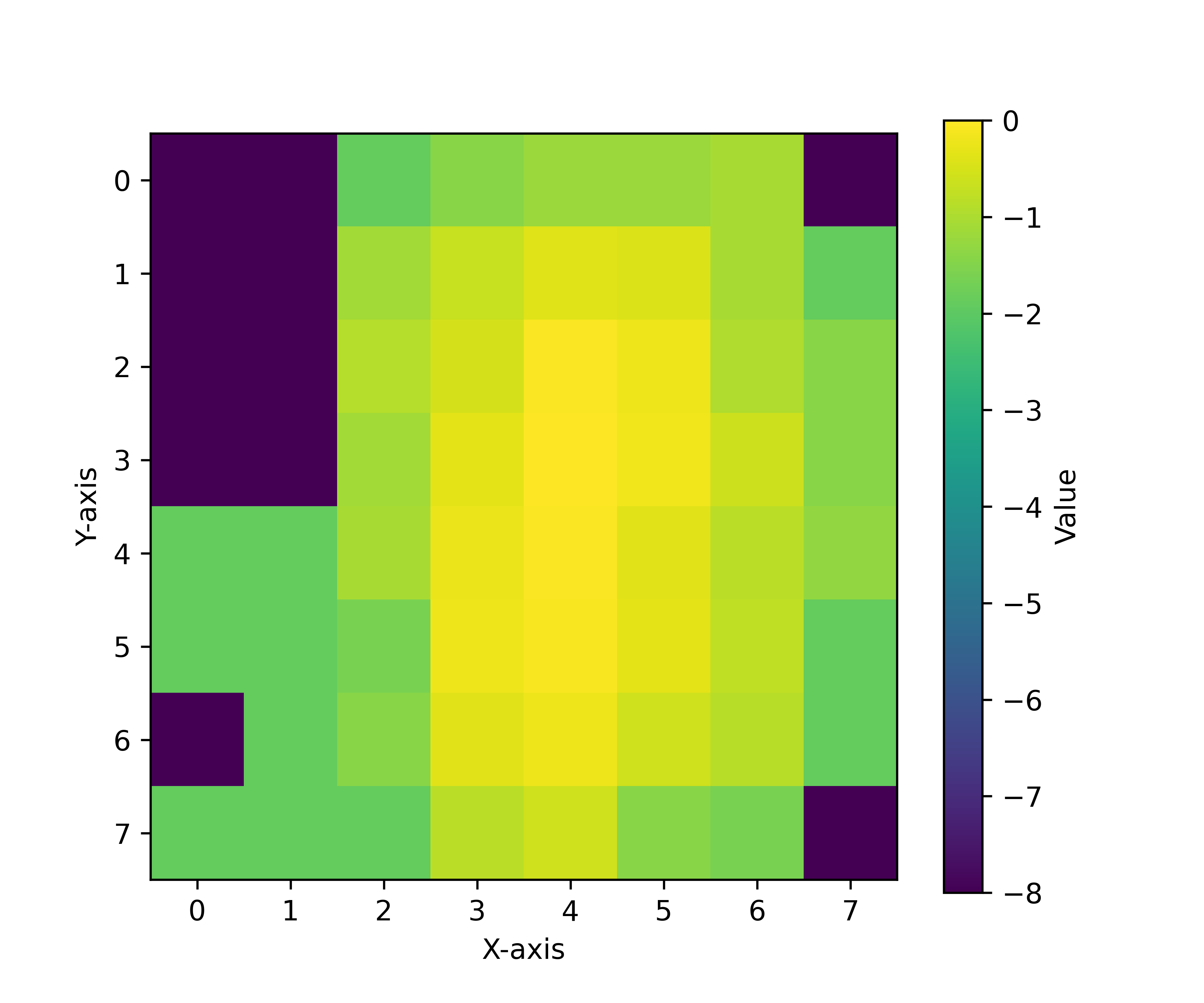}
        \caption{Causality Direction: 1}
    \end{subfigure}
    \hfill
    \begin{subfigure}[b]{0.3\textwidth}
        \includegraphics[width=\textwidth]{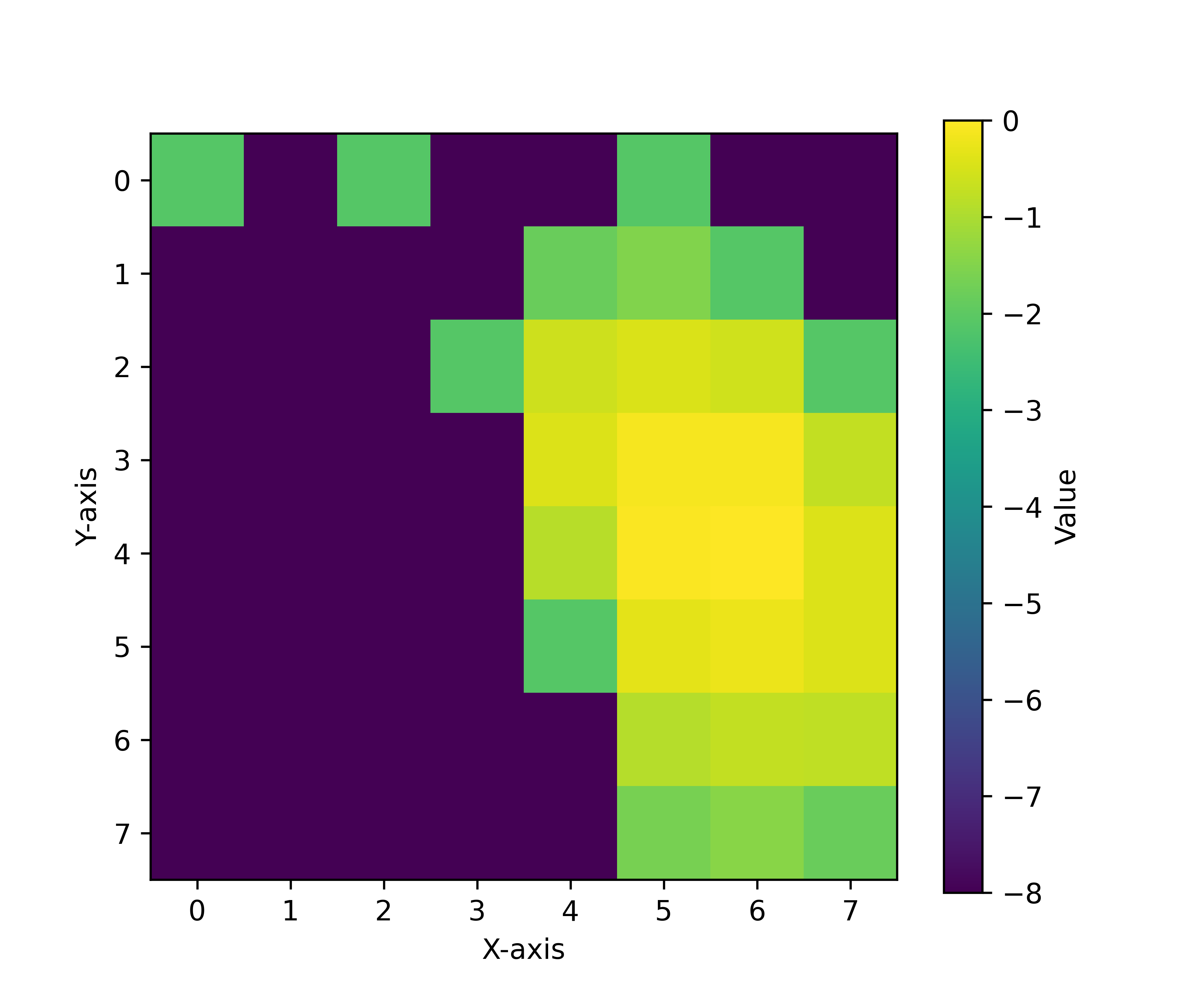}
        \caption{Causality Direction: 1}
    \end{subfigure}
    \caption{Heatmaps showing positive causality direction}
    \label{fig:causality_heatmaps_1}
\end{figure}
The data set is divided into three categories. Where data with positive causality is assigned number $1$, data with negative causality has label $-1$, and data where there is no causal relationship has label $0$.

\begin{figure}[H]
    \centering
    \begin{subfigure}[b]{0.3\textwidth}
        \includegraphics[width=\textwidth]{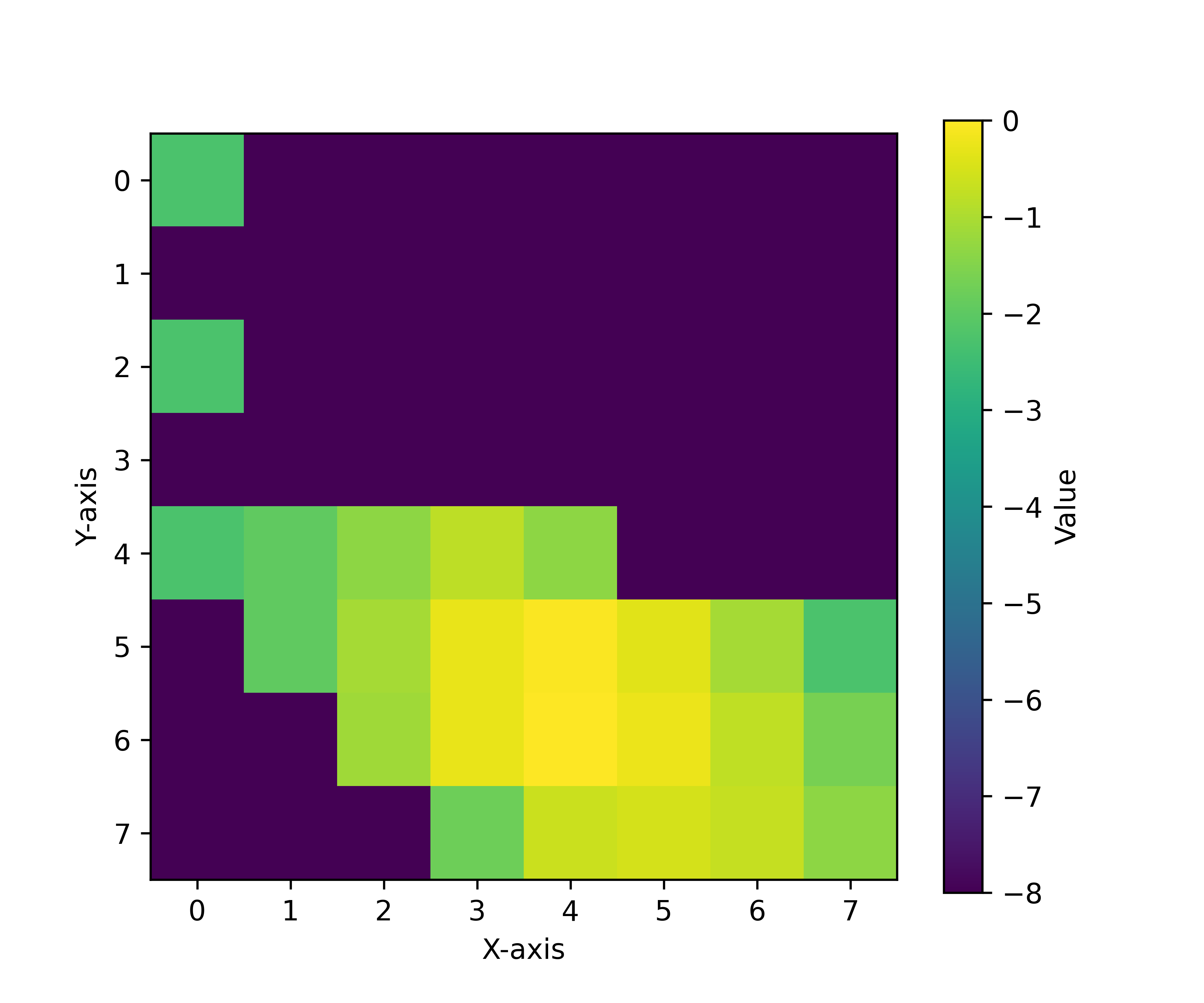}
        \caption{Causality Direction: -1}
    \end{subfigure}
    \hfill
    \begin{subfigure}[b]{0.3\textwidth}
        \includegraphics[width=\textwidth]{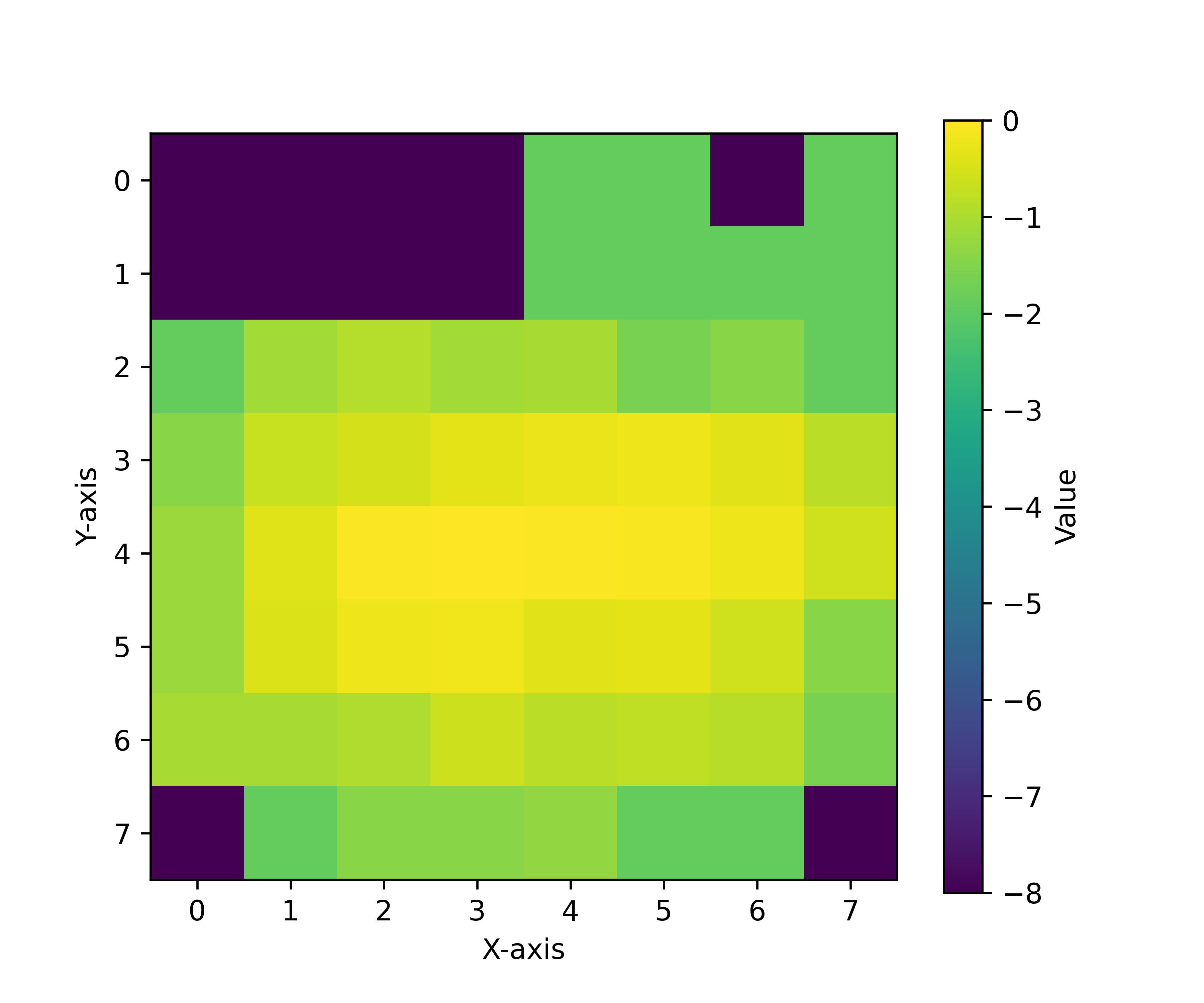}
        \caption{Causality Direction: -1}
    \end{subfigure}
    \hfill
    \begin{subfigure}[b]{0.3\textwidth}
        \includegraphics[width=\textwidth]{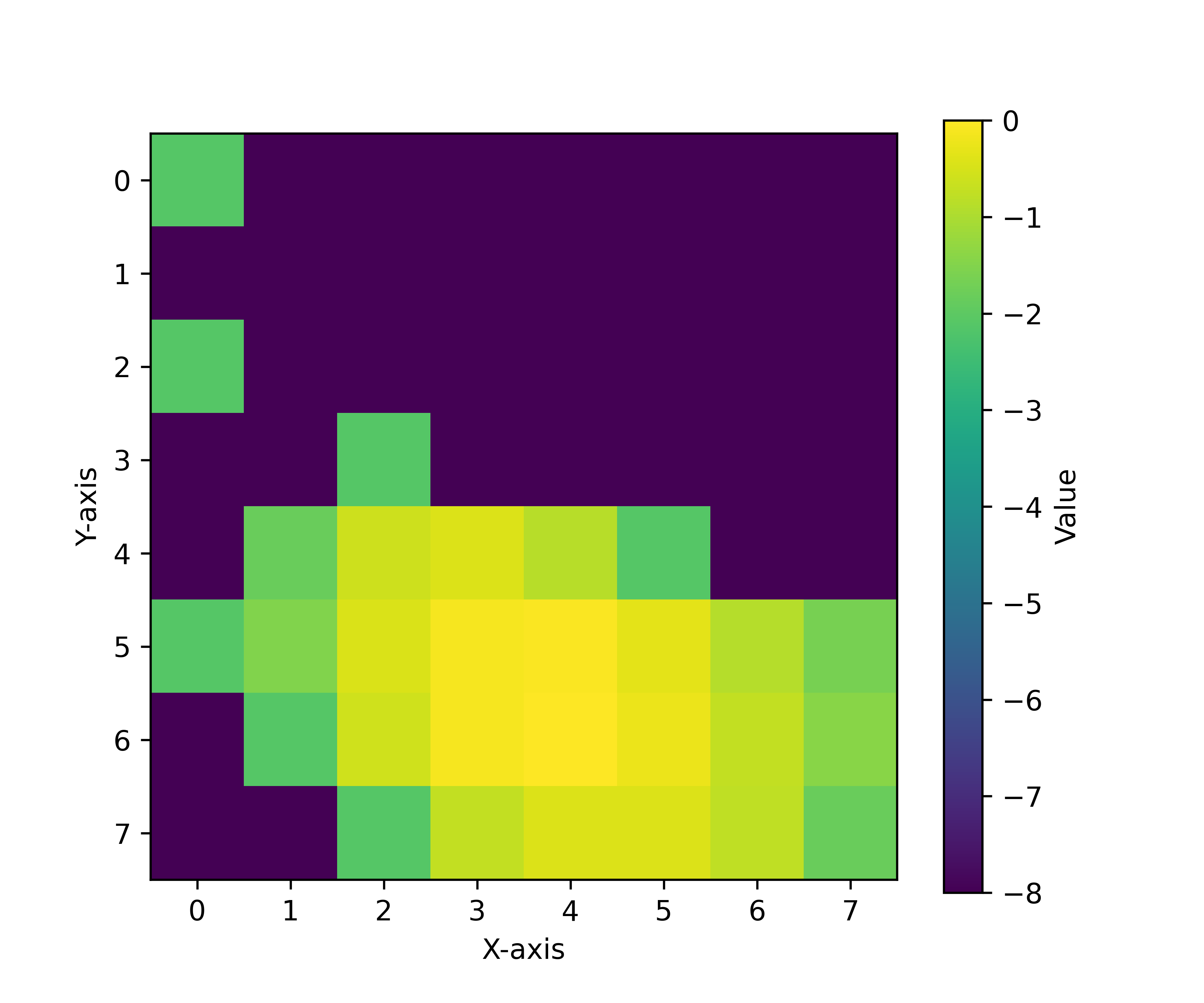}
        \caption{Causality Direction: -1}
    \end{subfigure}
    \caption{Heatmaps showing negative causality direction}
    \label{fig:causality_heatmaps_minus_1}
\end{figure}

\begin{figure}[H]
    \centering
    \begin{subfigure}[b]{0.3\textwidth}
        \includegraphics[width=\textwidth]{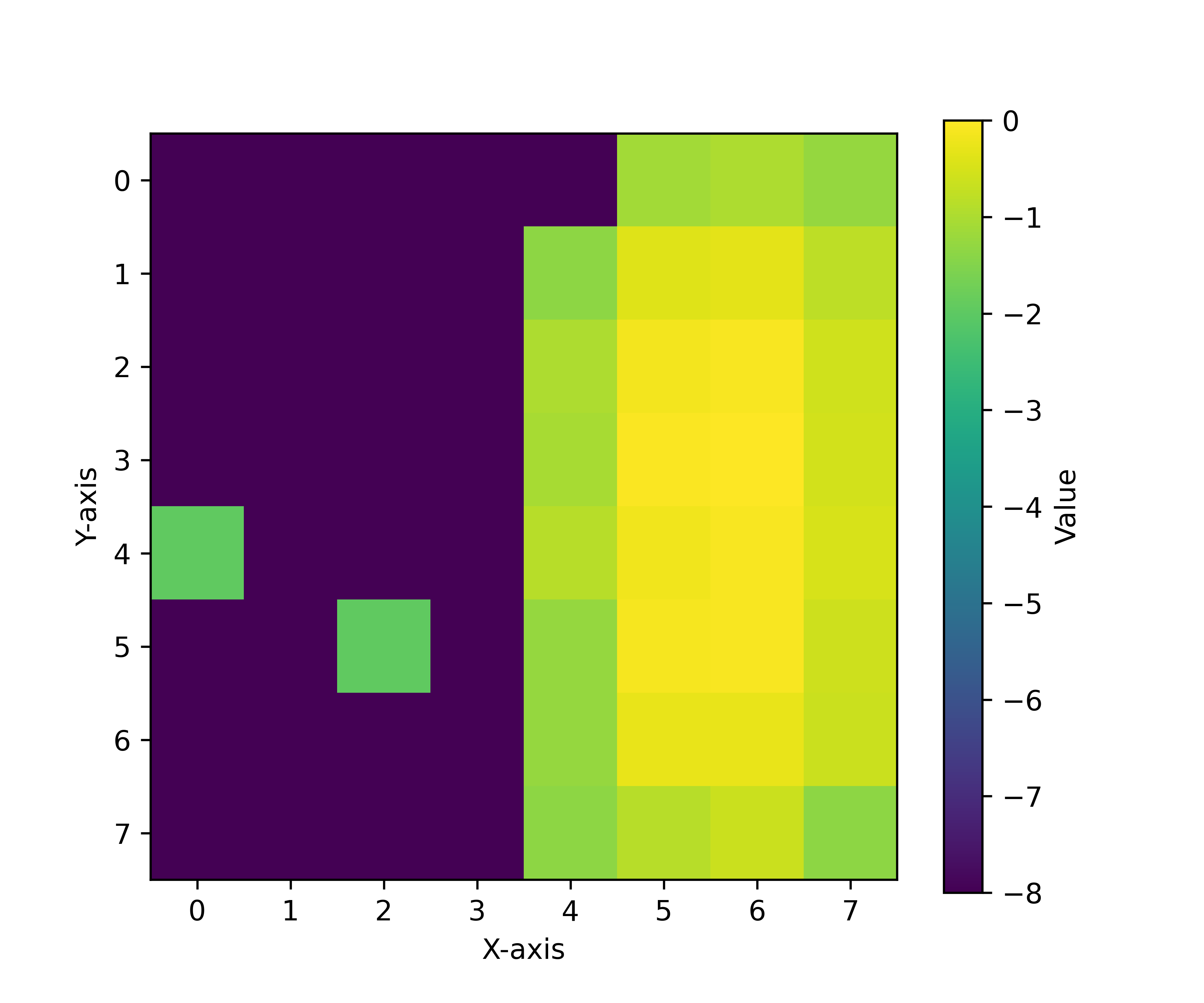}
        \caption{No Causality}
    \end{subfigure}
    \hfill
    \begin{subfigure}[b]{0.3\textwidth}
        \includegraphics[width=\textwidth]{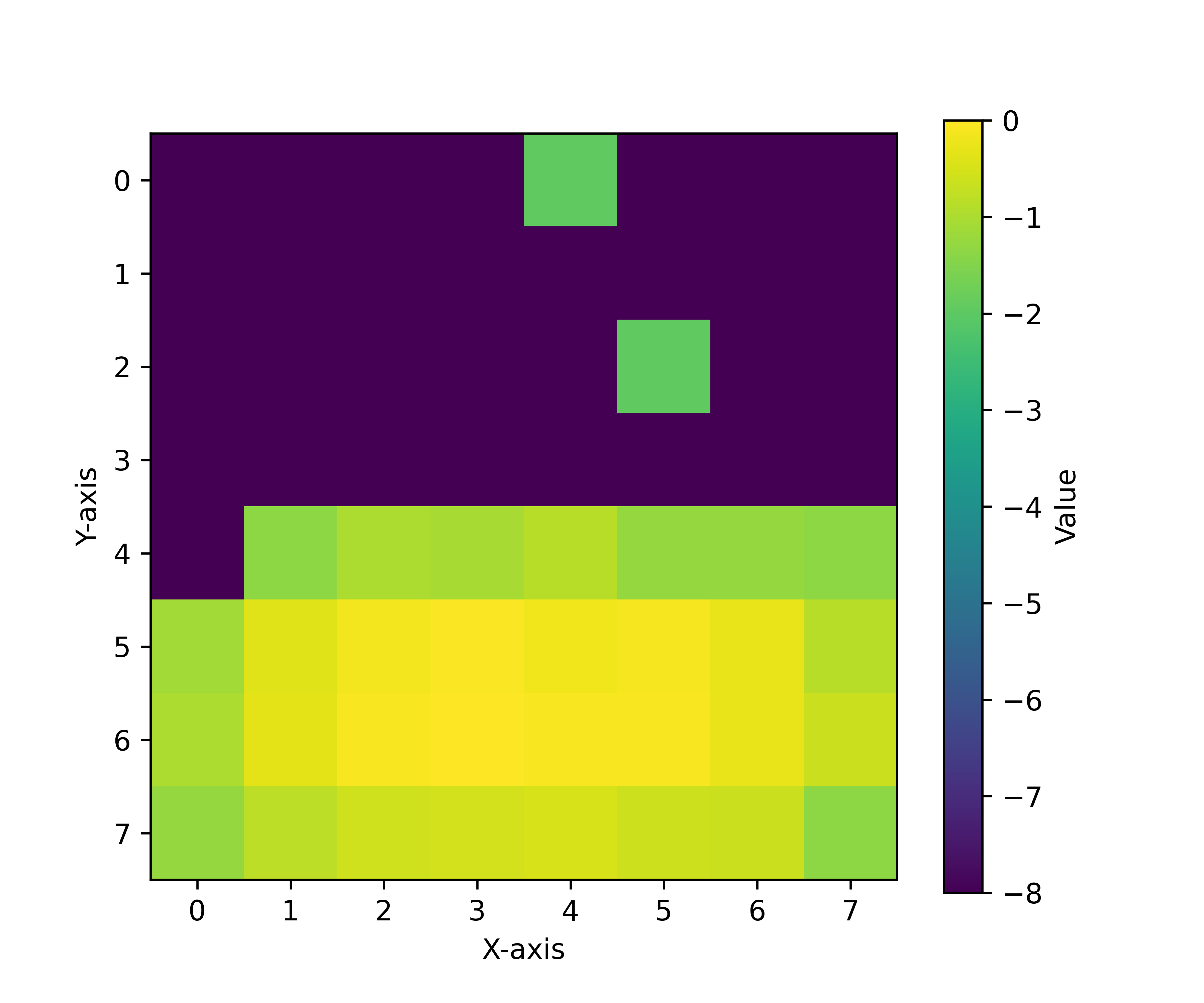}
        \caption{No Causality}
    \end{subfigure}
    \hfill
    \begin{subfigure}[b]{0.3\textwidth}
        \includegraphics[width=\textwidth]{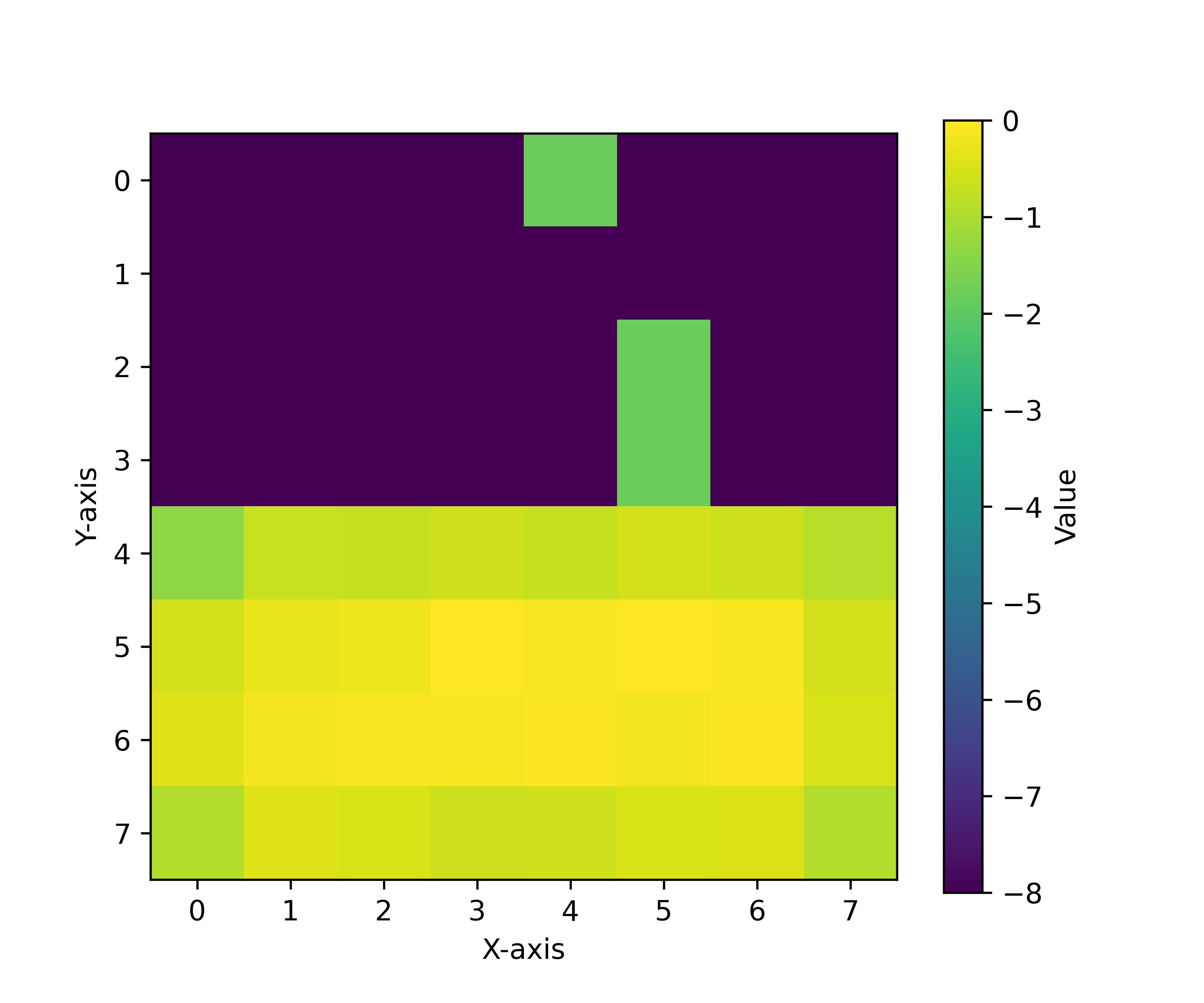}
        \caption{No Causality}
    \end{subfigure}
    \caption{Heatmaps showing data with no causality}
    \label{fig:causality_heatmaps_0}
\end{figure}

Examples of the data with positive causality direction are visible in \cref{fig:causality_heatmaps_1}, for negative causality look at \cref{fig:causality_heatmaps_minus_1}. Data with no causality is shown in \cref{fig:causality_heatmaps_0}.

\section{Ansatz Depth}
In this part of the work, an investigation of the impact of the size of the ansatz and thus the number of parameters in the quantum part of our hybrid neural network was done. The ansatz size was gradually increased from one to three repetitions, while all other settings remained the same throughout the whole training process. With all three different combinations, the training process was done from scratch, so that thorough analysis could be done and the impact of increasing the flexibility of quantum layers could be determined.
In the following text, we will provide a breakdown of how circuit depth affects different aspects of the training as well as the overall reliability of our model.

\begin{table}
\centering
\caption{Final Accuracy Metrics}
\begin{tabular}{lcc}
\toprule
\textbf{Repetitions} & \textbf{Training Accuracy} & \textbf{Validation Accuracy} \\
\midrule
1 Repetition  & 0.8467 & 0.8721 \\
2 Repetitions & 0.9121 & 0.8955 \\
3 Repetitions & 0.9004 & 0.9111 \\
\bottomrule
\end{tabular}
\label{tab:final_accuracy}
\end{table}
\begin{figure}[H]
    \centering
    \includegraphics[width=0.8\linewidth, trim=0 0 0 50, clip]{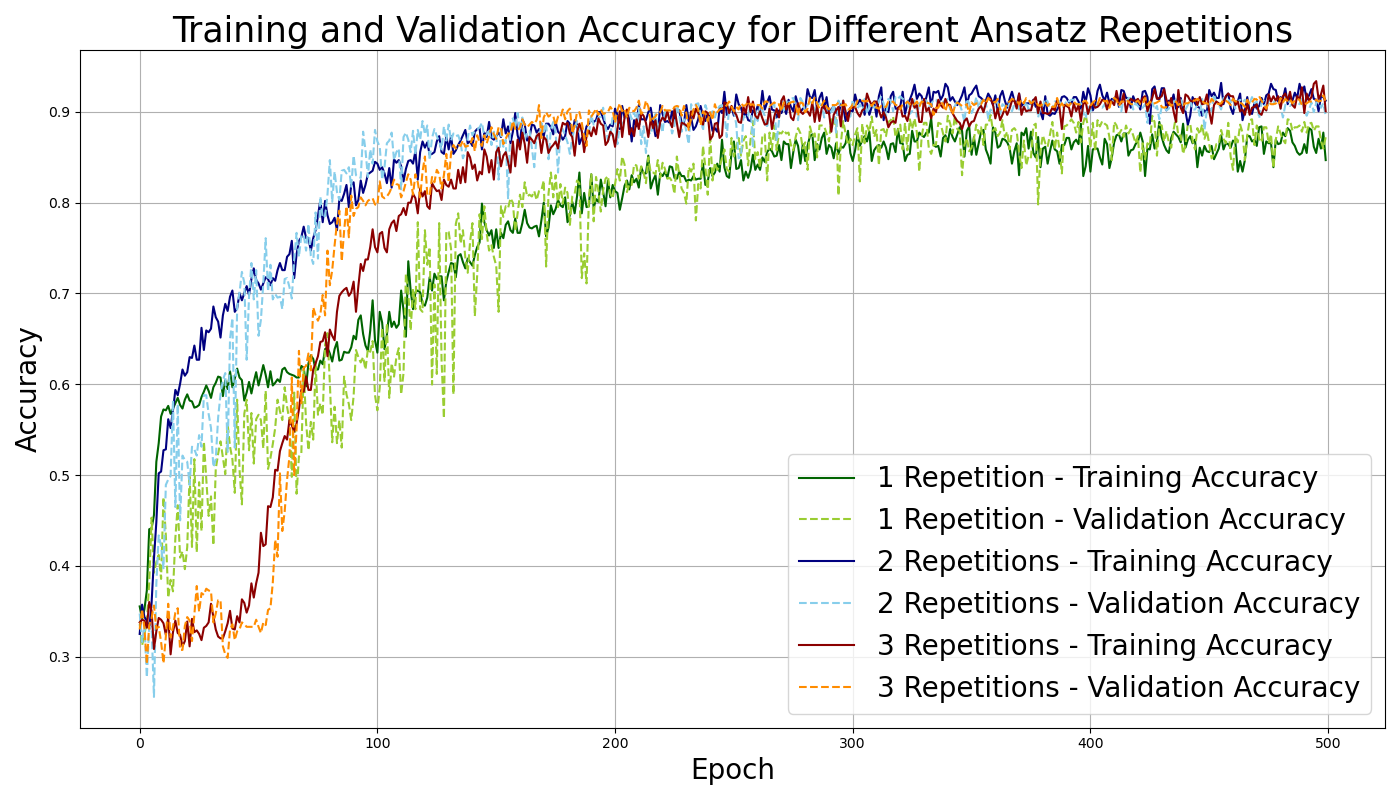}
    \caption{Training process for all combinations}
    \label{fig:ansatz_all}
\end{figure}
The first parameter we will have a look at is the overall accuracy of our trained method which is shown in \cref{tab:final_accuracy}. Here we can see that accuracy on both the training and validation data increases with the number of repetitions, with the biggest difference being between one repetition and two repetitions, while the improvement between two and three repetitions is modest, which suggests diminishing returns when increasing the size of the ansatz further. Still, the highest accuracy on the validation dataset was achieved with the three repetitions, which suggests that an ansatz with a larger number of parameters enhances the model's ability to generalize.
When we have a look at how the training and validation accuracies evolve during the training process shown in \cref{fig:ansatz_all}, where we can see these accuracies plotted across 500 epochs for all three models. In each case, the accuracies increase with increasing number of epochs. But if we have a closer look at the case-by-case plot, we see some interesting differences. 

\begin{figure}[H]
    \centering
    \includegraphics[width=0.8\linewidth, trim=0 0 0 22, clip]{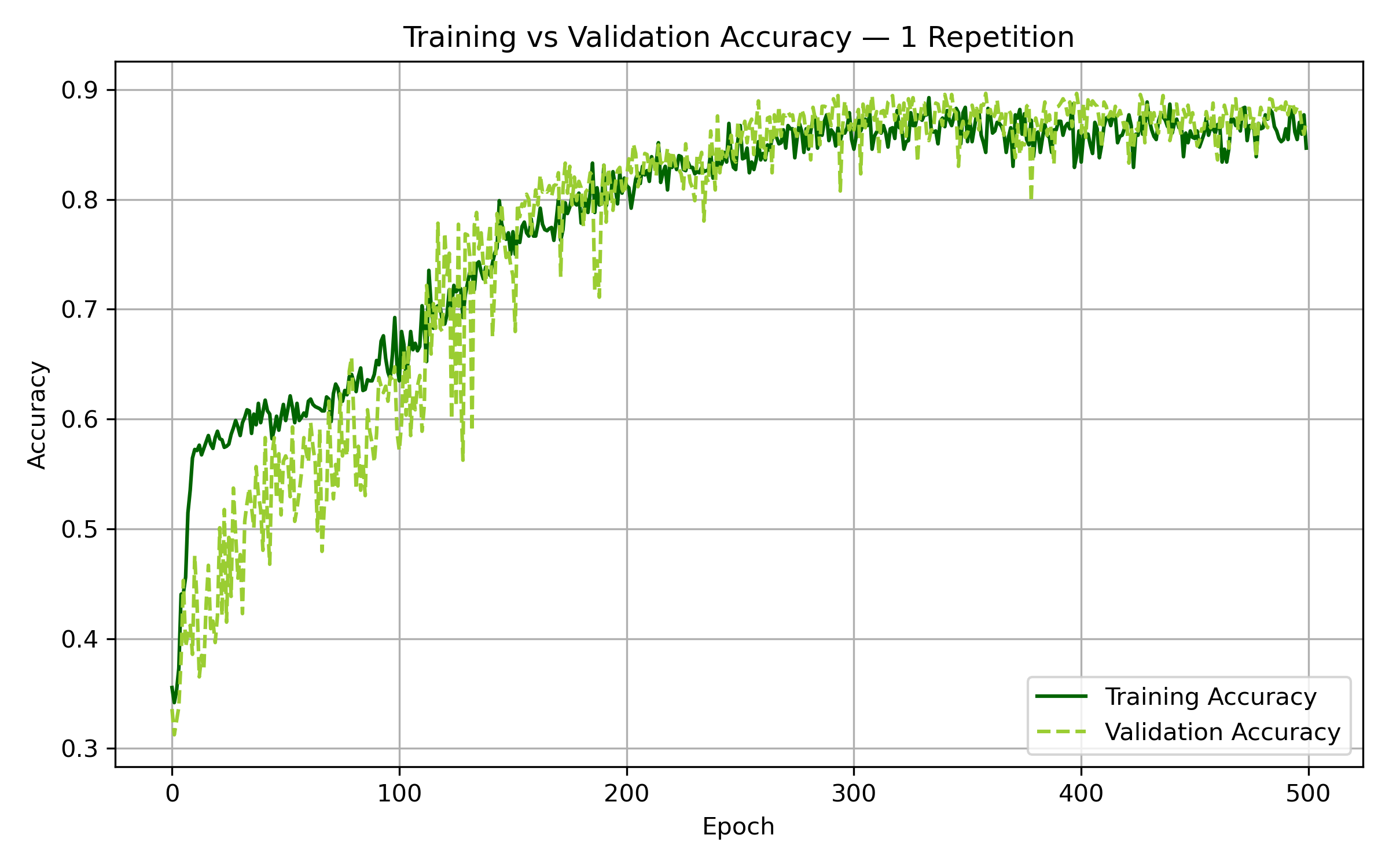}
    \caption{Training process with 1 ansatz repetition}
    \label{fig:ansatz_1}
\end{figure}

\begin{figure}[H]
    \centering
    \includegraphics[width=0.8\linewidth, trim=0 0 0 22, clip]{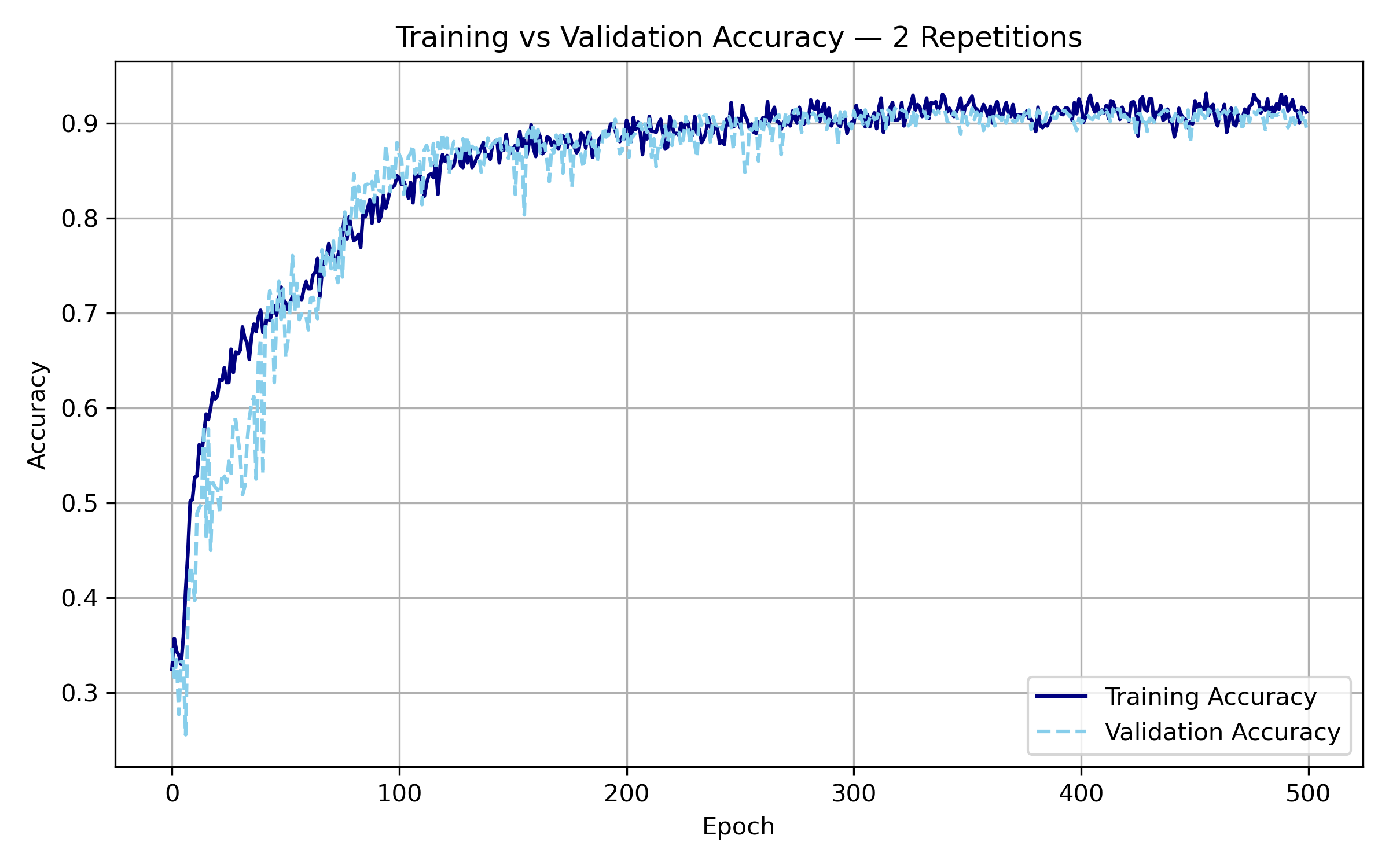}
    \caption{Training process with 2 ansatz repetitions}
    \label{fig:ansatz_2}
\end{figure}

For the case with just one ansatz repetition, shown in \cref{fig:ansatz_1}, we can see a steep initial increase in accuracy, but then there is a slight plateau, showing that the model has slight problems with further optimization. Also, we can see quite large fluctuations, especially in the accuracy computed on validation data.
When we have a look at \cref{fig:ansatz_2}, where the quantum layer has 2 ansatz repetitions, we still see the initial steep increase, and compared to the previous model, the decrease in learning speed is slower and more gentle. Also, the fluctuations appear to be smaller. 

\begin{figure}[H]
    \centering
    \includegraphics[width=0.8\linewidth, trim=0 0 0 22, clip]{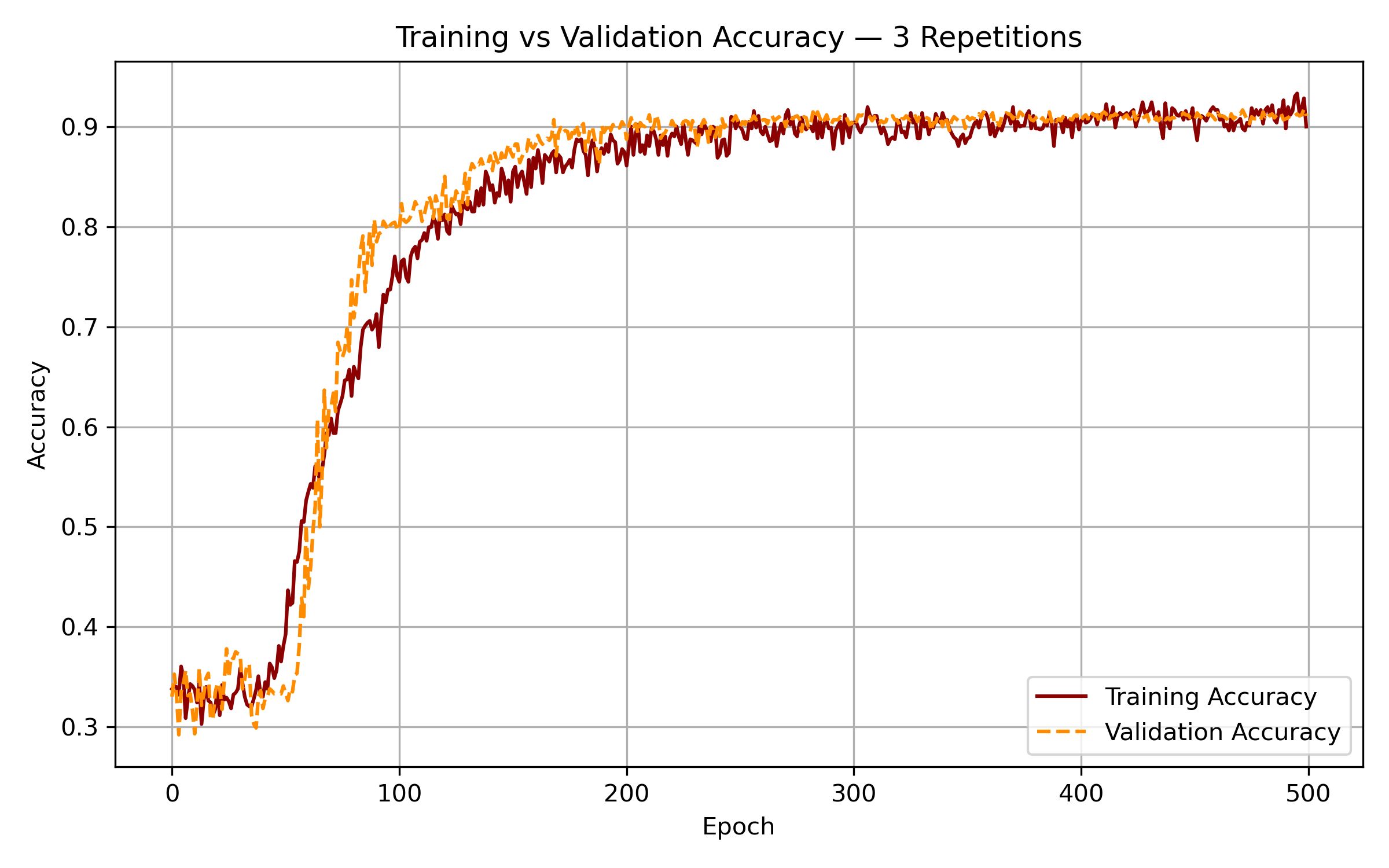}
    \caption{Training process with 3 ansatz repetitions}
    \label{fig:ansatz_3}
\end{figure}

In the case of three ansatz repetition in \cref{fig:ansatz_3}, we can see a noticeable difference in the first few epochs, where there is a plateau at the beginning of the training, pointing towards the fact, that when we increase the number of parameters, the optimization is starting to be a bit more difficult. But as we can see, after the initial problems, the increase in the model's flexibility proves its worth and both of the accuracies grow rapidly. What is also an interesting point, is the fact, that the fluctuations are the smallest in this case.

\begin{figure}
    \centering
    \includegraphics[width=0.8\linewidth]{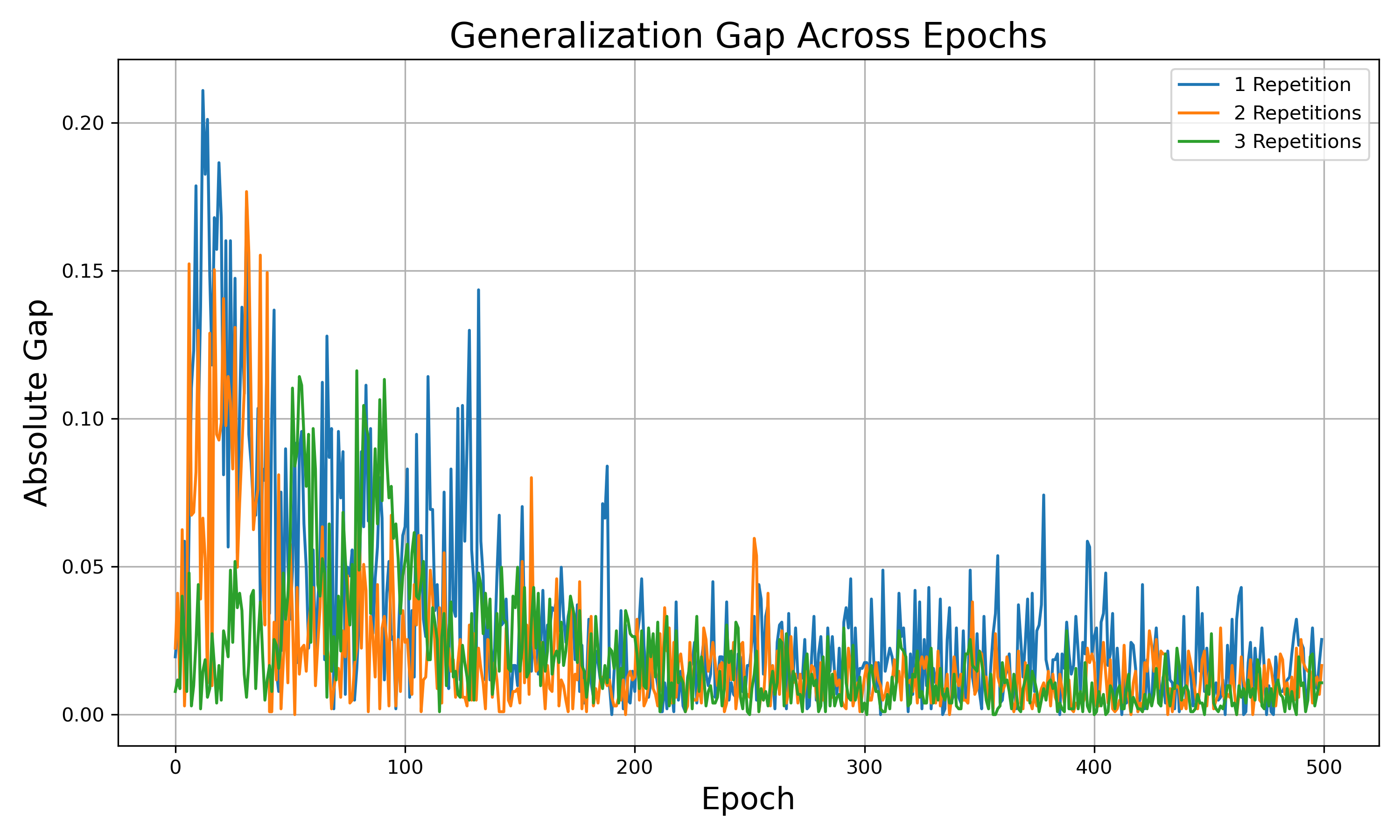}
    \caption{Generalization Gap for All Setups}
    \label{fig:gen_gap}
\end{figure}

Overall we can see that using multiple repetitions during the training leads to a more stable learning process at the cost of the initial plateau. Also, the smaller fluctuations especially in the validation data point to the fact, that the repetition of ansatz may act as some kind of a regularization, that improves the model's ability to generalize.
Considering the model's ability to generalize, we can have a more in-depth look at this, if we calculate the generalization gap, which we calculate like
\begin{equation}
G_{gap} = \left|A_{\text{train}} - A_{\text{test}}\right|,
\end{equation}
where  $A_{\textit{train}}$ is the training accuracy and $A_{\textit{test}}$ is the accuracy calculated on the validation data. 
The generalization gap describes the model's ability to predict the results of the date on which it was not trained. The generalization gaps for all three setups are plotted in \cref{fig:gen_gap}. Just by looking at the plot, we can see that for the case with just one ansatz repetition, even though the gap decreases, there is still a large fluctuation, showing that the model is unstable across subsequent epochs. For the cases with two and three repetitions, there is much more stability after the initial period of training.

Now, in \cref{tab:generalization_gap} we can have a look at two different metrics, describing the generalization gap, that is the final generalization gap, that is calculated at the end of the training period, where we can see that the gap decreases with the increasing number of repetitions, again pointing to the greater model stability with a larger number of repetitions.

The second metric is the mean generalization gap, that computed as an average across the whole training process, all 500 epochs. And again, in this case, the value decreases with an increase of ansatz parameters. This all suggests that increasing the number of parameters in the ansatz leads to a more robust model that has better generalization ability. The fact, that the generalization gap is the smallest in the case of the largest ansatz, tells us that a larger quantum layer can improve resistance to overfitting of our model.

\begin{table}[htbp]
\centering
\caption{Generalization Gap Metrics}
\begin{tabular}{lcc}
\toprule
\textbf{Repetitions} & \textbf{Final Generalization Gap} & \textbf{Mean Generalization Gap} \\
\midrule
1 Repetition  & 0.0254 & 0.0332 \\
2 Repetitions & 0.0166 & 0.0200 \\
3 Repetitions & 0.0107 & 0.0195 \\
\bottomrule
\end{tabular}
\label{tab:generalization_gap}
\end{table}

Now we will have a look at the overall learning efficiency. The first metric of this set shown in \cref{tab:learning_dynamics} is the epoch to reach $90\%$ validation accuracy, which we define as
\begin{equation}
    \text{Epoch}_{90\%}=\textit{min}\left\{ \text{epoch} \middle| \text{Validation Accuracy at epoch} \geq 90\% \right\},
\end{equation}
and it tells us the earliest epoch in which accuracy over $90\%$ was achieved. If that never happened over the course of the model's training, then we report that such accuracy was "Not reached". And as we can see, the scenario, where we are not able to reach accuracy over $90\%$, happened with the smallest ansatz. For the other two models, we see that for the case with two repetitions, we first reached validation accuracy over $90\%$ in epoch $195$, which further decreases for the model with three repetitions where we obtain such accuracy 26 epochs earlier. This indicates, that large ansatzes have better ability to learn our data, which we can attribute to the fact that they also have more points of freedom.

The next metric that is shown in \cref{tab:learning_dynamics} is the early learning slope, which we define as
\begin{equation}
    S_{EL} = \frac{A_{val}(t=N) - A_{val}(t=0)}{N},
\end{equation}
where again $A_{val}$ is the validation accuracy and $N$ is the number of epochs we consider in this calculation, for our case we chose $N=5$ as it reflects the initial trend in a training. Now we can see from the data, that the early learning slope is positive in the first case, showing that with the smallest number of parameters, we are able to immediately gain yields, as the optimization of the model is the easiest. On the other hand, we see that for the other two cases, the initial slope is negative, which suggests, that the optimizer needs at first to search the surroundings of our initial point and that it finds the right direction a bit later, and thus learns a bit more carefully early on.
The last metric shown in the \cref{tab:learning_dynamics} is the overfitting drop, which we again define as
\begin{equation}
    D_{overfit} = A_{peak} - A_{final},
\end{equation}
where $A_{peak}$ is the highest validation accuracy ever reached and $A_{final}$ is the validation accuracy at the end of the training process. This metric measures the drop in accuracy after reaching the peak value. As we can see, the overfitting drop significantly reduces in the case of three repetitions, from which we can again deduce that the ansatz repetitions act as regularization, making our model more robust with respect to overfitting. The change is significant, because, for the first two cases, the accuracy drops is $>2\%$, but in the third case it is just $0.68 \%$.
\begin{table}[htbp]
\centering
\caption{Learning Efficiency and Overfitting Analysis}
\begin{tabular}{lccc}
\toprule
\textbf{Repetitions} & \textbf{Epoch to Reach 90\% Val Accuracy} & \textbf{Early Learning Slope} & \textbf{Overfitting Drop} \\
\midrule
1 Repetition  & Not reached & 0.0117 & 0.0244 \\
2 Repetitions & 195         & -0.0068 & 0.0234 \\
3 Repetitions & 168         & -0.0051 & 0.0068 \\
\bottomrule
\end{tabular}
\label{tab:learning_dynamics}
\end{table}

Next, we will have a look at the fluctuations in our training accuracy. In \cref{tab:train_fluctuation} we report metrics that describe fluctuations in training accuracy. We examine this fluctuation by analyzing the sequence of training accuracies across all epochs. This sequence is denoted as
\begin{equation}
    \{ a_1, a_2, a_3, \dotsc, a_T \},
\end{equation}

where $ a_t \in [0,1]$ represents the training accuracy at epoch $t$, and $T$ is the total number of epochs. Now we want to measure local fluctuations, to do that, we first compute the absolute first-order differences of our sequence

\begin{equation}
    \Delta_t = |a_{t+1} - a_t|, \quad \text{for} \quad t = 1, \dotsc, T-1.
\end{equation}

These differences represent the magnitude of change in the accuracy computed on the training data between epochs. When we take $\{ \Delta_t \}_{t=1}^{T-1}$, two metrics are defined. The first one is the standard deviation of fluctuation, defined as
\begin{equation}
        \sigma_{\Delta} = \sqrt{ \frac{1}{T-1} \sum_{t=1}^{T-1} (\Delta_t - \mu_{\Delta})^2 }.
\end{equation}
Here $T$ is the total number of epochs. This metric measures the dispersion of the local changes. The second one is the mean of absolute fluctuations, which helps us to quantify the average size of local fluctuations. This is defined as
\begin{equation}
      \mu_{\Delta} = \frac{1}{T-1} \sum_{t=1}^{T-1} \Delta_t.
\end{equation}

Now we will have a look at the results itself. For single ansatz repetition again proves to be the worst case, as both the standard deviation and mean of absolute differences are the largest. Meaning that the fluctuations itself are the largest in the single repetition case. For the other two, we see that the values are really similar, which shows us, that the training of these two was more consistent and stable. Here we see that in terms of the magnitude of the fluctuations of training accuracy, there are no gains when using three repetitions, which tells us that there are some diminishing returns in this regard.

\begin{table}[htbp]
\centering
\caption{Training Fluctuation Metrics}
\begin{tabular}{lcc}
\toprule
\textbf{Repetitions} & \textbf{Training Std Dev} & \textbf{Training Mean Abs Diff} \\
\midrule
1 Repetition  & 0.0106 & 0.0122 \\
2 Repetitions & 0.0082 & 0.0103 \\
3 Repetitions & 0.0083 & 0.0105 \\
\bottomrule
\end{tabular}
\label{tab:train_fluctuation}
\end{table}

The last set of results for this analysis is displayed in \cref{tab:val_fluctuation}. Here we have the same two previous metrics but calculated with the accuracy of the validation dataset. But in this case, we also work with stability ratio, defined as
\begin{equation}
    R_{stability} = \frac{\mu_{\Delta}^{\text{val}}}{\mu_{\Delta}^{\text{train}}},
\end{equation}
where $\mu_{\Delta}^{\text{train}}$ and $\mu_{\Delta}^{\text{val}}$ denote the mean absolute fluctuations of the training and validation accuracy curves, respectively. When we have a stability ratio close to 1, it indicates comparable stability between training and validation. When the value is greater the one, it points to the fact, that the validation accuracy is more prone to fluctuations, indicating some problems with generalization instability or it points to potential overfitting. A value smaller than one, on the other hand, points to the fact that the model's ability to generalize is fine and stable.

In the \cref{tab:val_fluctuation} we see a slight difference compared to \cref{tab:train_fluctuation}, which describes fluctuation in training data. Because in the case of validation accuracy, we see improvement even when adding the third repetition. From this, we can say, that while adding a third repetition has no effect on the stability of the training with respect to the training data, we still benefit from it with the accuracy of validation data. This shows that the increase in repetitions is good for the generalization. 

When we have a look at the stability ratio, we can see that it improves quite dramatically when we add additional repetition. We even get to the territory of a stability ratio of less than one, where the network is really stable in terms of generalization, which is quite good, especially because it gives us more breathing room with respect to when we finish the training.

\begin{table}[htbp]
\centering
\caption{Validation Fluctuation and Stability Ratio}
\begin{tabular}{lccc}
\toprule
\textbf{Repetitions} & \textbf{Validation Std Dev} & \textbf{Validation Mean Abs Diff} & \textbf{Stability Ratio} \\
\midrule
1 Repetition  & 0.0288 & 0.0265 & 2.1791 \\
2 Repetitions & 0.0207 & 0.0158 & 1.5353 \\
3 Repetitions & 0.0131 & 0.0088 & 0.8335 \\
\bottomrule
\end{tabular}
\label{tab:val_fluctuation}
\end{table}

In summary, we can make several points. The first one is that there is a trade-off between the early training speed and overall stability. Where with a low number of repetitions we gain an initial training speedup, but this yield is then diminished by the fact, that models with more repetition tend to be more stable and reach convergence faster. 

The second point is that we need not just consider accuracy, as in such case, we would conclude, that adding the third repetition is unnecessary. But in fact, when we take into account also the robustness of the model, we see that the third repetition in ansatz grants us smaller overall fluctuations, a smaller generalization gap, and increased strength against overfitting.

The third point we can make is that the deeper quantum circuits seem to act as regularizers, where they on their own increase the model's ability to generalize and prevent overfitting. Also, higher repetitions give us a smoother training process, as is visible from both the training and the validation accuracy curves. This suggests that the landscape created by a higher number of ansatz repetitions is more structured and learnable.

The last point is that the increase in repetitions improves not just the outcome of the training, but the whole process of the learning, where models with a higher number of repetitions are overall more consistent and the models are more reliable.

\section{Different Feature Maps}
In this part, we will focus on the analysis of how different feature mapping affects the model and its abilities. In total 9 different feature mapping were tested. In \cref{tab:accuracies}, we can see that the feature mapping has a really large effect on the model and its learning ability. Because, from all nine variants, only one of them proved to be efficient, while the other 8 hindered the model. Considering the final accuracies for both training and validation data, we can make one reasonable conclusion, and that is the fact that the choice of suitable feature mapping is crucial in hybrid machine learning as it can make the difference between a well-trained model and an ill-trained one. But let's have a look at this in more detail.

\begin{table}[h!]
\centering
\begin{tabular}{|l|c|c|}
\hline
\textbf{Model} & \textbf{Training Accuracy} & \textbf{Validation Accuracy} \\
\hline
zz\_feature\_map\_reps\_2\_linear & 0.2832 & 0.3340 \\
z\_feature\_map\_reps\_2 & 0.3193 & 0.3291 \\
pauli\_z\_yy\_zxz\_linear & 0.3145 & 0.3320 \\
pauli\_xyz\_1\_rep & 0.9082 & 0.9014 \\
zz\_feature\_map\_reps\_3\_full & 0.3350 & 0.3235 \\
zz\_feature\_map\_reps\_1\_linear\_entanglement & 0.3262 & 0.3174 \\
pauli\_z\_yy\_zxz\_rep\_2 & 0.3271 & 0.3301 \\
z\_feature\_map\_reps\_1 & 0.3301 & 0.3408 \\
z\_feature\_map\_reps\_3 & 0.3125 & 0.3535 \\
\hline
\end{tabular}
\caption{Training and Validation Accuracies for Different Models}
\label{tab:accuracies}
\end{table}

In the next part of this analysis, \ac{pca} \cite{abdi2010principal} was used as a tool to further study how the feature mapping affects the model in different stages. \ac{pca} reduces high-dimensional data into a few principal components that describe the majority of the variance. The \ac{pca} was done at three different stages on the trained model, the first stage was just after the classical part, the second one was after the feature mapping was done and the last was done for the output of the quantum layer. Alongside the \ac{pca}, silhouette scores \cite{shahapure2020cluster} were computed. The silhouette score measures how different is a point from points in different clusters. When we have a silhouette score close to 1, then the clusters are well-separated, while scores that are closer to 0 or negative indicate overlapping or ill-defined clusters. Both the \ac{pca} and the silhouette score were used as a tool to provide us with more insight into how feature mapping affects the model in different stages.

\begin{figure}[htbp]
    \centering

    \begin{subfigure}[t]{0.45\textwidth}
        \centering
        \includegraphics[width=\textwidth]{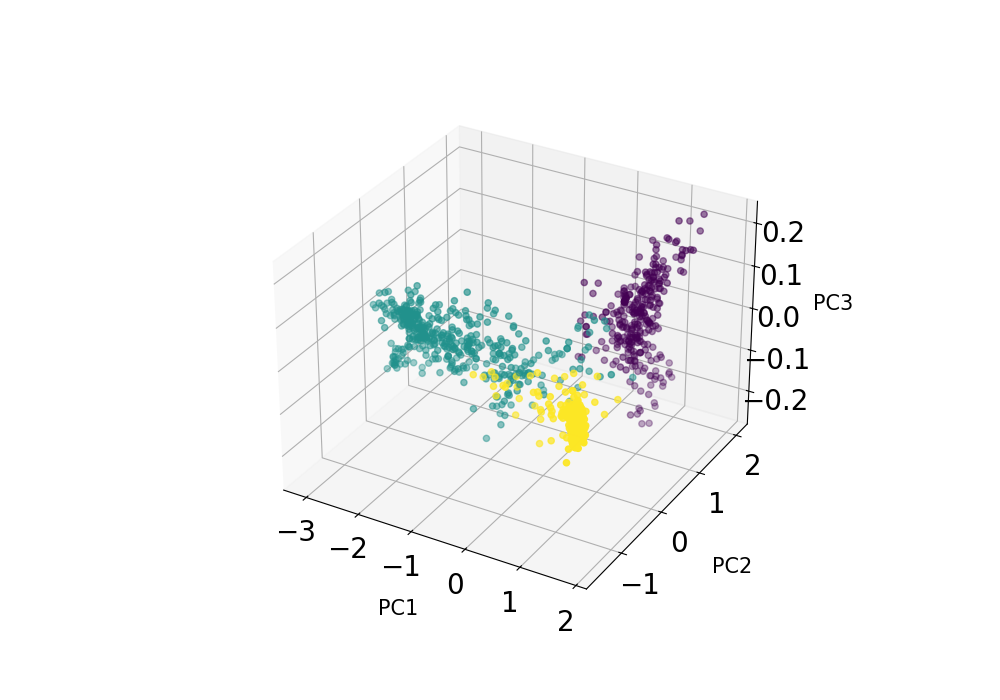}
        \caption{Principal Component Analysis After Classical Part, Color-coded by Fitted Values}
    \end{subfigure}
    \hfill
    \begin{subfigure}[t]{0.45\textwidth}
        \centering
        \includegraphics[width=\textwidth]{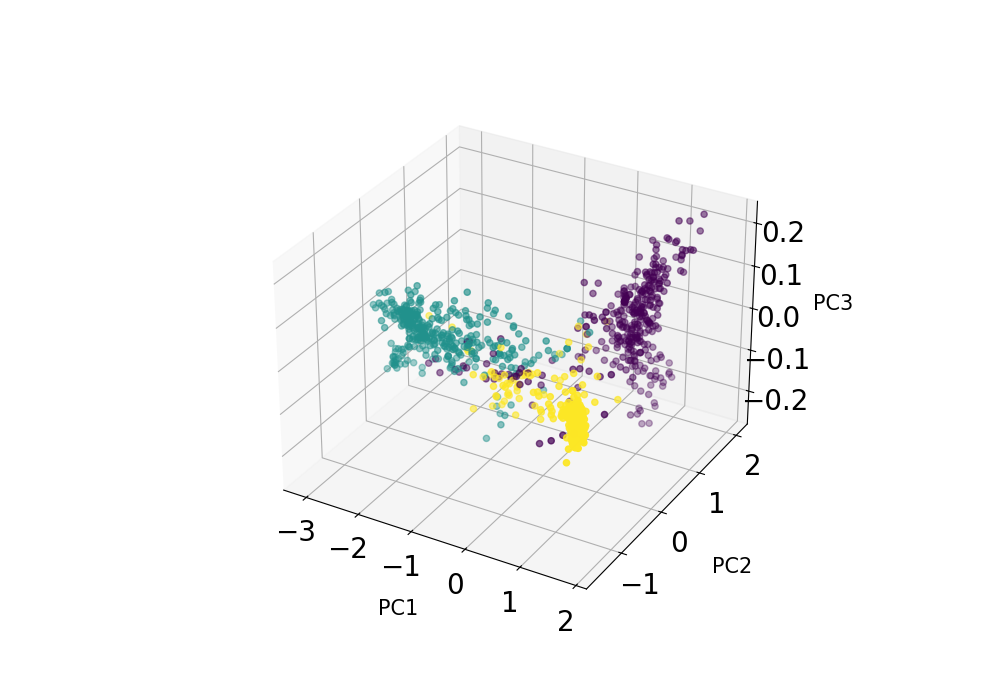}
        \caption{Principal Component Analysis After Classical Part, Color-coded by Training Values}
    \end{subfigure}

    \vspace{1em}

    \begin{subfigure}[t]{0.45\textwidth}
        \centering
        \includegraphics[width=\textwidth]{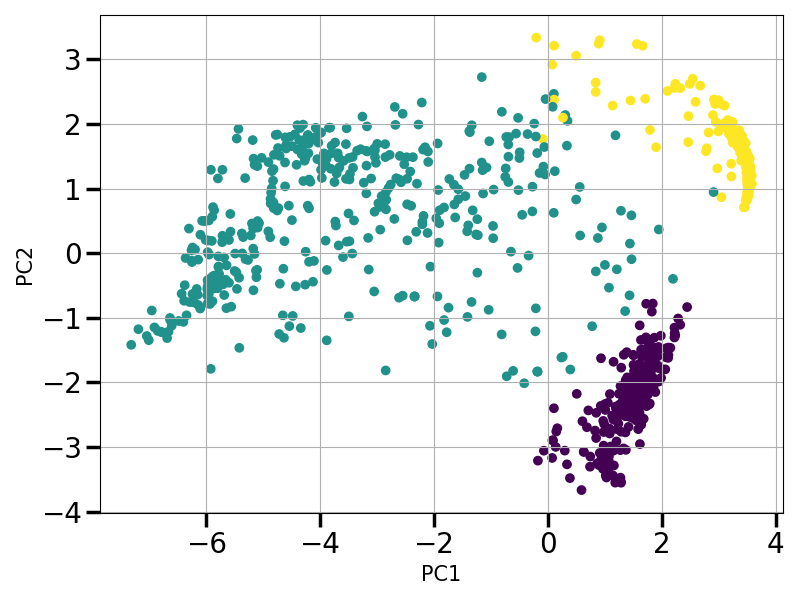}
        \caption{Principal Component Analysis After Feature Mapping, Color-coded by Fitted Values}
    \end{subfigure}
    \hfill
    \begin{subfigure}[t]{0.45\textwidth}
        \centering
        \includegraphics[width=\textwidth]{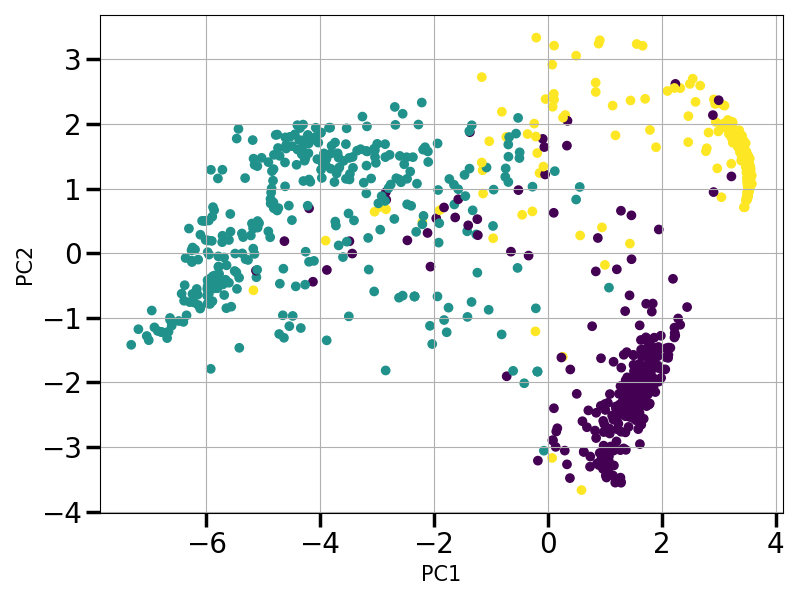}
        \caption{Principal Component Analysis After Feature Mapping, Color-coded by Training Values}
    \end{subfigure}

    \vspace{1em}

    \begin{subfigure}[t]{0.45\textwidth}
        \centering
        \includegraphics[width=\textwidth]{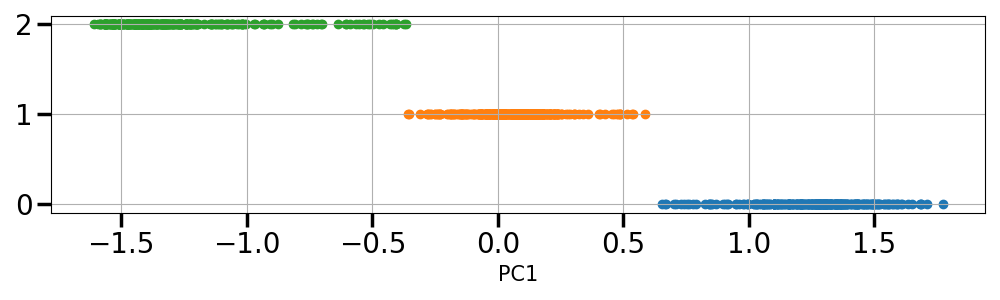}
        \caption{Principal Component Analysis After QNN, Color-coded by Fitted Values}
    \end{subfigure}
    \hfill
    \begin{subfigure}[t]{0.45\textwidth}
        \centering
        \includegraphics[width=\textwidth]{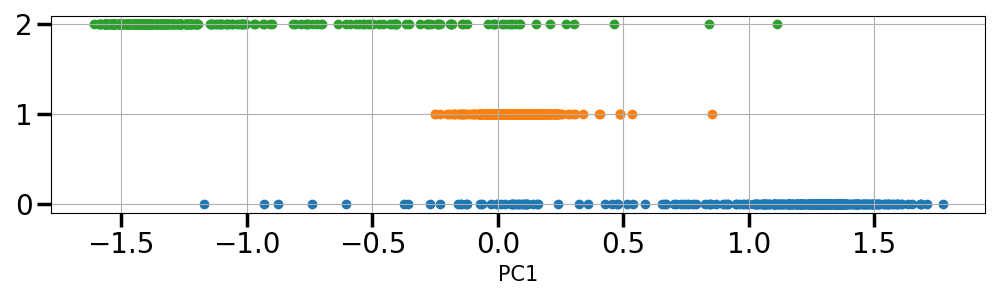}
        \caption{Principal Component Analysis After QNN, Color-coded by Training Values}
    \end{subfigure}

    \caption{PCA Progression for \texttt{pauli\_xyz\_1\_rep} — \textbf{Training Data}}
    \label{fig:pauli_xyz_1_rep_train}
\end{figure}

To start, we will have a look at the training data and the \ac{pca} done at the end of the classical layer, just before feature mapping is applied. The data is summarized in \cref{tab:pca_class_train}. The first thing we can see is that in general, a large proportion of the principal component variance is captured by the first two components, in all cases exceeding $95\%$. The issue comes with the silhouette score, where for all but one case, they are close to zero or even negative. This indicates, that the classical part of the neural network effectively compresses the data into lower dimensional representation, it oftentimes fails to introduce meaningful separation of the data into correct clusters based on the classes. The only case, where the silhouette scores are higher for both validation and training data, is the case of \textit{Pauli XYZ} mapping, but even there with values of $0.7$ for fitted values and $0.62$ for training data, the clusters are not yet well separated. This suggests that further quantum transformations are necessary to induce meaningful data separability.
\begin{table}[ht]
\centering
\caption{PCA Results After Classical Part Training Data}
\label{tab:pca_class_train}
\begin{tabular}{p{9cm} l p{3cm} p{2cm}}
\toprule
Feature Map & Stage & PC Variance & Silhouette Score \\
\midrule
ZZ Feature Map Reps 2 Linear Training & Fitted values & [0.84, 0.16, 0.01] & 0.20 \\
ZZ Feature Map Reps 2 Linear Training & Training values & [0.84, 0.16, 0.01] & 0.02 \\
Z Feature Map Reps 2 Training & Fitted values & [0.97, 0.02, 0.01] & 0.05 \\
Z Feature Map Reps 2 Training & Training values & [0.97, 0.02, 0.01] & -0.02 \\
Pauli Z YY ZXZ Linear Training & Fitted values & [1.00, 0.00, 0.00] & -0.00 \\
Pauli Z YY ZXZ Linear Training & Training values & [1.00, 0.00, 0.00] & -0.07 \\
Pauli XYZ 1 Rep Training & Fitted values & [0.67, 0.33, 0.00] & 0.70 \\
Pauli XYZ 1 Rep Training & Training values & [0.67, 0.33, 0.00] & 0.62 \\
ZZ Feature Map Reps 3 Full Training & Fitted values & [1.00, 0.00, 0.00] & 0.00 \\
ZZ Feature Map Reps 3 Full Training & Training values & [1.00, 0.00, 0.00] & -0.03 \\
ZZ Feature Map Reps 1 No Entanglement Training & Fitted values & [0.78, 0.15, 0.07] & -0.01 \\
ZZ Feature Map Reps 1 No Entanglement Training & Training values & [0.78, 0.15, 0.07] & -0.01 \\
Pauli Z YY ZXZ Rep 2 Training & Fitted values & [1.00, 0.00, 0.00] & -0.02 \\
Pauli Z YY ZXZ Rep 2 Training & Training values & [1.00, 0.00, 0.00] & -0.04 \\
Z Feature Map Reps 1 Training & Fitted values & [0.95, 0.04, 0.01] & -0.02 \\
Z Feature Map Reps 1 Training & Training values & [0.95, 0.04, 0.01] & -0.02 \\
Z Feature Map Reps 3 Training & Fitted values & [0.96, 0.03, 0.00] & 0.32 \\
Z Feature Map Reps 3 Training & Training values & [0.96, 0.03, 0.00] & 0.01 \\
\bottomrule
\end{tabular}
\end{table}
Now as we look at the values in \cref{tab:pca_feat_train}, obtained after the selected feature mapping was performed, we notice several things. The first one is that there is a decrease in the variance of the first two components, which suggests that there is an increase in the feature complexity. The second one is that the silhouette scores still remain relatively low, meaning that immediate separability was not achieved. The one exception is the \textit{Pauli XYZ} mapping, which retains moderate silhouette scores, indicating, that there is still some degree of cluster structure. All of this tells us that quantum feature mapping alone is not sufficient to achieve high-quality clustering and that further training is needed. This is the expected scenario, as following the feature mapping is the quantum layer with trainable parameters.

\begin{figure}[htbp]
    \centering

    \begin{subfigure}[t]{0.45\textwidth}
        \centering
        \includegraphics[width=\textwidth]{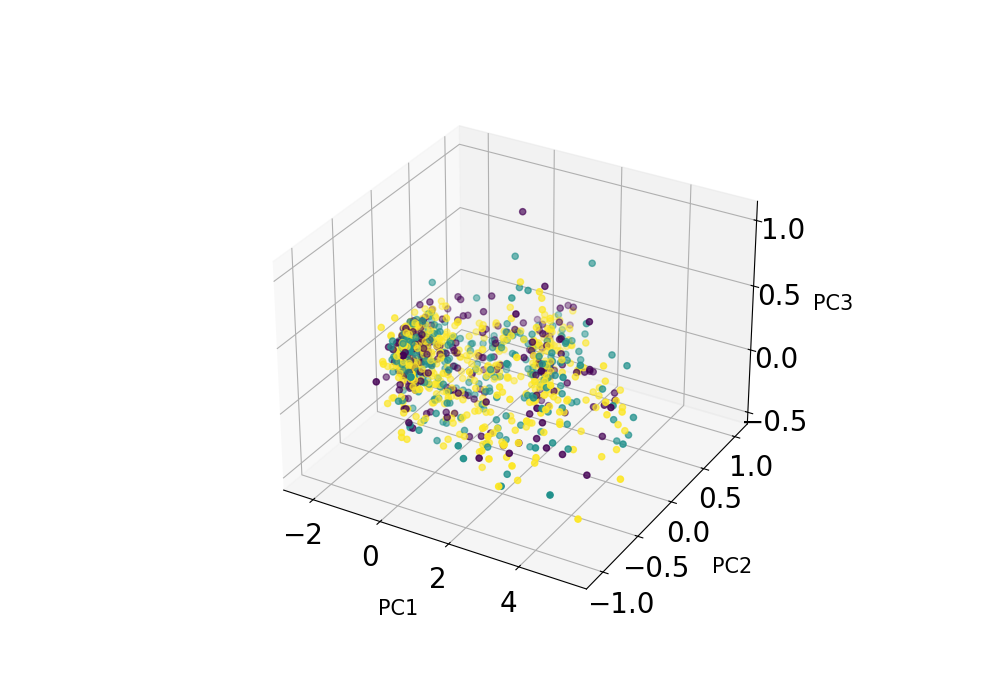}
        \caption{Principal Component Analysis After Classical Part, Color-coded by Fitted Values}
    \end{subfigure}
    \hfill
    \begin{subfigure}[t]{0.45\textwidth}
        \centering
        \includegraphics[width=\textwidth]{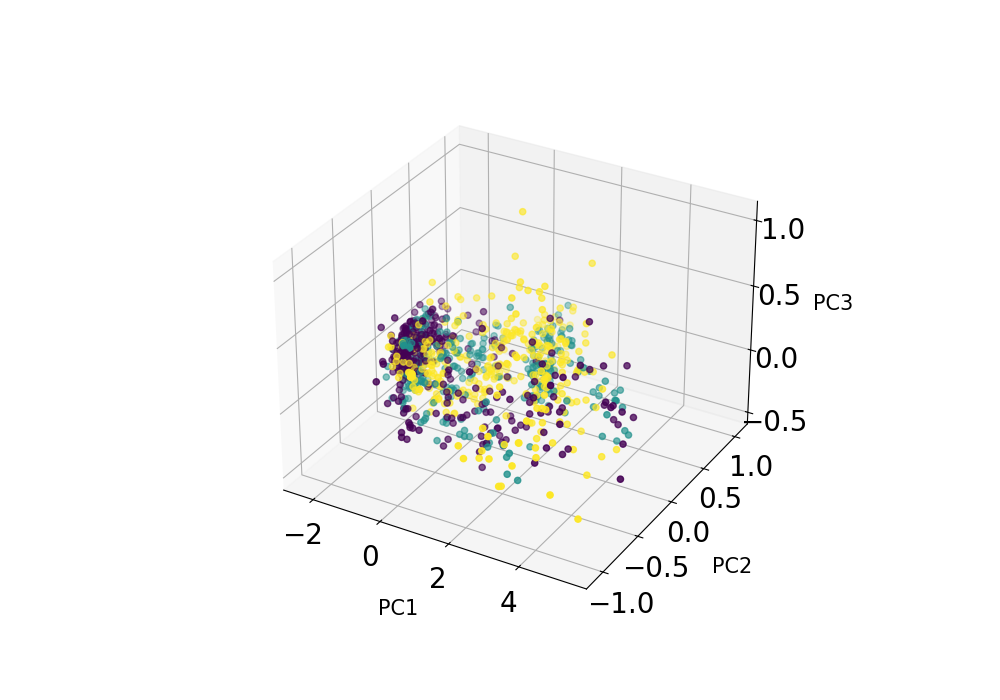}
        \caption{Principal Component Analysis After Classical Part, Color-coded by Training Values}
    \end{subfigure}

    \vspace{1em}

    \begin{subfigure}[t]{0.45\textwidth}
        \centering
        \includegraphics[width=\textwidth]{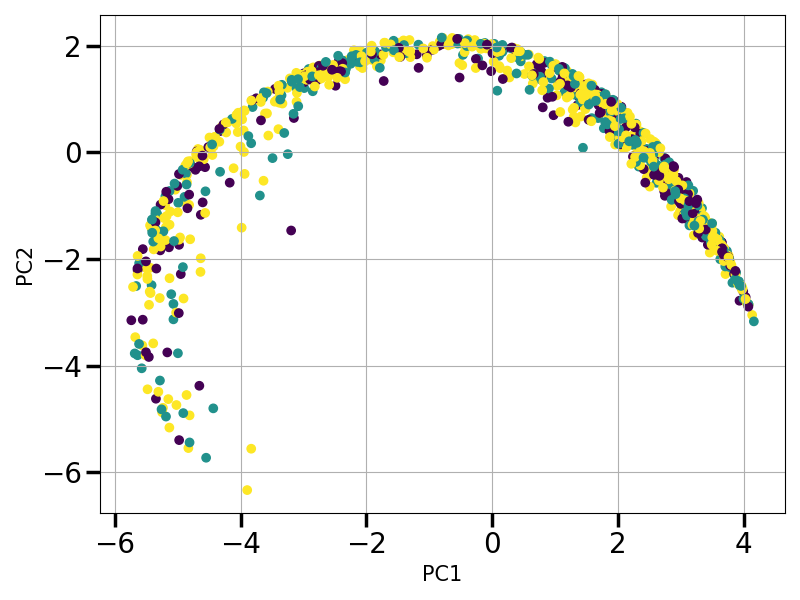}
        \caption{Principal Component Analysis After Feature Mapping, Color-coded by Fitted Values}
    \end{subfigure}
    \hfill
    \begin{subfigure}[t]{0.45\textwidth}
        \centering
        \includegraphics[width=\textwidth]{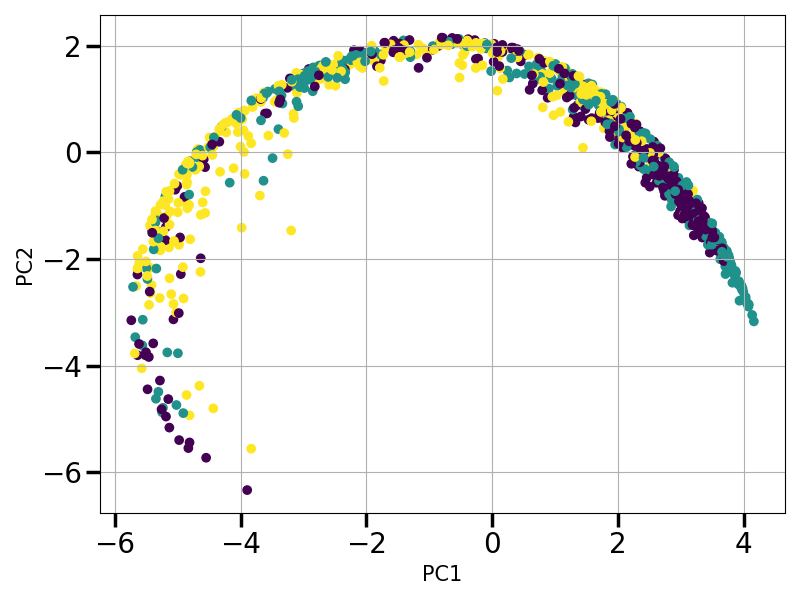}
        \caption{Principal Component Analysis After Feature Mapping, Color-coded by Training Values}
    \end{subfigure}

    \vspace{1em}

    \begin{subfigure}[t]{0.45\textwidth}
        \centering
        \includegraphics[width=\textwidth]{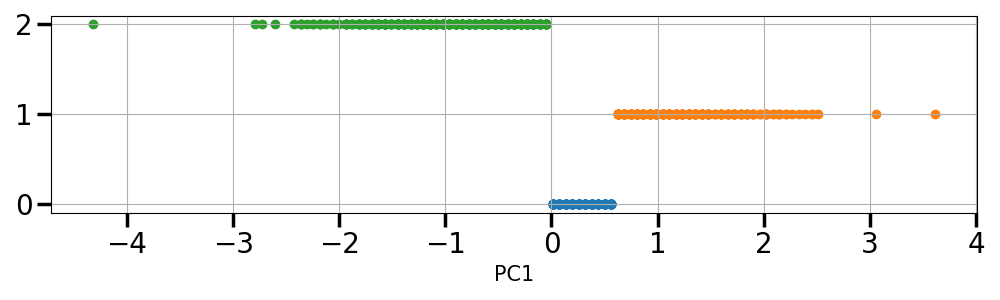}
        \caption{Principal Component Analysis After QNN, Color-coded by Fitted Values}
    \end{subfigure}
    \hfill
    \begin{subfigure}[t]{0.45\textwidth}
        \centering
        \includegraphics[width=\textwidth]{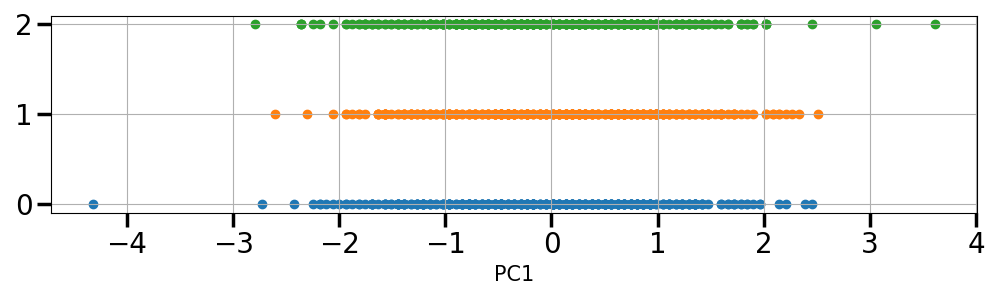}
        \caption{Principal Component Analysis After QNN, Color-coded by Training Values}
    \end{subfigure}

    \caption{PCA Progression for \texttt{z\_feature\_map\_reps\_1} — \textbf{Training Data}}
    \label{fig:z_feature_map_reps_1_train}
\end{figure}

The last table showing \ac{pca} values and silhouette score is \cref{tab:pca_qnn_train}. Here the variance collapsed into a single component, which indicates that the quantum layer has done significant restructuring of the feature space. We also see that silhouette scores have increased significantly for several different feature mappings. We can see the feature mappings ranked in \cref{tab:fm_sil_rank}. This shows that the \ac{qnn} effectively learns to sort the data into clusters and maps them into more separable spaces. Now, even when we see an increase in silhouette scores across multiple mappings, we have to go back to \cref{tab:accuracies}, where we have already seen that only one feature mapping was effective. To see what is happening here, let's have a look at the feature mappings and their \ac{pca}'s one by one.

\begin{table}[ht]
\centering
\caption{PCA Results After Feature Mapping on Training Data}
\label{tab:pca_feat_train}
\begin{tabular}{p{9cm} l p{3cm} p{2cm}}
\toprule
Feature Map & Stage & PC Variance & Silhouette Score \\
\midrule
ZZ Feature Map Reps 2 Linear Training & Fitted values& [0.18, 0.16] & 0.04 \\
ZZ Feature Map Reps 2 Linear Training &Training values& [0.18, 0.16] & -0.01 \\
Z Feature Map Reps 2 Training & Fitted values& [0.23, 0.20] & 0.08 \\
Z Feature Map Reps 2 Training &Training values& [0.23, 0.20] & 0.01 \\
Pauli Z YY ZXZ Linear Training & Fitted values& [0.08, 0.07] & 0.00 \\
Pauli Z YY ZXZ Linear Training &Training values& [0.08, 0.07] & -0.00 \\
 Pauli XYZ 1 Rep Training & Fitted values& [0.68, 0.18] & 0.55 \\
 Pauli XYZ 1 Rep Training &Training values& [0.68, 0.18] & 0.48 \\
ZZ Feature Map Reps 3 Full Training & Fitted values& [0.12, 0.11] & 0.00 \\
ZZ Feature Map Reps 3 Full Training &Training values& [0.12, 0.11] & -0.00 \\
ZZ Feature Map Reps 1 No Entanglement Training & Fitted values& [0.68, 0.26] & -0.02 \\
ZZ Feature Map Reps 1 No Entanglement Training &Training values& [0.68, 0.26] & -0.02 \\
Pauli Z YY ZXZ Rep 2 Training & Fitted values& [0.08, 0.07] & -0.00 \\
Pauli Z YY ZXZ Rep 2 Training &Training values& [0.08, 0.07] & -0.00 \\
Z Feature Map Reps 1 Training & Fitted values& [0.69, 0.18] & -0.01 \\
Z Feature Map Reps 1 Training &Training values& [0.69, 0.18] & -0.00 \\
Z Feature Map Reps 3 Training & Fitted values& [0.28, 0.25] & -0.02 \\
Z Feature Map Reps 3 Training &Training values& [0.28, 0.25] & -0.00 \\
\bottomrule
\end{tabular}
\end{table}

\begin{table}[ht]
\centering
\caption{PCA Results After QNN on Training Data}
\label{tab:pca_qnn_train}
\begin{tabular}{p{9cm} l p{2cm} p{2cm}}
\toprule
Feature Map & Stage & PC Variance & Silhouette Score \\
\midrule
 ZZ Feature Map Reps 2 Linear Training & Fitted values& [1.00] & 0.57 \\
ZZ Feature Map Reps 2 Linear Training & Training values& [1.00] & -0.04 \\
  Z Feature Map Reps 2 Training & Fitted values& [1.00] & 0.65 \\
Z Feature Map Reps 2 Training & Training values& [1.00] & -0.05 \\
Pauli Z YY ZXZ Linear Training & Fitted values& [1.00] & 0.49 \\
Pauli Z YY ZXZ Linear Training & Training values& [1.00] & -0.04 \\
 Pauli XYZ 1 Rep Training & Fitted values& [1.00] & 0.80 \\
 Pauli XYZ 1 Rep Training & Training values& [1.00] & 0.62 \\
 ZZ Feature Map Reps 3 Full Training & Fitted values& [1.00] & 0.57 \\
ZZ Feature Map Reps 3 Full Training & Training values& [1.00] & -0.02 \\
ZZ Feature Map Reps 1 No Entanglement Training & Fitted values& [1.00] & 0.38 \\
ZZ Feature Map Reps 1 No Entanglement Training & Training values& [1.00] & -0.01 \\
Pauli Z YY ZXZ Rep 2 Training & Fitted values& [1.00] & 0.27 \\
Pauli Z YY ZXZ Rep 2 Training & Training values& [1.00] & -0.02 \\
Z Feature Map Reps 1 Training & Fitted values& [1.00] & 0.42 \\
Z Feature Map Reps 1 Training & Training values& [1.00] & -0.02 \\
 Z Feature Map Reps 3 Training & Fitted values& [1.00] & 0.75 \\
Z Feature Map Reps 3 Training & Training values& [1.00] & -0.11 \\
\bottomrule
\end{tabular}
\end{table}

\begin{table}[ht]
\centering
\caption{Ranking of Feature Maps by Silhouette Score (QNN Fit, Validation)}
\label{tab:fm_sil_rank}
\begin{tabular}{p{1cm} p{6cm} c}

\toprule
\textbf{Rank} & \textbf{Feature Map} & \textbf{Silhouette Score (Validation)} \\
\midrule
1 & Pauli XYZ 1 Rep & 0.80 \\
2 & Z Feature Map Reps 3 & 0.75 \\
3 & Z Feature Map Reps 2 & 0.65 \\
4 & ZZ Feature Map Reps 2 Linear & 0.57 \\
5 & ZZ Feature Map Reps 3 Full & 0.57 \\
6 & Pauli Z YY ZXZ Linear & 0.49 \\
7 & Z Feature Map Reps 1 & 0.42 \\
8 & ZZ Reps 1 No Entanglement & 0.38 \\
9 & Pauli Z YY ZXZ Rep 2 & 0.27 \\
\bottomrule
\end{tabular}
\end{table}

The first feature mapping that we will focus on is the \textit{Pauli XYZ}, the \ac{pca} of this mapping is plotted in \cref{fig:pauli_xyz_1_rep_train}. Here we can see that after the classical stage, there is a moderate class separation, but the overlap between the different classes is still noticeable. Here the application of feature mapping improves the separation in a way. What we mainly see is the effect of non-linear transformations, that unfold the structure of the data a bit, even though the mixing is still visible by the right column in the figure, where the data color represents the training values, whereas the left column is color-coded by fitted values.
\begin{figure}[htbp]
    \centering

    \begin{subfigure}[t]{0.45\textwidth}
        \centering
        \includegraphics[width=\textwidth]{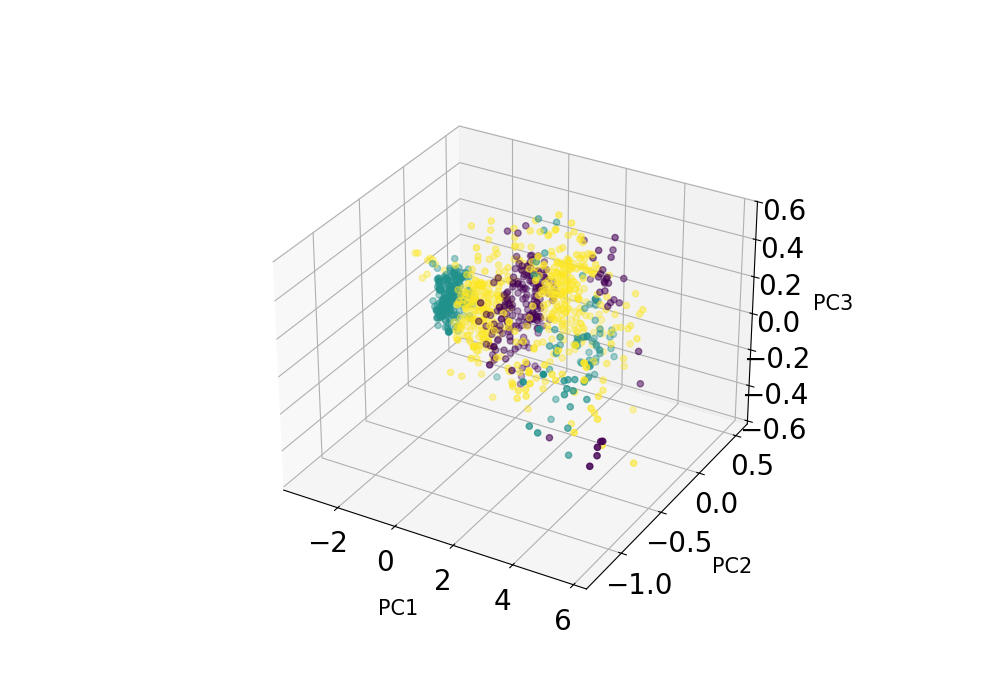}
        \caption{Principal Component Analysis After Classical Part, Color-coded by Fitted Values}
    \end{subfigure}
    \hfill
    \begin{subfigure}[t]{0.45\textwidth}
        \centering
        \includegraphics[width=\textwidth]{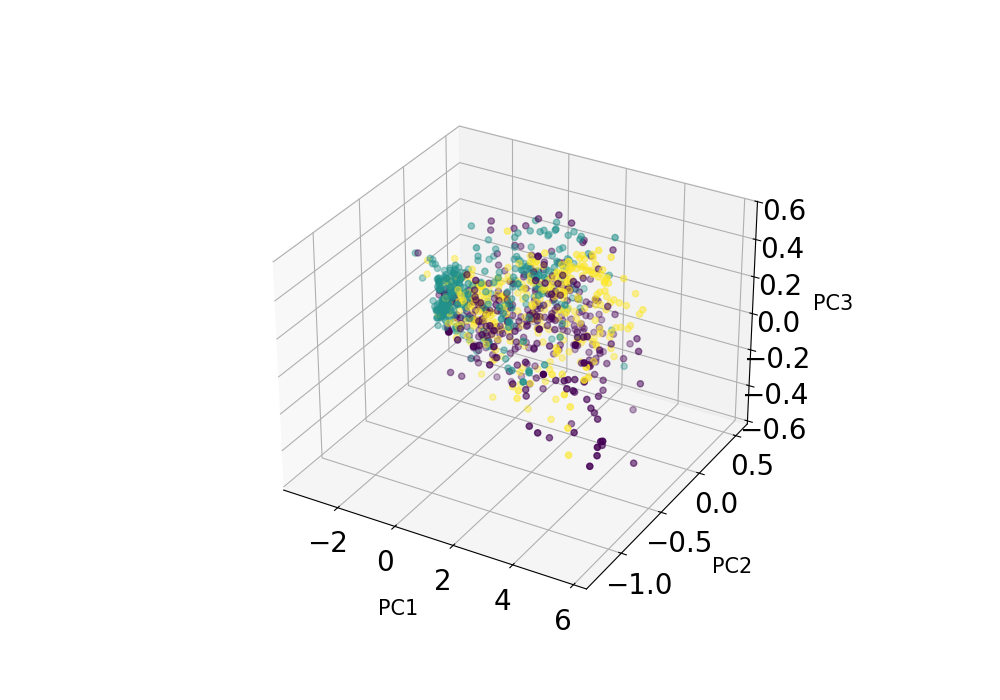}
        \caption{Principal Component Analysis After Classical Part, Color-coded by Training Values}
    \end{subfigure}

    \vspace{1em}

    \begin{subfigure}[t]{0.45\textwidth}
        \centering
        \includegraphics[width=\textwidth]{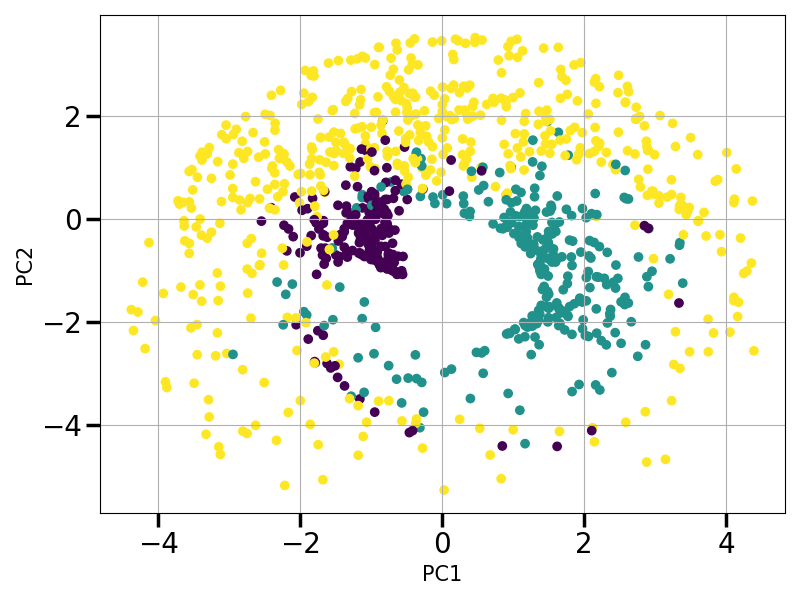}
        \caption{Principal Component Analysis After Feature Mapping, Color-coded by Fitted Values}
    \end{subfigure}
    \hfill
    \begin{subfigure}[t]{0.45\textwidth}
        \centering
        \includegraphics[width=\textwidth]{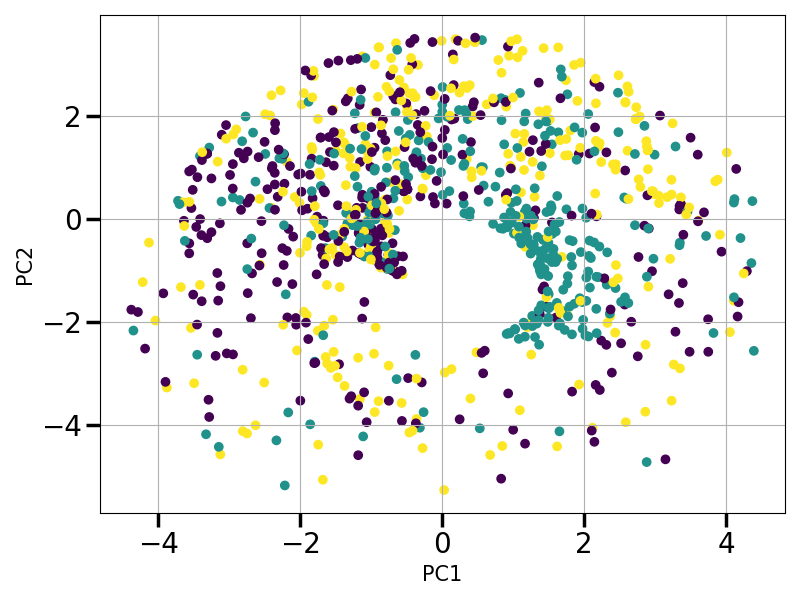}
        \caption{Principal Component Analysis After Feature Mapping, Color-coded by Training Values}
    \end{subfigure}

    \vspace{1em}

    \begin{subfigure}[t]{0.45\textwidth}
        \centering
        \includegraphics[width=\textwidth]{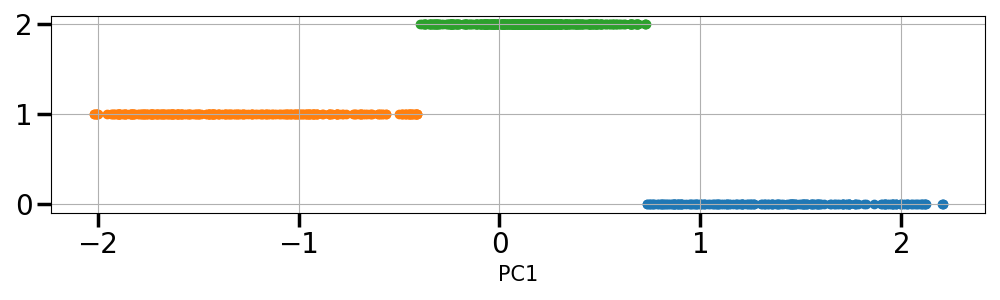}
        \caption{Principal Component Analysis After QNN, Color-coded by Fitted Values}
    \end{subfigure}
    \hfill
    \begin{subfigure}[t]{0.45\textwidth}
        \centering
        \includegraphics[width=\textwidth]{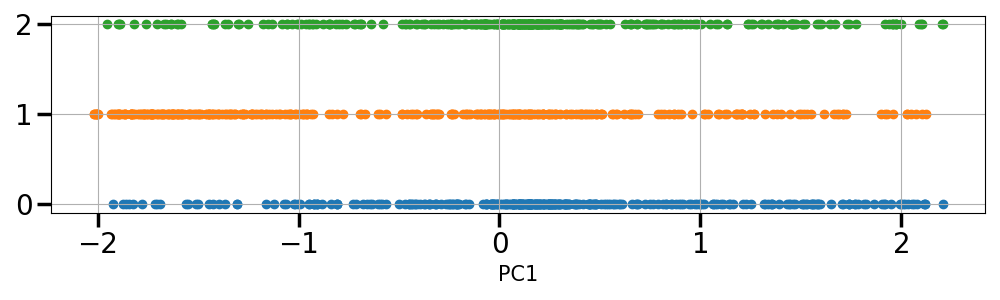}
        \caption{Principal Component Analysis After QNN, Color-coded by Training Values}
    \end{subfigure}

    \caption{PCA Progression for \texttt{z\_feature\_map\_reps\_2} — \textbf{Training Data}}
    \label{fig:z_feature_map_reps_2_train}
\end{figure}

However, after the quantum layer, we can see the separation more clearly in the bottom row, and that is the fact that the data was compressed into a single component. Overall we can see that correctly chosen feature mapping enhances the class separability, and the whole model culminates in a structured, easily classifiable feature space.

Next, let's have a look at \cref{fig:z_feature_map_reps_1_train}, where we see the first feature mapping that consists of just Z rotations. We see that after the classical part, there are no distinct clusters, and after the application of the feature mapping, we see that the data has a specific shape. This is because when only Z rotations are present, described by
\begin{equation}
    U_Z(x) = \exp\left(-i \sum_j \phi_j(x) Z_j\right).
\end{equation}
This means that the qubit is rotated just around the Z-axis, meaning that the Z-coordinate, when looking at the projection on the Bloch sphere remains the same. When working with just one rotation, during the \ac{pca} we see a ring-like structure, which is something we will also observe in further mappings with just Z rotations. This shows, that the data is not well separated as it is clustered into small space. When we have a look at the plot on the bottom right, we see that when plotted with respect to the training values, the clusters are not separated at all, showing that the model with just Z rotations and one repetition has failed to learn.

Now, we shall have a look, at what difference adding a repetition to the feature mapping make, because in \cref{fig:z_feature_map_reps_2_train}, we see the same \textit{Z Feature Mapping} but with \textit{2 repetitions}. The repetition is still not enough to not hinder the model's training as the classical part, the first row fails to separate the data. But in this case, we can see that after the feature mapping was applied the data is more spread through the plots in the second row, but still, we observe the telltale ring-like structure of feature mapping with only Z rotations. And again, this mapping has hindered the model's training, this is evident from the bottom right plot.
\begin{figure}[htbp]
    \centering

    \begin{subfigure}[t]{0.45\textwidth}
        \centering
        \includegraphics[width=\textwidth]{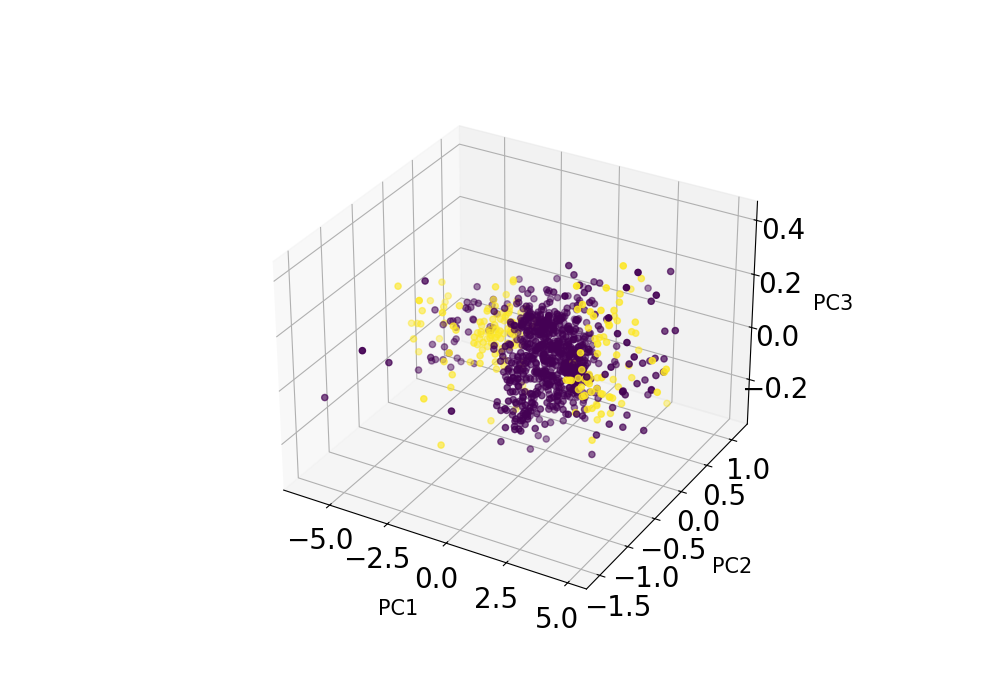}
        \caption{Principal Component Analysis After Classical Part, Color-coded by Fitted Values}
    \end{subfigure}
    \hfill
    \begin{subfigure}[t]{0.45\textwidth}
        \centering
        \includegraphics[width=\textwidth]{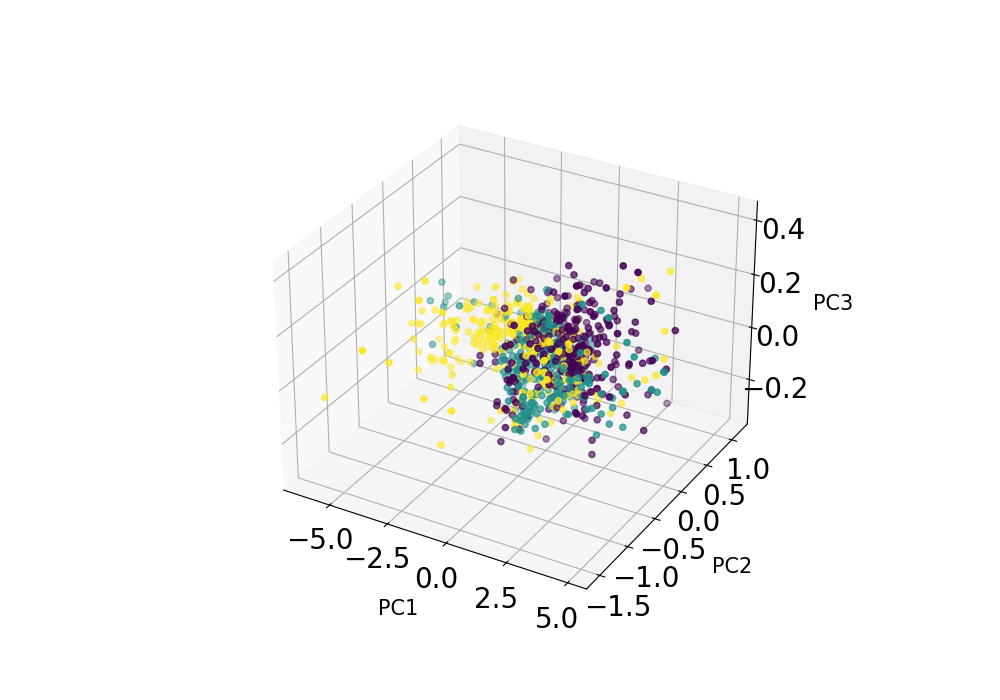}
        \caption{Principal Component Analysis After Classical Part, Color-coded by Training Values}
    \end{subfigure}

    \vspace{1em}

    \begin{subfigure}[t]{0.45\textwidth}
        \centering
        \includegraphics[width=\textwidth]{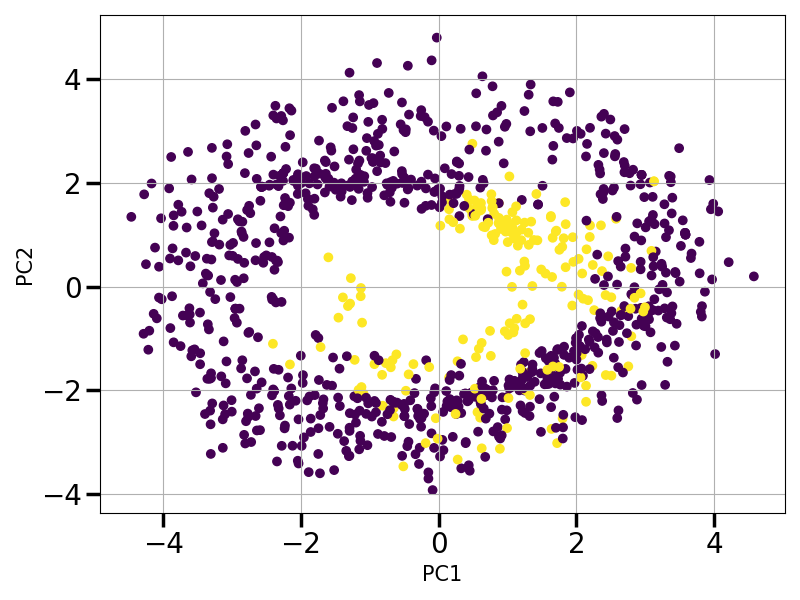}
        \caption{Principal Component Analysis After Feature Mapping, Color-coded by Fitted Values}
    \end{subfigure}
    \hfill
    \begin{subfigure}[t]{0.45\textwidth}
        \centering
        \includegraphics[width=\textwidth]{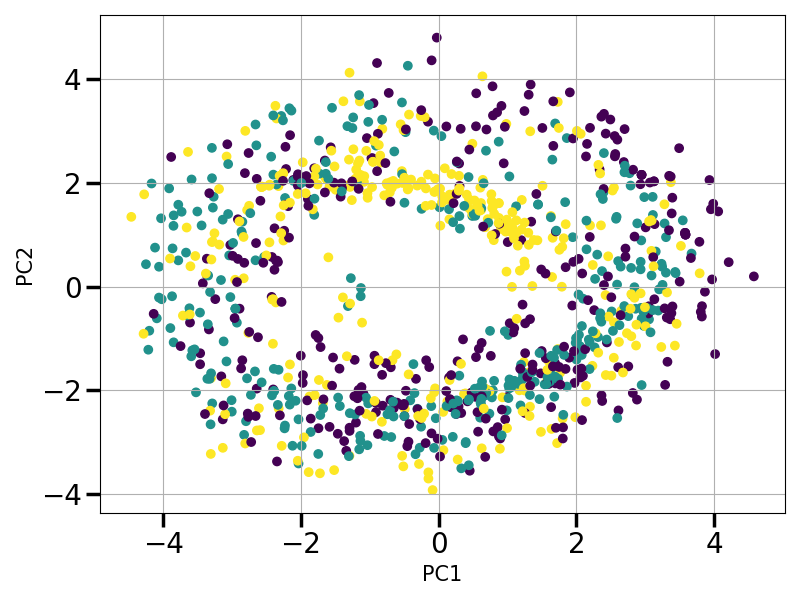}
        \caption{Principal Component Analysis After Feature Mapping, Color-coded by Training Values}
    \end{subfigure}

    \vspace{1em}

    \begin{subfigure}[t]{0.45\textwidth}
        \centering
        \includegraphics[width=\textwidth]{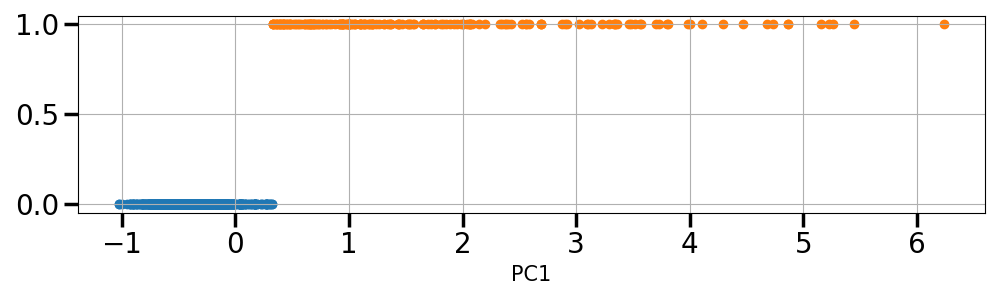}
        \caption{Principal Component Analysis After QNN, Color-coded by Fitted Values}
    \end{subfigure}
    \hfill
    \begin{subfigure}[t]{0.45\textwidth}
        \centering
        \includegraphics[width=\textwidth]{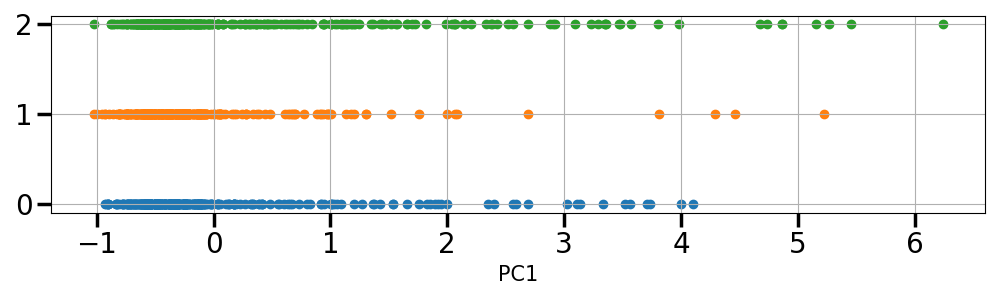}
        \caption{Principal Component Analysis After QNN, Color-coded by Training Values}
    \end{subfigure}

    \caption{PCA Progression for \texttt{z\_feature\_map\_reps\_3} — \textbf{Training Data}}
    \label{fig:z_feature_map_reps_3_train}
\end{figure}
So, we saw that adding repetition helped with spreading the data after the application of feature mapping. From this, it seems it would be a good idea to add one more repetition, to see if it will not finally make this type of mapping usable. But sadly, this is not the case as is evident from \cref{fig:z_feature_map_reps_3_train}. Where the outcome is even worse than in the previous two cases, as now, the model has failed to distinguish between all three classes and collapsed the output to just two. While the ring-like structure is still present, we see that this mapping really does not work, as is evident when the data is color-coded by their training labels, visible in the right corner. From this we can conclude, that feature mapping consisting of just single Z rotations is not really viable for our use case.

Well, simple feature mapping with just Z rotations did not work, but in Qiskit there is also ZZ feature mapping available, let's try them then. The first attempt is visible in \cref{fig:zz_feature_map_reps_1_linear_entanglement_train}. In this case, we selected just one repetition and linear entanglement. Again we see similar problems as in \ref{fig:z_feature_map_reps_1_train}, as the data is not separable after the classical layer, creates a half-ring structure after feature mapping, where the data is compressed into a thin shape, and after the quantum layer, the clusters that were fitted are in no way equal to the training labels. This all tells us that even when using one repetition of feature mapping consisting of a second-order Pauli-Z evolution circuit the model is unable to learn.

\begin{figure}[htbp]
    \centering

    \begin{subfigure}[t]{0.45\textwidth}
        \centering
        \includegraphics[width=\textwidth]{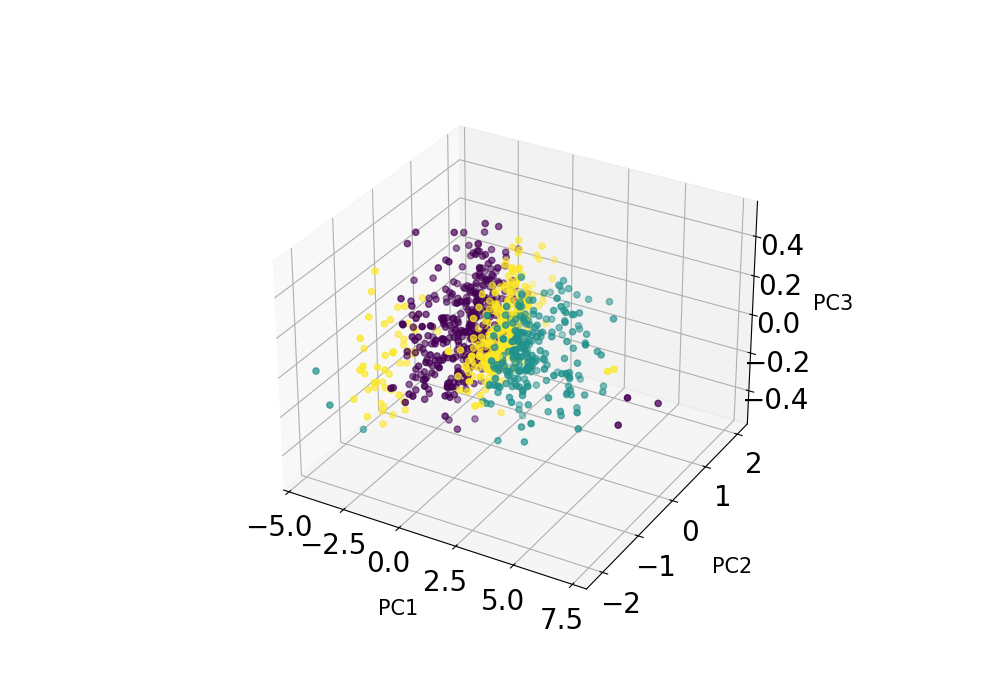}
        \caption{Principal Component Analysis After Classical Part, Color-coded by Fitted Values}
    \end{subfigure}
    \hfill
    \begin{subfigure}[t]{0.45\textwidth}
        \centering
        \includegraphics[width=\textwidth]{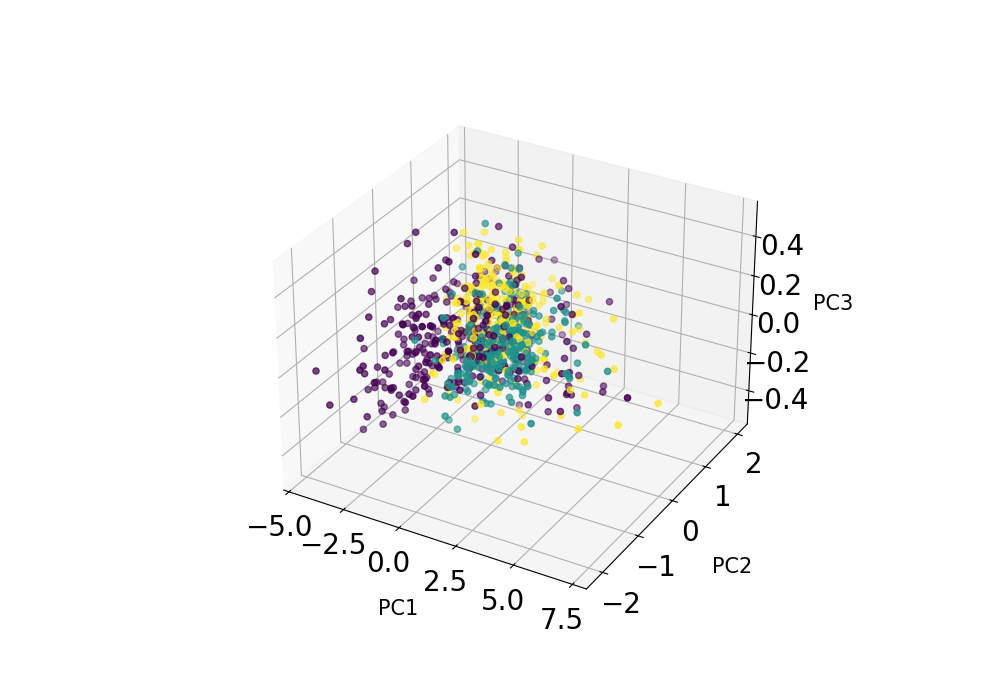}
        \caption{Principal Component Analysis After Classical Part, Color-coded by Training Values}
    \end{subfigure}

    \vspace{1em}

    \begin{subfigure}[t]{0.45\textwidth}
        \centering
        \includegraphics[width=\textwidth]{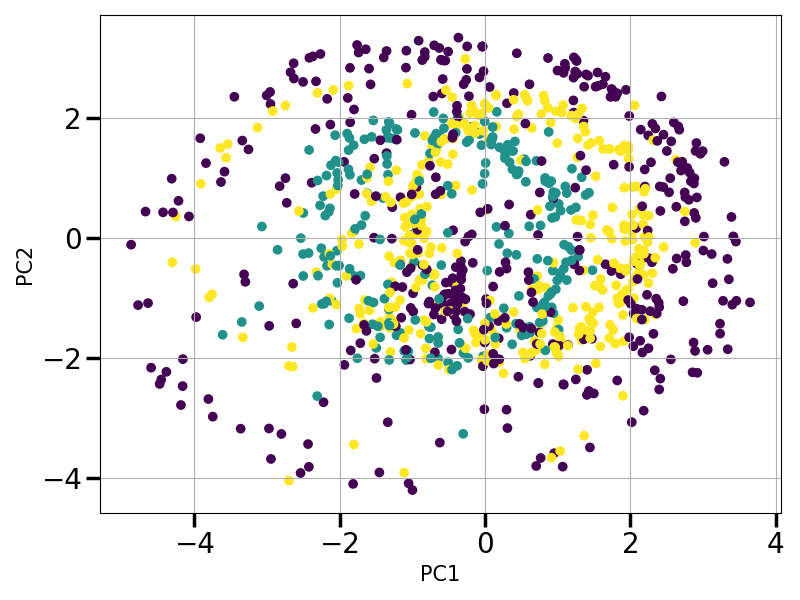}
        \caption{Principal Component Analysis After Feature Mapping, Color-coded by Fitted Values}
    \end{subfigure}
    \hfill
    \begin{subfigure}[t]{0.45\textwidth}
        \centering
        \includegraphics[width=\textwidth]{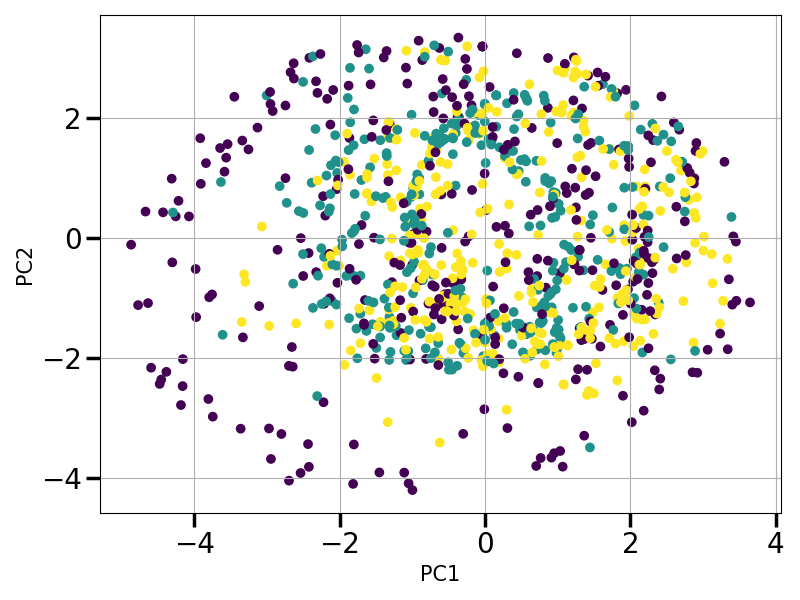}
        \caption{Principal Component Analysis After Feature Mapping, Color-coded by Training Values}
    \end{subfigure}

    \vspace{1em}

    \begin{subfigure}[t]{0.45\textwidth}
        \centering
        \includegraphics[width=\textwidth]{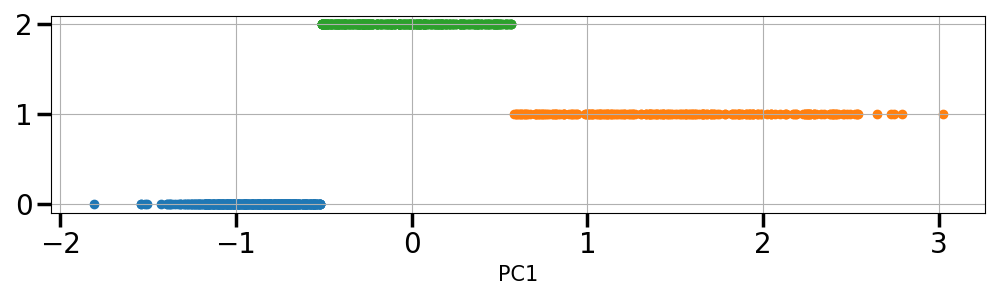}
        \caption{Principal Component Analysis After QNN, Color-coded by Fitted Values}
    \end{subfigure}
    \hfill
    \begin{subfigure}[t]{0.45\textwidth}
        \centering
        \includegraphics[width=\textwidth]{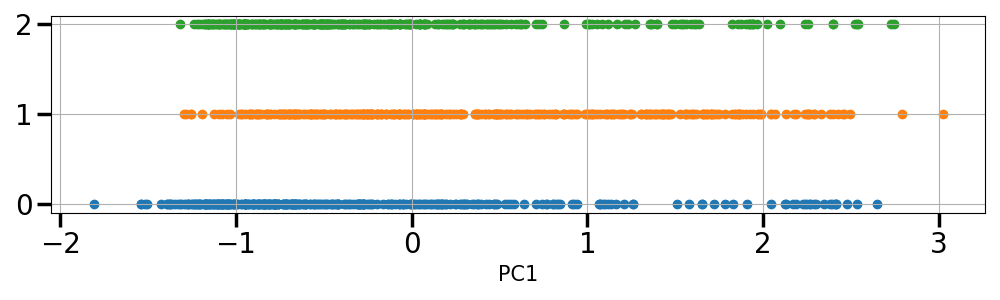}
        \caption{Principal Component Analysis After QNN, Color-coded by Training Values}
    \end{subfigure}

    \caption{PCA Progression for \texttt{zz\_feature\_map\_reps\_2\_linear} — \textbf{Training Data}}
    \label{fig:zz_feature_map_reps_2_linear_train}
\end{figure}

\begin{figure}[htbp]
    \centering

    \begin{subfigure}[t]{0.45\textwidth}
        \centering
        \includegraphics[width=\textwidth]{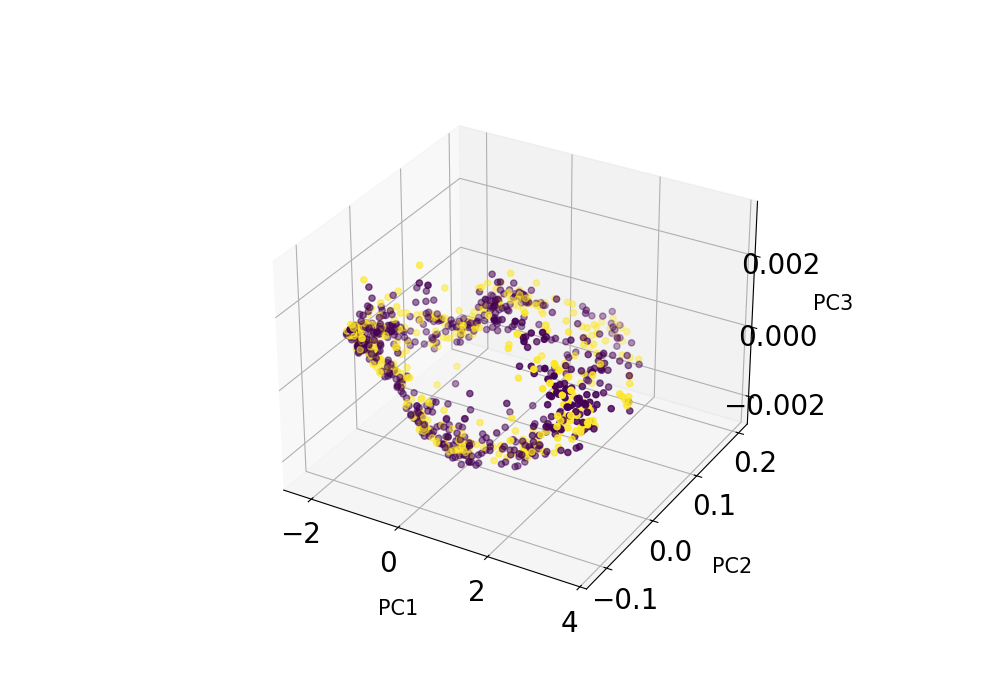}
        \caption{Principal Component Analysis After Classical Part, Color-coded by Fitted Values}
    \end{subfigure}
    \hfill
    \begin{subfigure}[t]{0.45\textwidth}
        \centering
        \includegraphics[width=\textwidth]{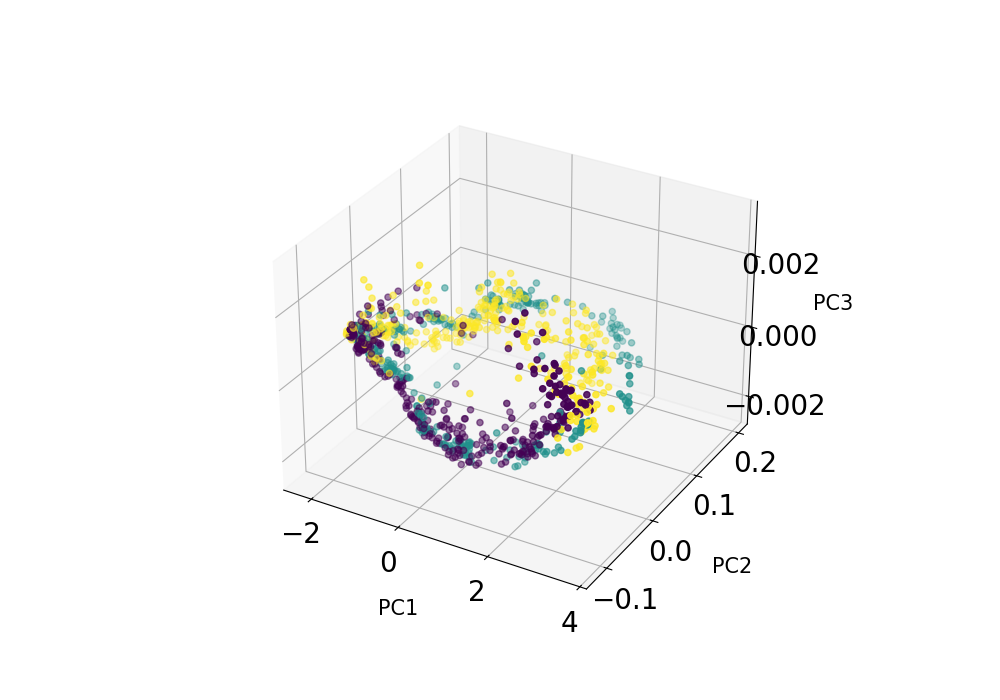}
        \caption{Principal Component Analysis After Classical Part, Color-coded by Training Values}
    \end{subfigure}

    \vspace{1em}

    \begin{subfigure}[t]{0.45\textwidth}
        \centering
        \includegraphics[width=\textwidth]{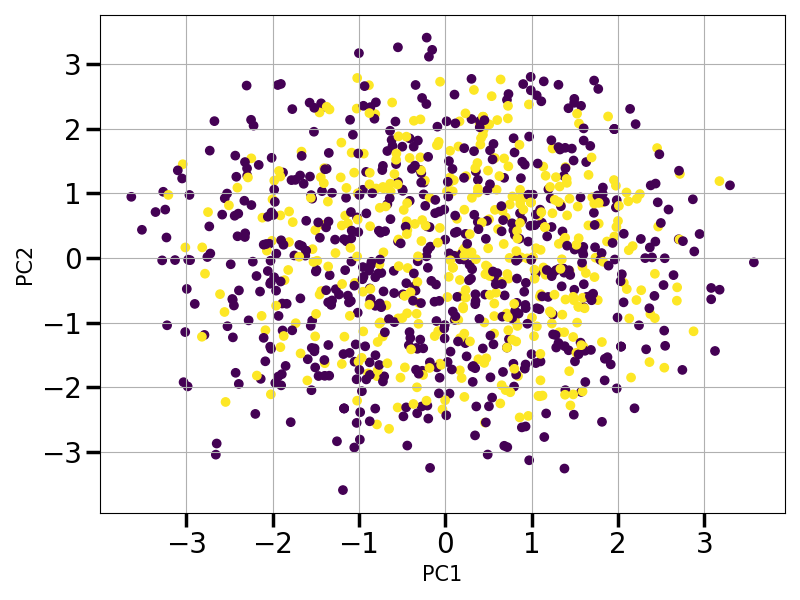}
        \caption{Principal Component Analysis After Feature Mapping, Color-coded by Fitted Values}
    \end{subfigure}
    \hfill
    \begin{subfigure}[t]{0.45\textwidth}
        \centering
        \includegraphics[width=\textwidth]{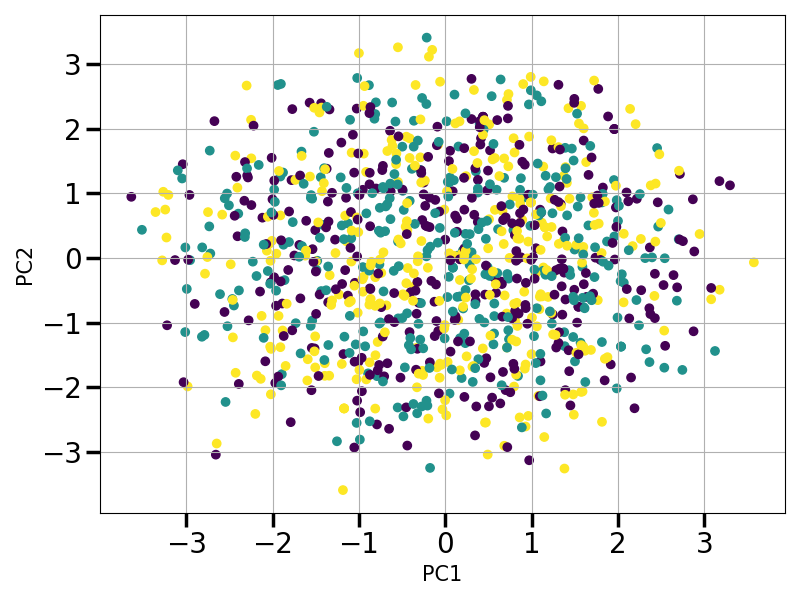}
        \caption{Principal Component Analysis After Feature Mapping, Color-coded by Training Values}
    \end{subfigure}

    \vspace{1em}

    \begin{subfigure}[t]{0.45\textwidth}
        \centering
        \includegraphics[width=\textwidth]{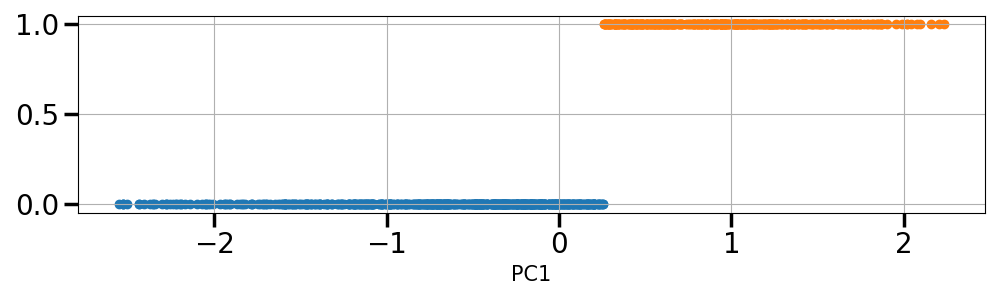}
        \caption{Principal Component Analysis After QNN, Color-coded by Fitted Values}
    \end{subfigure}
    \hfill
    \begin{subfigure}[t]{0.45\textwidth}
        \centering
        \includegraphics[width=\textwidth]{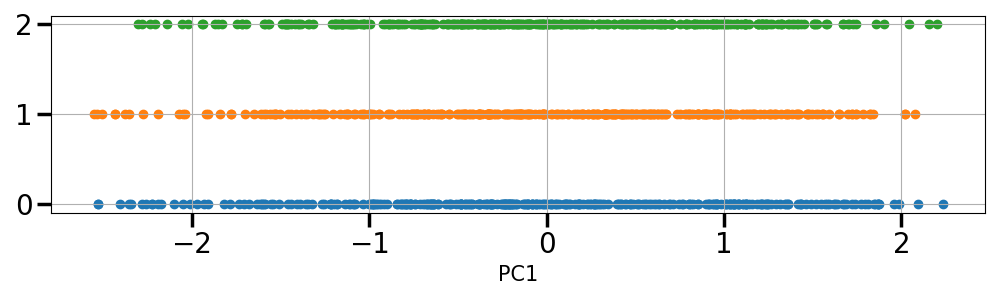}
        \caption{Principal Component Analysis After QNN, Color-coded by Training Values}
    \end{subfigure}

    \caption{PCA Progression for \texttt{zz\_feature\_map\_reps\_3\_full} — \textbf{Training Data}}
    \label{fig:zz_feature_map_reps_3_full_train}
\end{figure}

\begin{figure}[htbp]
    \centering

    \begin{subfigure}[t]{0.45\textwidth}
        \centering
        \includegraphics[width=\textwidth]{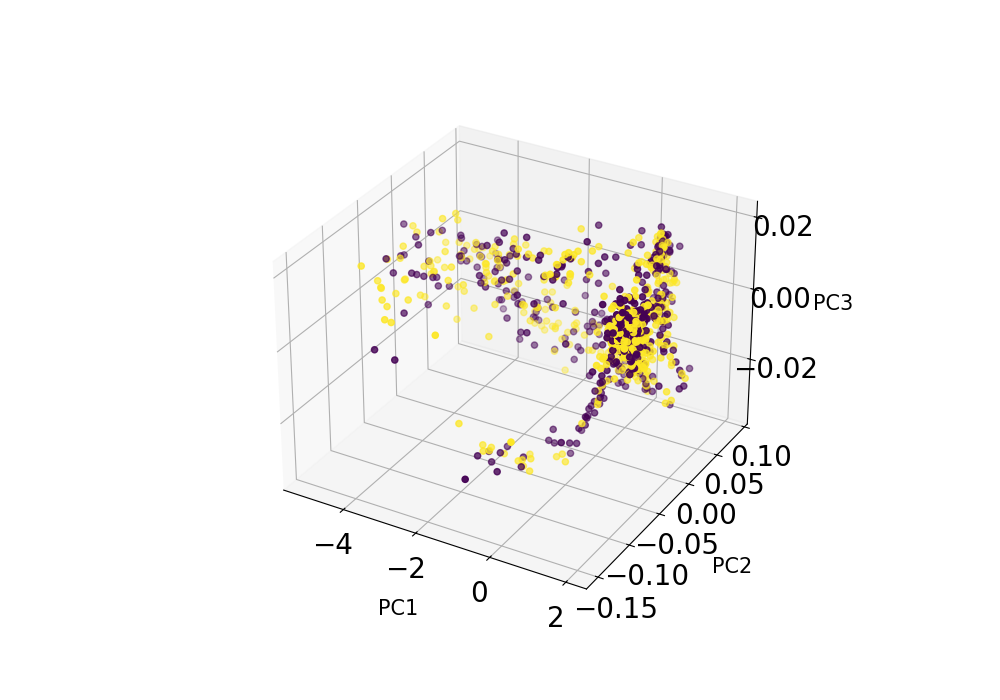}
        \caption{Principal Component Analysis After Classical Part, Color-coded by Fitted Values}
    \end{subfigure}
    \hfill
    \begin{subfigure}[t]{0.45\textwidth}
        \centering
        \includegraphics[width=\textwidth]{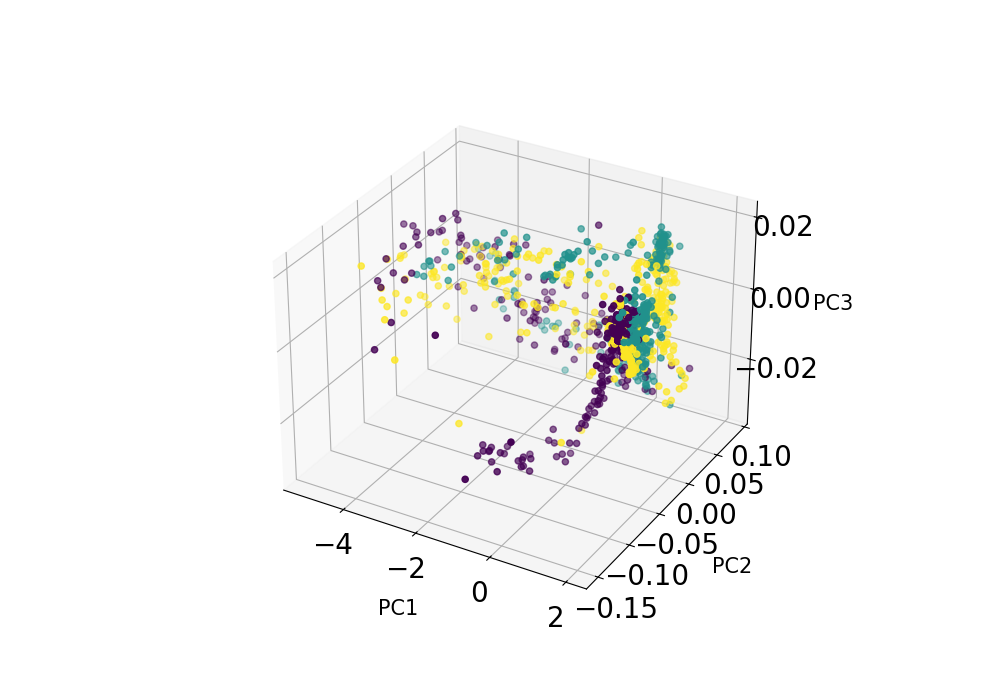}
        \caption{Principal Component Analysis After Classical Part, Color-coded by Training Values}
    \end{subfigure}

    \vspace{1em}

    \begin{subfigure}[t]{0.45\textwidth}
        \centering
        \includegraphics[width=\textwidth]{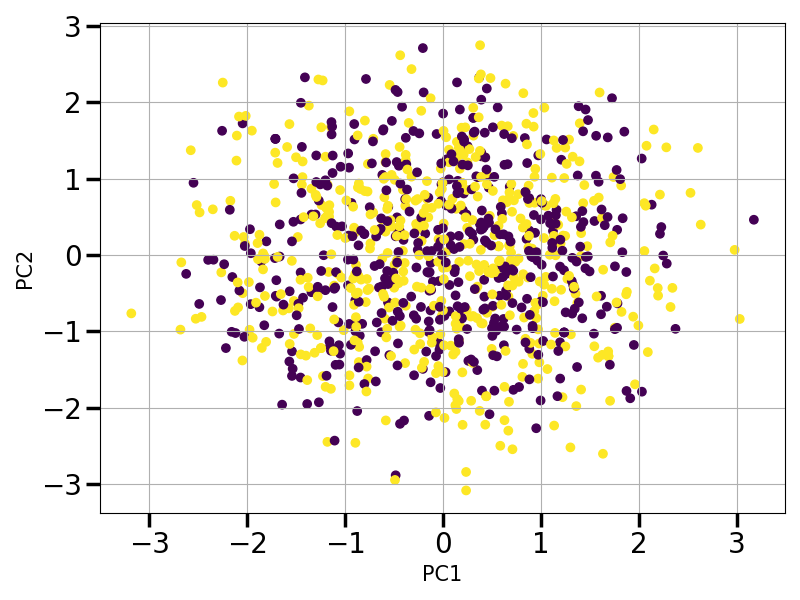}
        \caption{Principal Component Analysis After Feature Mapping, Color-coded by Fitted Values}
    \end{subfigure}
    \hfill
    \begin{subfigure}[t]{0.45\textwidth}
        \centering
        \includegraphics[width=\textwidth]{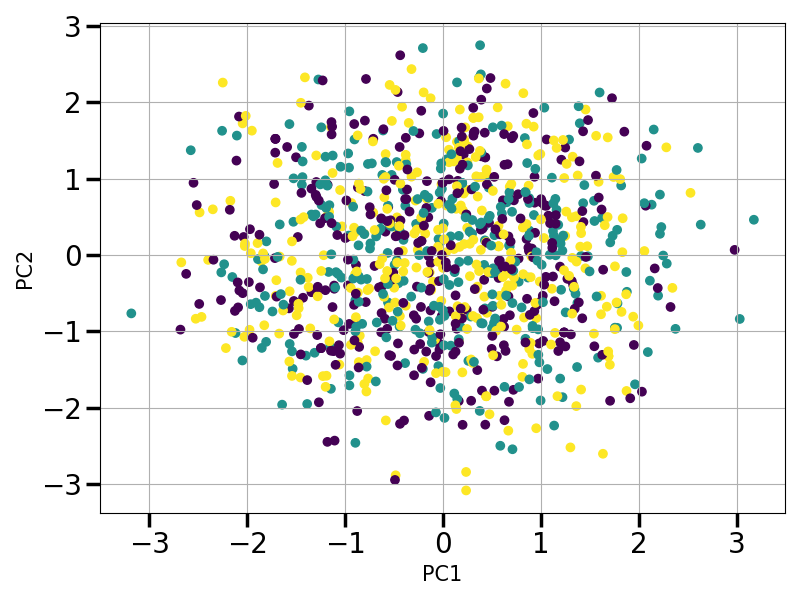}
        \caption{Principal Component Analysis After Feature Mapping, Color-coded by Training Values}
    \end{subfigure}

    \vspace{1em}

    \begin{subfigure}[t]{0.45\textwidth}
        \centering
        \includegraphics[width=\textwidth]{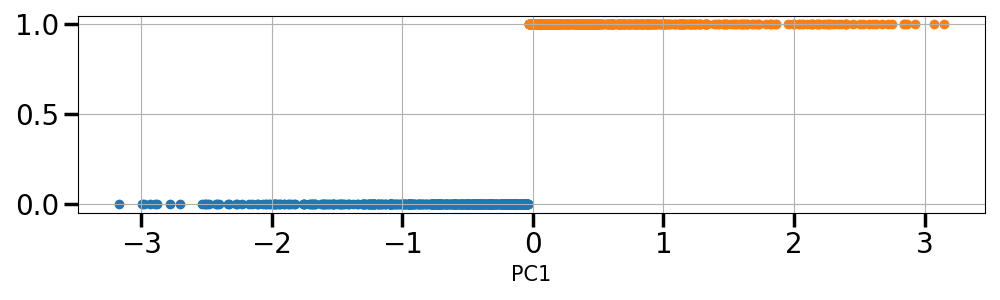}
        \caption{Principal Component Analysis After QNN, Color-coded by Fitted Values}
    \end{subfigure}
    \hfill
    \begin{subfigure}[t]{0.45\textwidth}
        \centering
        \includegraphics[width=\textwidth]{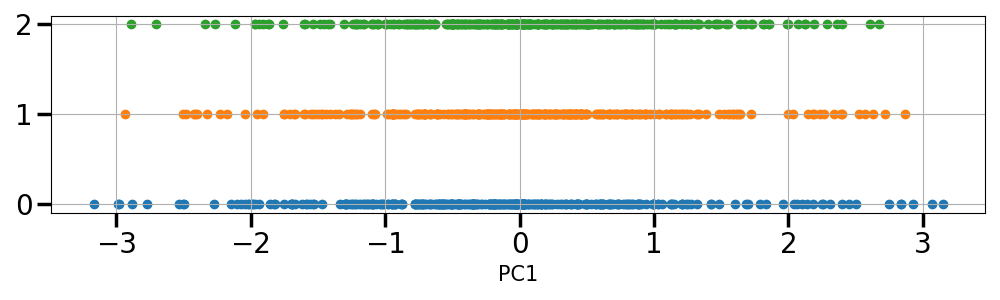}
        \caption{Principal Component Analysis After QNN, Color-coded by Training Values}
    \end{subfigure}

    \caption{PCA Progression for \texttt{pauli\_z\_yy\_zxz\_linear} — \textbf{Training Data}}
    \label{fig:pauli_z_yy_zxz_linear_train}
\end{figure}

But let's try a similar approach as before, let's increase the number of repetitions to two. Here, in \cref{fig:zz_feature_map_reps_2_linear_train}, we see one slight improvement and that is the fact, that when comparing the fitted values and training labels, some similarity can be seen. While the data still forms the typical ring-like structure, one class is mapped into a ring with a larger diameter. But in the end, this approach still failed, as even two repetitions are not enough to work in the model.
\begin{figure}[htbp]
    \centering

    \begin{subfigure}[t]{0.45\textwidth}
        \centering
        \includegraphics[width=\textwidth]{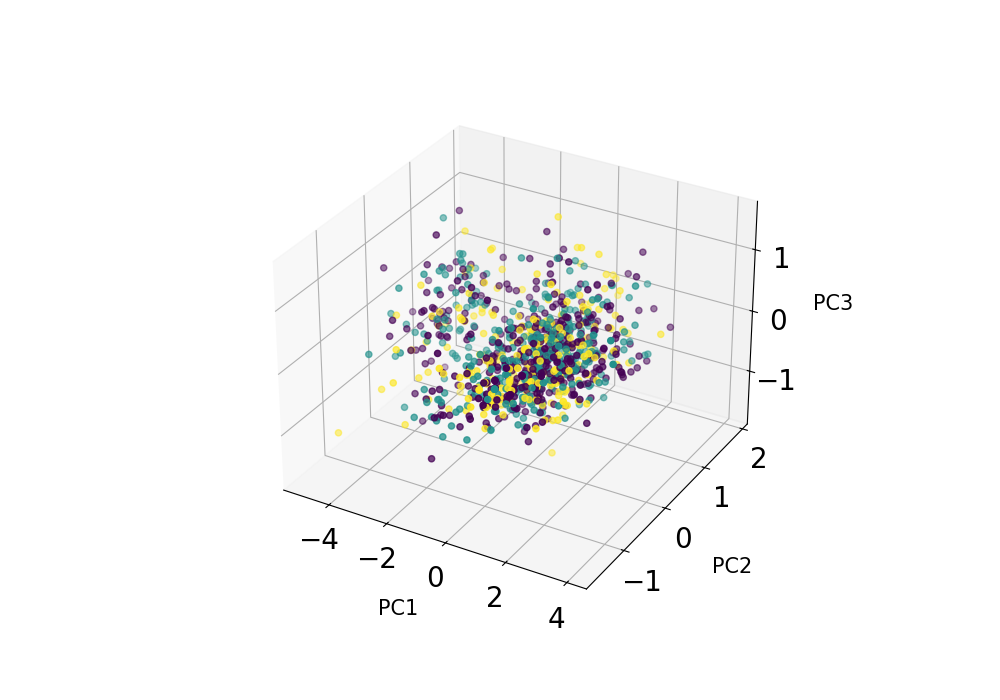}
        \caption{Principal Component Analysis After Classical Part, Color-coded by Fitted Values}
    \end{subfigure}
    \hfill
    \begin{subfigure}[t]{0.45\textwidth}
        \centering
        \includegraphics[width=\textwidth]{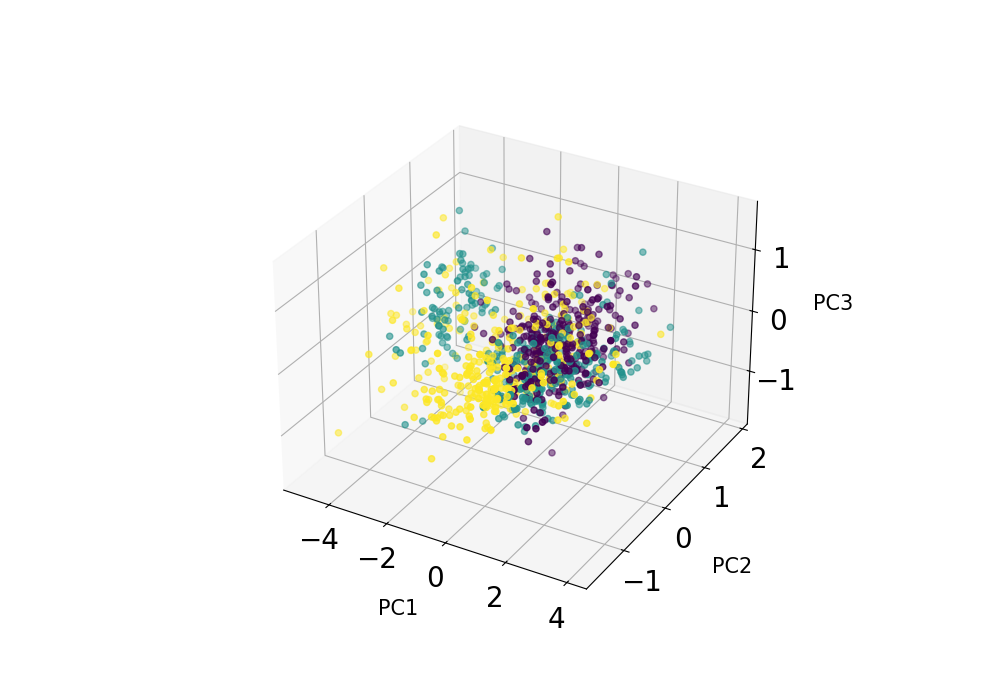}
        \caption{Principal Component Analysis After Classical Part, Color-coded by Training Values}
    \end{subfigure}

    \vspace{1em}

    \begin{subfigure}[t]{0.45\textwidth}
        \centering
        \includegraphics[width=\textwidth]{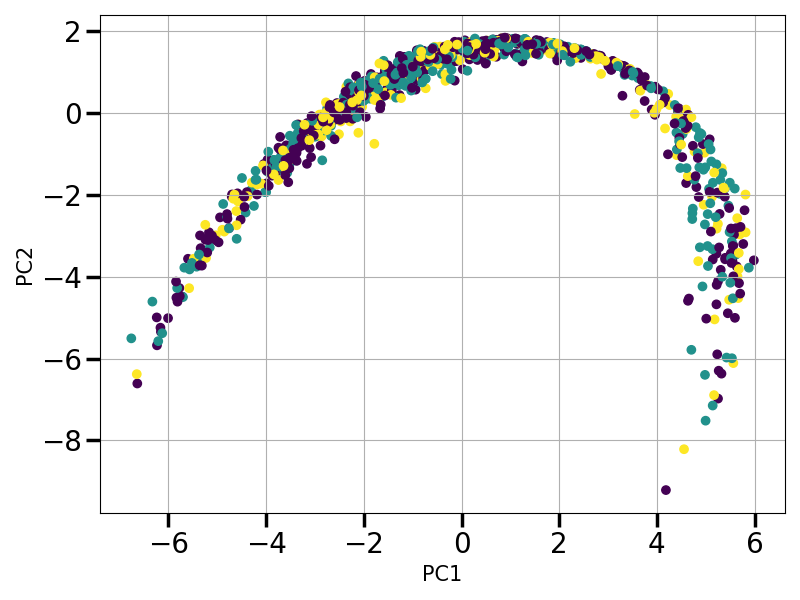}
        \caption{Principal Component Analysis After Feature Mapping, Color-coded by Fitted Values}
    \end{subfigure}
    \hfill
    \begin{subfigure}[t]{0.45\textwidth}
        \centering
        \includegraphics[width=\textwidth]{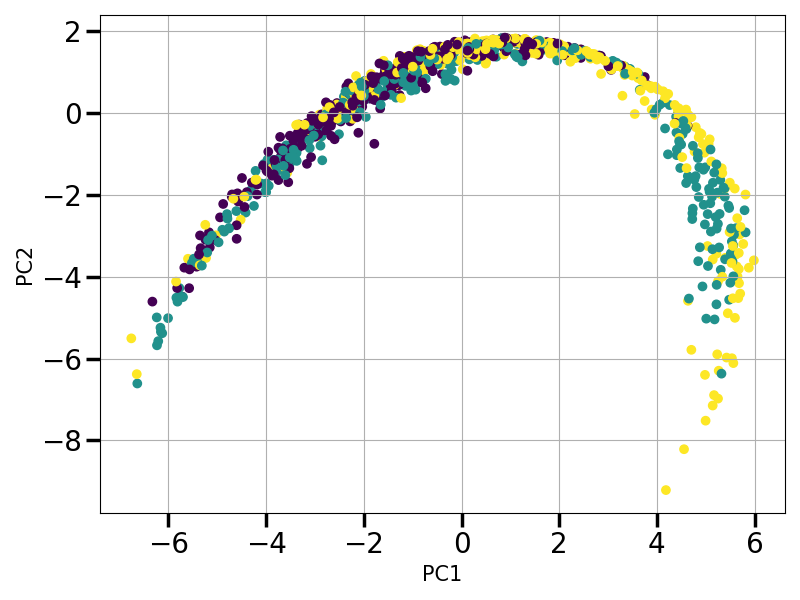}
        \caption{Principal Component Analysis After Feature Mapping, Color-coded by Training Values}
    \end{subfigure}

    \vspace{1em}

    \begin{subfigure}[t]{0.45\textwidth}
        \centering
        \includegraphics[width=\textwidth]{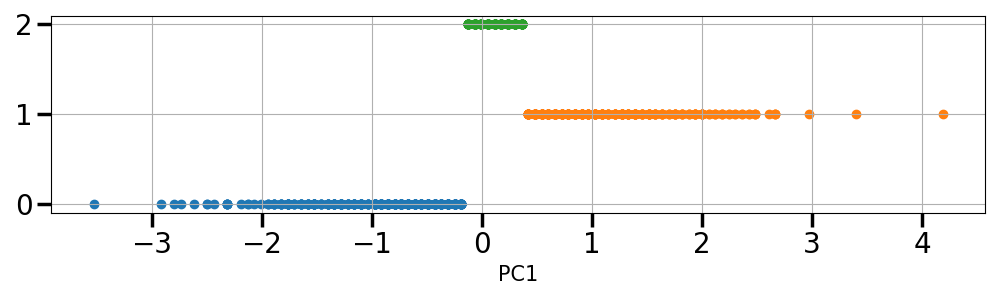}
        \caption{Principal Component Analysis After QNN, Color-coded by Fitted Values}
    \end{subfigure}
    \hfill
    \begin{subfigure}[t]{0.45\textwidth}
        \centering
        \includegraphics[width=\textwidth]{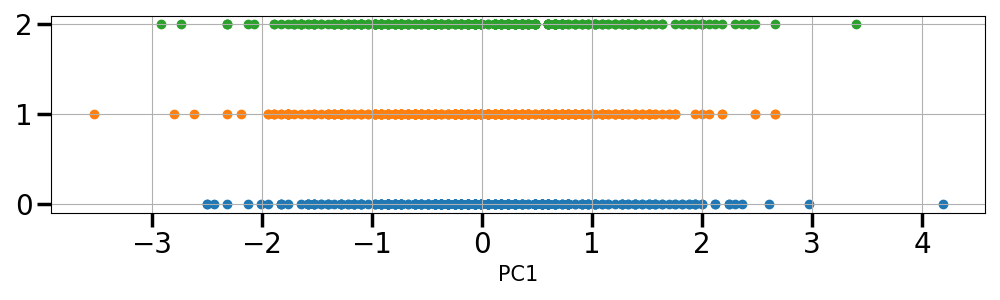}
        \caption{Principal Component Analysis After QNN, Color-coded by Training Values}
    \end{subfigure}

    \caption{PCA Progression for \texttt{zz\_feature\_map\_reps\_1\_linear\_entanglement} — \textbf{Training Data}}
    \label{fig:zz_feature_map_reps_1_no_entanglement_train}
\end{figure}
For the last step with the ZZ feature mapping, we examined the case with three repetitions and in this case full entanglement. The initial assumption was, that the fully entangled circuit would allow for more complex relations between the data to be detected, but that ended up not being the case, as can be seen in \cref{fig:zz_feature_map_reps_3_full_train}. Here we see a similarity with \cref{fig:z_feature_map_reps_3_train}, because again, the model collapsed in dimensionality, detecting just two distinct classes instead of three. And even when we examine the data in the right column, color-coded by training labels, we realize, that there is no separability of the data and that the model failed completely, meaning that the increase in complexity of the encoding ended up being harmful, and thus points to the fact, that blind increase in complexity has no beneficial yields.
\begin{figure}[htbp]
    \centering

    \begin{subfigure}[t]{0.45\textwidth}
        \centering
        \includegraphics[width=\textwidth]{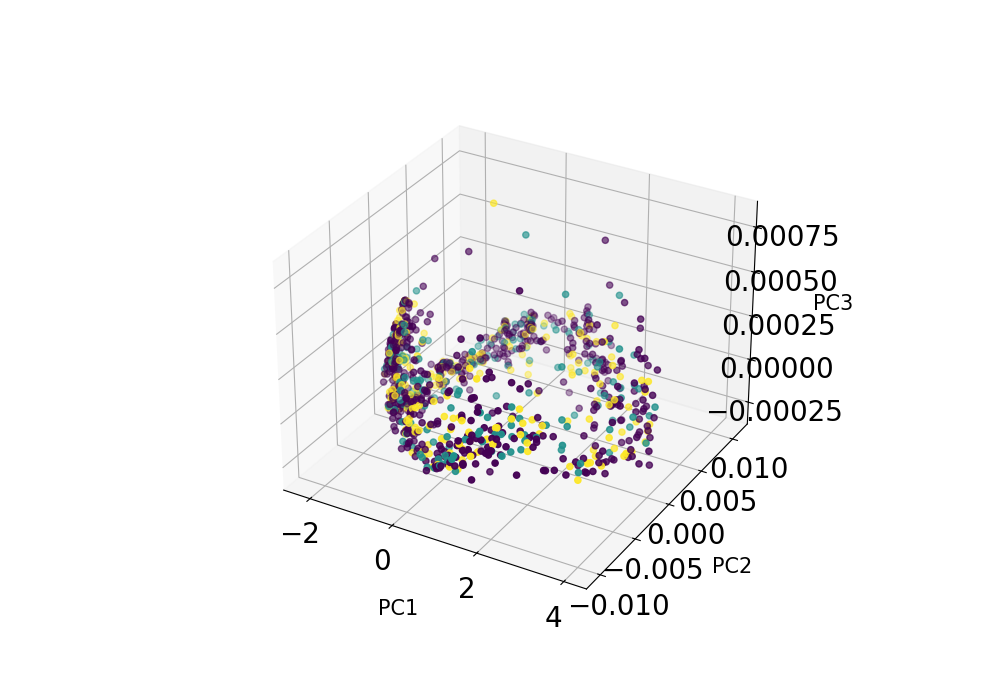}
        \caption{Principal Component Analysis After Classical Part, Color-coded by Fitted Values}
    \end{subfigure}
    \hfill
    \begin{subfigure}[t]{0.45\textwidth}
        \centering
        \includegraphics[width=\textwidth]{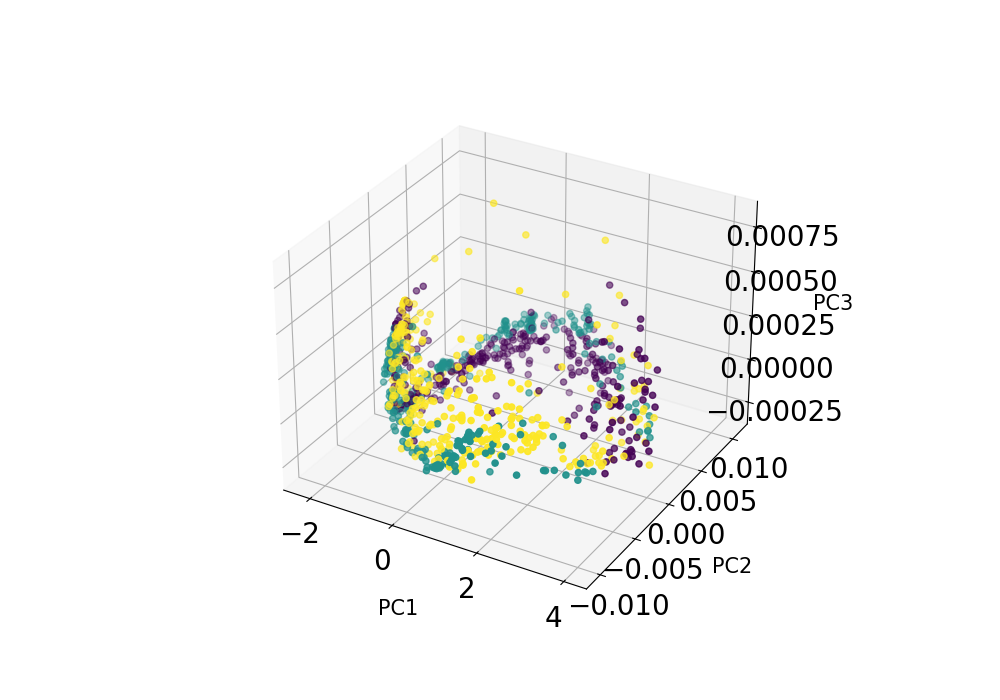}
        \caption{Principal Component Analysis After Classical Part, Color-coded by Training Values}
    \end{subfigure}

    \vspace{1em}

    \begin{subfigure}[t]{0.45\textwidth}
        \centering
        \includegraphics[width=\textwidth]{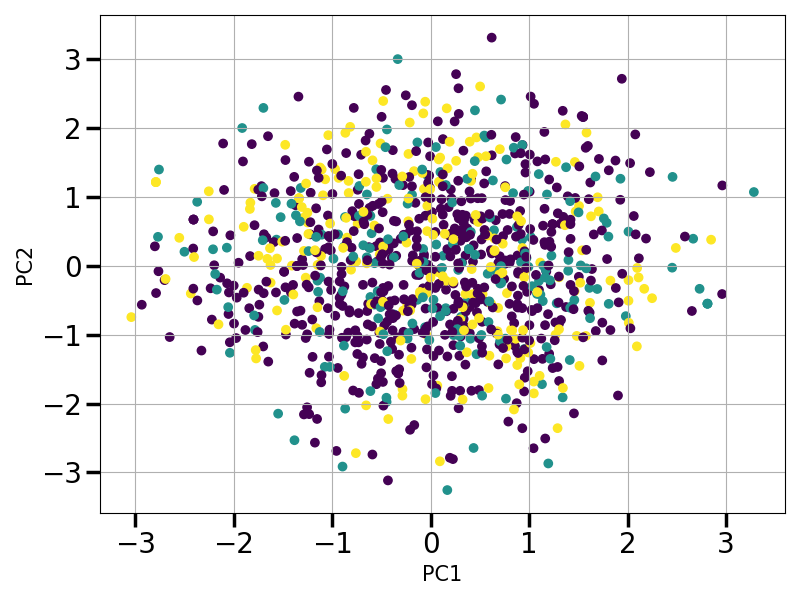}
        \caption{Principal Component Analysis After Feature Mapping, Color-coded by Fitted Values}
    \end{subfigure}
    \hfill
    \begin{subfigure}[t]{0.45\textwidth}
        \centering
        \includegraphics[width=\textwidth]{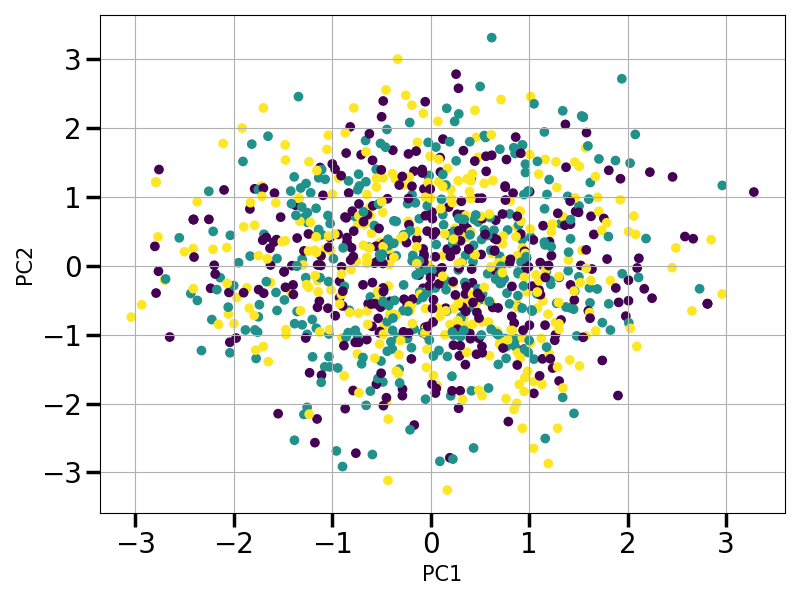}
        \caption{Principal Component Analysis After Feature Mapping, Color-coded by Training Values}
    \end{subfigure}

    \vspace{1em}

    \begin{subfigure}[t]{0.45\textwidth}
        \centering
        \includegraphics[width=\textwidth]{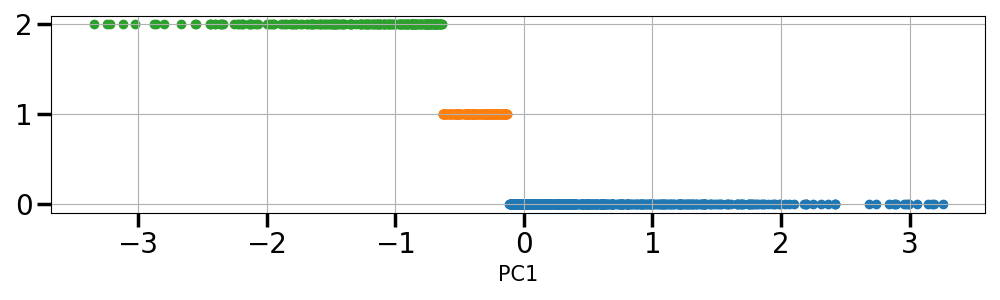}
        \caption{Principal Component Analysis After QNN, Color-coded by Fitted Values}
    \end{subfigure}
    \hfill
    \begin{subfigure}[t]{0.45\textwidth}
        \centering
        \includegraphics[width=\textwidth]{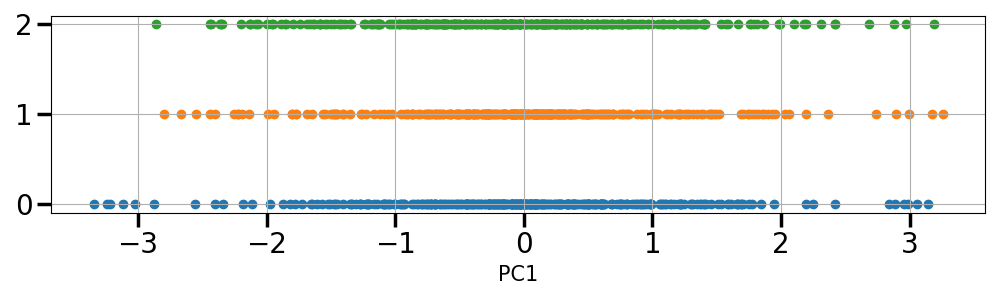}
        \caption{Principal Component Analysis After QNN, Color-coded by Training Values}
    \end{subfigure}

    \caption{PCA Progression for \texttt{pauli\_z\_yy\_zxz\_rep\_2} — \textbf{Training Data}}
    \label{fig:pauli_z_yy_zxz_rep_2_train}
\end{figure}
The penultimate feature mapping that we examined was the Pauli mapping with Z, YY, and ZXZ rotations, here we tried to introduce rotations along all axes, as in the first and only successful case. While using only one repetition, the model again failed to differentiate all three classes and collapsed to two dimensions only as seen in \cref{fig:pauli_z_yy_zxz_linear_train}. Once again we must conclude, that an incorrectly chosen feature map hinders the whole model.

And lastly, we examined whether we can at least improve the model a bit if we introduce another repetition of the same feature mapping. As displayed in \cref{fig:pauli_z_yy_zxz_rep_2_train}, we see one improvement, and that is that in the case of the addition of the repetition the model is at least able to correctly learn, that there are three output classes instead of just one, but alas, the accuracy remains very low.

This part of thorough analysis tells us, that feature mapping in hybrid neural networks must be carefully selected, and that one can't blindly increase its complexity and expect better results.

\section{Fisher Discriminant ratio}

But for the sake of completion let's have a look at one more metric, by which we can evaluate the different feature mappings. That is the Fisher Discriminant Ratio \cite{kim2005robust}, which is a statistical measure that we use to evaluate the separability between two classes based on one selected feature.
Between two classes, it is defined by
\begin{equation}
\text{FDR} = \frac{(\mu_1 - \mu_2)^2}{\sigma_1^2 + \sigma_2^2},
\end{equation}
where $\mu_1$ and $\mu_2$ are the means of the feature values for class 1 and class 2, respectively, $\sigma_1^2$ and $\sigma_2^2$ are the variances of the feature values for class 1 and class 2. A higher score suggests that the class separability is higher, indicating it is this feature that gives us more value when we want to determine how to distinguish between the classes. 

In \cref{tab:fdr} we present the different Fisher Discriminant ratios between all possible pairs of classes. The one feature map that stands out as the best, is the only feature map that was well-suited for our scenario and was beneficial to the training of the model, and that is the Pauli map with X, Y, and Z rotations. On the other hand, three maps showcase values of 1.0 consistently, indicating poor or no separability, while others have higher values, but from our previous analysis, we know that there were serious problems with them, from which we can concur, that the separability they claim to have based on the Fisher Discriminant ratio is rather happenstance than the systematic result.
\begin{table}[h]
\centering
\resizebox{\textwidth}{!}{%
\begin{tabular}{|l|c|c|c|c|c|c|c|}
\hline
\textbf{Feature Map} & \textbf{0 vs 1} & \textbf{0 vs 2} & \textbf{1 vs 0} & \textbf{1 vs 2} & \textbf{2 vs 0} & \textbf{2 vs 1} & \textbf{Avg FDR} \\
\hline
zz\_feature\_map\_reps\_1\_linear & 16.9932 & 16.8899 & 12.2353 & 11.7945 & 13.8974 & 13.4108 & 14.2035 \\
zz\_feature\_map\_reps\_2\_linear & 23.8764 & 22.9970 & 33.3721 & 36.3299 & 29.2346 & 32.3274 & 29.6896 \\
zz\_feature\_map\_reps\_3\_full & 1.0000 & 1.0000 & 1.0000 & 1.0000 & 1.0000 & 1.0000 & 1.0000 \\
z\_feature\_map\_reps\_1 & 17.0916 & 17.1477 & 17.8302 & 17.8370 & 15.4481 & 15.3909 & 16.7909 \\
z\_feature\_map\_reps\_2 & 6.4512 & 7.4356 & 5.6529 & 5.6185 & 7.6743 & 6.5337 & 6.5610 \\
z\_feature\_map\_reps\_3 & 33.3170 & 30.8994 & 30.6517 & 29.1383 & 28.2103 & 28.4005 & 30.1029 \\
pauli\_xyz\_1\_rep & 168.7070 & 69.1871 & 40.4918 & 80.1410 & 23.2032 & 112.6916 & 82.4036 \\
pauli\_z\_yy\_zxz\_linear & 1.0000 & 1.0000 & 1.0000 & 1.0000 & 1.0000 & 1.0000 & 1.0000 \\
pauli\_z\_yy\_zxz\_rep\_2 & 1.0000 & 1.0000 & 1.0000 & 1.0000 & 1.0000 & 1.0000 & 1.0000 \\
\hline
\end{tabular}%
}
\caption{Feature map comparison by FDR}
\label{tab:fdr}
\end{table}

Based on the Fisher Discriminant ratios we calculated, we can make this conclusion. The Pauli mapping with X, Y, and Z rotations has the largest discriminative power from all of the tested feature mappings, with an average of 82.4, which indicates excellent separation between all classes.

\chapter*{Conclusion}\label{chap:conc}\addcontentsline{toc}{chapter}{\protect\numberline{}Conclusion}
Based on the previous analysis of our hybrid neural network, we can deduce several interesting conclusions that will provide insights for those who want to incorporate quantum layers into their machine-learning models for classification.

The first point is that when we increase the depth of the ansatz, we gain several advantages, such as better generalization ability of our model and greater stability. An increase in the ansatz depth with more parameters consistently improves validation accuracy, reduces the generalization gap, and affects the overall stability of the training process. These advantages, come at a slight cost of initially reduced learning speed, which is negligible, as after a while, the learning speed increases, and the model with most parameters in the quantum layer was the first to achieve an accuracy of over $90\%$.

One thing we have to keep in mind is the fact that the returns we get from increasing the number of ansatz repetitions are diminishing, as the largest leap occurs when we increase the number of repetitions from one to two, while the addition of the third repetition shows overall smaller improvements. This suggests, that if we continued to increase the number of parameters in the quantum layer, we would reach the territory of no advantage gained, or even the lands, where the increase in the dimensionality of the problem, would decrease the model's reliability. 

What is one advantage that we observe when adding the third repetition is the fact, that the longer quantum circuits act as regularizers, that implicitly smooth the learning process and reduce the overfitting without any need to employ explicit regularization. The third repetition of the ansatz also led to the smoothest training curve, with the smallest fluctuation for both the validation and training data. This points to the fact that greater flexibility in quantum circuits avoids erratic learning processes.

Also, when we realize that the stability ratio drops below 1 when using three repetitions in the quantum circuit, we gain a well-generalizing model, as the model is more stable on the validation data than on the training ones.

The next large point we can make is the fact that the selection of feature mapping is crucial for the model's learning process. From the nine tested feature mappings only one, the Pauli mapping with X, Y, and Z rotations, resulted in success. This observation highlights the critical role that feature mapping plays in hybrid neural networks, and that the choice of unsuitable feature mapping can hinder the model completely.

Considering the different feature mappings, we observed that the simple Z-Rotation feature mapping is ineffective, as when we rely on just Z rotations, we do not obtain good enough data structures for a successful learning process, sometimes we even fail to distinguish between all the classes that are present in the training data. Even when we add multiple repetitions, the results are not improving and the models are showing very poor generalization ability.

Another point to make is the fact, that using more complex feature maps is not always better, as by increasing the entanglement or number of repetitions without proper thought and design, we can worsen the performance of our model instead of causing improvement. This shows, that when we want to introduce further complexity, we must do so carefully and with intent. The choice of incorrect feature mapping can lead to dimensional collapse. This is related to the curse of dimensionality, where an unnecessarily high-dimensional feature space can lead to sparsity, making it more difficult for classical optimizers to converge efficiently. Instead of capturing useful structure, the model may become over-parameterized, and the optimization landscape can become flat or highly non-convex. Therefore, careful feature map design is crucial to ensure both expressive power and trainability of the quantum model.

When testing different models, one has to be careful to allow for a sufficient number of epochs before making the decision whether this model is well designed or not. As in models with a larger number of ansatz parameters, we have a slower early learning slope, sometimes even we can also observe an initial decrease in accuracy. Ultimately, the models with more trainable parameters outperform the more shallow models, as they make use of the larger flexibility that they have thanks to the increased number of points of freedom. So one should be patient at the early training stage, to avoid unrightfully dismissing well-defined models.

When we perform \ac{pca} and calculate the Silhouette score, we can see that successful models create well-separated and structured feature space after the quantum layer, while models that are poorly performing show either chaotic data distribution or on the other hand tightly compressed data clusters, without any class separation. This again tells us that if we want our model to be successful, we need to choose feature mapping, that allows enough flexibility to separate the distinct data clusters.

We can also use Fisher Discriminant Ratio to confirm how well are the data separable between different classes, but we have to be careful not to depend just on this one metric, as it can show good enough separability even in the case where the model is ill-trained. Thus it is better in the case of a hybrid classifier to consider this as a secondary metric and not to make an important decision just based on it without further context.

Overall, we want to stress that the learning efficiency and the final performance of a hybrid, quantum-classical model are strongly linked to the choices we make when designing the quantum part. And it is wise to keep in mind, that the structure of quantum ansatz and choice of feature mapping does not affect just the final accuracies, but also the learning dynamics of the hybrid model. When choosing the feature mapping, we should aim to include multi-axis rotation, which seems to be essential for well trained model.

\section*{Future Outlook}

Building upon the results of this thesis, further work is planned to expand the study of hybrid quantum-classical neural networks. These models, which combine classical machine learning components with quantum layers, have shown potential in controlled experimental settings, but warrant further investigation to assess their scalability and versatility. The next steps include presenting these findings at a quantum computing symposium in Poland, where feedback from the community will help refine the methodology and highlight directions for future improvements.

Additional experiments will be conducted to explore more complex quantum circuits and larger datasets. This will allow for a deeper evaluation of how different quantum circuit architectures influence model performance, particularly in terms of expressivity, training stability, and generalization. Increasing dataset complexity will also test the robustness of the hybrid approach in more realistic learning scenarios and may uncover the practical limits of the current methods.

It is also intended to publish the extended results in conference proceedings, followed by the preparation of a full research paper. These dissemination efforts will contribute to the broader academic conversation surrounding quantum machine learning. Through these activities, the research aims to contribute further to the development of quantum-enhanced learning models, providing insights that support the long-term goal of integrating quantum computing into the toolbox of modern computational science.

\printbibliography[heading=bibintoc]


\end{document}